\documentclass{aastex61}

\newcommand{\awds}{AWDs}
\newcommand{\cvs}{CVs}
\newcommand{\ssss}{SSSs}
\newcommand{\syss}{symbiotic~stars}

\newcommand{\heao}{{\sl HEAO-1\/}}
\newcommand{\einstein}{{\sl Einstein\/}}
\newcommand{\exosat}{{\sl EXOSAT\/}}
\newcommand{\rosat}{{\sl ROSAT\/}}
\newcommand{\euve}{{\sl EUVE\/}}
\newcommand{\integral}{{\sl INTEGRAL\/}}
\newcommand{\swift}{{\sl Swift\/}}
\newcommand{\ginga}{{\sl Ginga\/}}
\newcommand{\asca}{{\sl ASCA\/}}
\newcommand{\xte}{{\sl RXTE\/}}
\newcommand{\chandra}{{\sl Chandra\/}}
\newcommand{\xmm}{{\sl XMM-Newton\/}}
\newcommand{\suzaku}{{\sl Suzaku\/}}
\newcommand{\nustar}{{\sl NuSTAR\/}}
\newcommand{\hitomi}{{\sl Hitomi\/}}
\newcommand{\erosita}{{\sl eROSITA\/}}
\newcommand{\gaia}{{\sl Gaia\/}}
\newcommand{\athena}{{\sl Athena\/}}

\newcommand{\porb}{$P_{\rm o}$}
\newcommand{\pspin}{$P_{\rm s}$}
\newcommand{\pbeat}{$P_{\rm b}$}

\newcommand{\msun}{M$_\odot$}
\newcommand{\kt}{$kT$}
\newcommand{\nh}{N$_{\rm H}$}
\newcommand{\cps}{ct\,s$^{-1}$}
\newcommand{\eps}{ergs\,s$^{-1}$}
\newcommand{\epcs}{ergs\,cm$^{-2}$s$^{-1}$}

\newcommand{\kms}{km\,s$^{-1}$}
\newcommand{\mwd}{M$_{wd}$}
\newcommand{\rwd}{R$_{wd}$}
\newcommand{\smdot}{$\dot m$}
\turnoffedit

\received{July 1, 2016}
\revised{September 27, 2016}
\accepted{\today}
\submitjournal{PASP}

\shorttitle{X-rays from Accreting White Dwarfs}
\shortauthors{Mukai}

\begin{document}

\title{X-ray Emissions from Accreting White Dwarfs: a Review}

\correspondingauthor{Koji Mukai}
\email{Koji.Mukai@nasa.gov}

\author[0000-0002-8286-8094]{K. Mukai}
\affil{CRESST and X-ray Astrophysics Laboratory, NASA/GSFC, Greenbelt,
       MD 20771, USA}
\affiliation{Also Department of Physics, University of Maryland,
       Baltimore County, 1000 Hilltop Circle, Baltimore, MD 21250, USA}



\begin{abstract}
Interacting binaries in which a white dwarf accretes material from
a companion --- cataclysmic variables (CVs) in which the mass donor
is a Roche-lobe filling star on or near the main sequence, and \syss\ in
which the mass donor is a late type giant --- are relatively commonplace.
They display a wide range of behaviors in the optical, X-rays, and other
wavelengths, which still often baffle observers and theorists alike.
Here I review the existing body of research on X-ray emissions from
these objects for the benefits of both experts and newcomers to the
field. I provide introductions to the past and current X-ray observatories,
the types of known X-ray emissions from these objects, and the data analysis
techniques relevant to this field. I then summarize of our knowledge
regarding the X-ray emissions from magnetic \cvs, non-magnetic \cvs\ and
\syss, and novae in eruption. I also discuss space density and the X-ray
luminosity functions of these binaries and their contribution to the
integrated X-ray emission from the Galaxy. I then discuss open questions
and future prospects.
\end{abstract}




\section{Introduction}

White dwarf is the most common endpoint of stellar evolution, and many
stars are born in binary systems with sufficiently small initial separations
so that they go through one or more phases of mass exchange. As a result,
there are numerous accreting white dwarfs (\awds) in the solar neighborhood
(see \S 8.1).  The wide array of phenomena exhibited by \awds\ make
them interesting objects of study on their own right. Their proximity
and relative simplicity (relativistic effects can be neglected for
most purposes) make them an excellent labolatory to study accretion
physics, on their own or in comparison with other accretion-driven systems.
An understanding of the long-term evolution of \awds\ of all types is
necessary in studying the progenitor channels of Supernovae of Type
Ia (SNIa), which is one of the outstanding questions in astrophysics and
cosmology. Finally, \awds\ turn up in significant numbers in transient searches
and in X-ray surveys, and are an important contributor to the
unresolved X-ray emissions from the Galactic ridge and bulge. Thus,
researchers who are interested in the X-ray sky in general, including
those that are engaged in variability and X-ray surveys not specifically
targeted to study \awds, must pay attention to \awds, and learn
to recognize them.

Accretion and nuclear fusion are the two primary energy sources in
\awds.  When fusion is not taking place, the accretion luminosity
often dominates over those of component stars, and a significant fraction
is emitted as X-rays. Nuclear fusion can take place either explosively
in the form of nova eruptions, or in the form of continuous shell burning.
While in progress, nuclear fusion dominates over accretion; in addition,
it is also possible that total energy emitted by \awds\ integrated over
its life time is dominated by nuclear fusion, depending on the fraction
of the accreted hydrogen burned. When the surface of the nuclear-burning
white dwarf is exposed, it is a powerful source of soft X-rays
(predominantly below 0.5 keV) as supersoft sources (\ssss). Finally,
both accretion and fusion can lead to mass loss, with velocities as
high as the escape velocity. Such flows also have kinetic energy
sufficient to result in X-ray emission, if they encounter
a strong shock in a sufficiently dense environment.

It is often the case that a large fraction of the luminosity
of \awds\ is emitted in the X-ray range. The X-ray emission mechanism
is usually tied to the processes on, or immediately around, the white
dwarf, while optical observations reveal the processes further out.
Therefore, X-ray observations are an essential tool for us to advance
our understanding of \awds, but if and only if interpreted in the context
of multiwavelength observations.

For the purpose of this review, I define X-rays as photons in the energy
range 0.1--100 keV, further divided into soft (0.1--2 keV), medium energy
(2--10 keV), and hard (10--100 keV) bands. The extreme ultraviolet (EUV)
range, if defined as 60--1000\AA, overlaps with the softest range of the
soft X-rays at 0.1--0.2 keV where it is least susceptible to interstellar
absorption. I include EUV observations in this band in this article,
implicitly at times and explicitly at other times. Here I review the state
of the art of X-ray observations of \awds\ as of late 2016. I include
both major classes of \awds, cataclysmic variables (\cvs) with a Roche-lobe
filling donor, and symbiotic stars with a late type giant donor, in this review.

This review is intended to provide a summary of subclasses,
the tools of trade for X-ray studies of \awds, the
major findings obtained to date and questions, and puzzles that remain.
The intended audience includes both researchers familiar with \awds\ in
general but not with X-ray observations, and X-ray astronomers not intimately
familiar with \awds. If this review also proves useful to experts of this
field as a reference, and provide a chance for them to reflect on larger
issues that can be addressed by X-ray observations of \awds, I will have
succeeded well beyond the minimum requirements I set myself.

The rest of this review is structured as follows. In the next two subsections,
I present a brief overview of X-ray observatories and the major subclasses of
\awds\ for the non-experts. In \S 2, I present
a brief overview of physical processes leading to X-ray emission in \awds.
In \S 3, I present different methods for extracting information from
X-ray data. I discuss the problem of identifying \awds\ and its subclasses
in \S4. I present the physical and astrophysical inferences for magnetic
systems, non-magnetic systems, and novae, respectively, in \S 5, 6, and 7.
I then discuss the collective properties of \awds\ in \S 8, followed by
a discussion of the problems and questions raised in the preceding sections
in \S 9. I conclude with a brief summary and future prospects in \S 10.

\subsection{Brief Overview of X-ray Observatories}

The earliest type of technology used for X-ray astronomy are the
non-imaging, collimated proportional counters, operating primarily
in the medium energy band. While many such instruments had a large
collecting area, they also suffered from large background rates,
both due to particles and to celestial X-rays from objects other
than the intended targets, and hence had limited sensitivities.
Non-imaging instruments are therefore useful only for bright X-ray sources (of
order 10$^{-11}$ \epcs\ or higher), or just a few dozen X-ray brightest
\awds.  Nevertheless, important results on \awds\ have been obtained with
non-imaging instruments, notably the large area counter (LAC) on-board
\edit2{\ginga}
(1987--1991; \citealt{Tetal1989}) and the proportional counter array
(PCA) on-board \xte\ (1995--2012; \citealt{Jetal1996}).

True imaging in X-rays started with \einstein\ (1978--1981;
\citealt{Getal1979}) in the 0.1--4 keV band. Imaging soft X-ray observation
continued with \rosat\ (1990--1999; \citealt{T1983}), which also carried
the Wide Filed Camera (WFC), which was sensitive to EUV photons.
\euve\ (1992--2001; \citealt{BM1991}) was a mission dedicated
to EUV observations.  In the medium energy X-ray range, \asca\ (1993--2000;
\citealt{TIH1994}) was the first satellite to carry out imaging
observations in energies up to 10 keV.  \xmm\ (1999--present;
\citealt{XMMREF}) and \chandra\ (1999--present; \citealt{CHANREF})
can reach flux levels below 10$^{-14}$ \epcs\ in the 0.3--10 keV band.
Other recent instruments with imaging capabilities in the soft and medium
energy ranges include the X-ray Imaging Spectrometer (XIS) on-board \suzaku\
(2005--2015; \citealt{SUZAKUREF}) and the X-Ray Telescope (XRT)
on-board \swift\ (2004--present; \citealt{SWIFTREF}).

The technology used to build soft and medium-energy X-ray telescopes
alone is not sufficient to build hard X-ray imaging optics. One successful
alternative is the coded mask aperture. Instruments based on this technology
have a large field of view (FOV), require long integration times, and sources
can only be ``imaged'' after specialized and computer-intensive
processing. Coded-aperture mask instruments can then localize
individual sources to several arcminutes. Two prominent examples
are \integral\  (2002--present; \citealt{INTEGRALREF}) and the burst
alert telescope (BAT) on-board \swift. More recently, X-ray telescope
technology has been extended to hard X-rays using multi-layer coating.
The first satellite to achieve this is the \nustar\ mission (2012--present;
\citealt{NUSTARREF}).

In the soft and medium energy X-rays, CCD-based detectors have
become the standard. CCD data typically have energy resolution of 2--6\%,
which is very low in the standards of optical spectroscopy.  Grating
instruments provide high resolution, but have relatively low effective
areas, once the efficiency of the grating is taken into account. Thus
one can only observe the X-ray brightest \awds. In the soft X-ray/EUV,
the spectrometer instrument on \euve\ and the low energy transmission
grating (LETG) on \chandra\ have produced notable results on \awds.
At higher energies, results on \awds\ have been published using
the reflection grating spectrometers (RGS) on \xmm\ in the 0.3--2 keV
range, and using the high energy transmission grating (HETG) on \chandra\
in the 0.5--8 keV range.

\subsection{Brief Overview of Many Object Types}

\cvs\ and \syss\ can be sub-classified in myriad ways. While many of the
detailed divisions are beyond the scope of this review, it is impossible
to describe \awds\ without referring to at least some of the major subclasses.

\cvs\ are systems in which the mass donor is a Roche-lobe filling late
type star on or near the main sequence. Those with hydrogen-deficient
donors are called the AM~CVn stars \citep{AMCVN}: they are sometimes treated
as a subclass of \cvs, and sometimes considered a parallel class. Excluding
AM~CVn stars, the majority of \cvs\ have an orbital period (\porb) in the 80
min--10 hr range, and their evolution is driven by angular momentum loss.
Gravitational radiation is a mechanism that is believed to operate in all
\cvs\ \citep{KMG1962}, while an additional mechanism, almost certainly magnetic
braking, operates mostly for systems with periods longer than $\sim$3 hrs
\citep{KBP2011}. \cvs\ evolve from long period to short, temporarily
detaching when the magnetic braking mechanism becomes less effective
and resuming accretion at \porb\ $\sim$ 2 hrs. Therefore, there are
fewer \cvs\ with 2--3 hr orbital periods, referred to as the period gap.

In \syss, the mass donor is a late type giant. While nuclear evolution
of the giant ultimately drives the mass transfer in \syss, it is not clear
how many systems operate purely by capturing the wind of the donor, how
many donors overflow their Roche-lobe, and how important the hybrid
``wind Roche-lobe overflow'' mechanism \citep{MP2007} is.

Traditional classification schemes of \cvs\ relied largely on their
variability properties. I will use the term ``nova eruptions'' to refer
to high amplitude brightening caused by violent ejection of mass powered by
thermonuclear runaways. Perhaps the only way for \awds\ to avoid
nova eruptions over evolutionary time scales is if the accreted matter is
continuously burned: such systems are observed as \ssss, while nova
eruptions lead to a temporary (weeks--years) super-soft source stage.
Variability unrelated to nuclear burning includes dwarf nova outbursts,
which are widely believed to be due to disk instability that operates
when the mass transfer rate from the donor is low (see \S 6.2). Other
\cvs\ exhibit occasional low states, whose origins are less clear.

In the context of this review, the most important classification
scheme is based on the magnetic field of the white dwarf primary.
\cvs\ are said to be non-magnetic when the field is not strong enough
to alter the accretion flow, often taken to mean that the accretion
disk extends all the way down to the surface of the primary. In magnetic
\cvs, the magnetic field controls the flow at least near the primary.
The latter are divided into two major subclasses.  In intermediate polars
(IPs, also known as DQ~Her type systems; \citealt{Joe1994}), a partial
accretion disk is usually present, and the the spin period (\pspin)
of the primary is significantly shorter than \porb.  In polars (or
AM~Her type systems; \citealt{Mark1990}), the magnetic field is so
strong as to prevent the formation of an accretion disk, optical and
infrared light contains a strong contribution by the cyclotron emission
from the accretion region and is therefore polarized, and the spin is
usually synchronized to the
orbit (\pspin = \porb). Asynchronous polars are objects in which \pspin\ and
\porb\ differ by $<$1\% but otherwise have the properties of a polar.
In \syss, the equivalent of polars should not exist due to the much larger
binary separation, but there is no obvious reason why IP-like systems should
not exist.

For further discussion of CV subclasses, their inter-relationships,
and evolutionary scenarios, readers are directed to \cite{tome} and
the upcoming review by \cite{PK2017}. 

\section{Types of X-ray Emissions and Basic Physical Processes}

X-rays from \awds\ are mostly, if not exclusively, emitted through
thermal processes.  Thermal X-ray emissions are usually divided into
optically thick (optical depth, $\tau$ much greater than 1)
and optically thin ($\tau <1$).  This is clearly a simplification:
many \awds\ may have regions of intermediate optical depth ($\tau \sim 1$).

When the accretion flow hits the white dwarf surface and \edit2{is} shocked
above the surface, the plasma must cool and further decelerate
before it can settle onto the white dwarf surface. For the highest
temperature, fully ionized plasma, the Bremsstrahlung continuum
dominates over line emission.  As the plasma cools to lower temperatures
($kT \sim 1$ keV), when many ions retain some electrons, line cooling becomes
important and then dominant \citep{GW1993}.

\begin{figure}[t]
\plotone{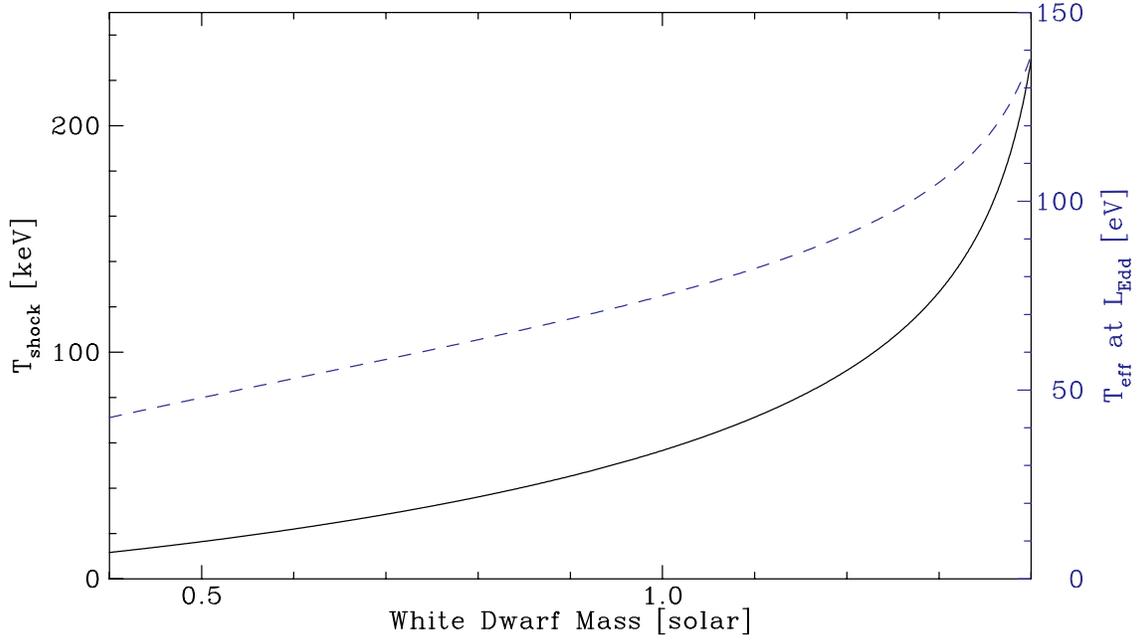}
\caption{Expected shock temperature as a function of the white dwarf
         mass, assuming a radial accretion with free-fall from infinity
	 (solid black line, left axis). Also plotted is the maximum possible
	 effective temperature at local Eddington limit as a function of the
	 white dwarf mass (dashed blue line, right scale). \label{tshock}}
\end{figure}

In the case of radial accretion (either spherically symmetric or
magnetically confined ``accretion column\footnote{Here I use this
terminology without necessarily implying a specific shape of the accretion
region.}''), accretion flow is initially cool and supersonic, and at the
free-fall velocity that approaches the final value of
$$	v_{ff} = \sqrt{\frac{2GM_{wd}}{R_{wd}}}	$$
where \mwd\ and \rwd\ are the mass and radius of the white dwarf primary,
respectively, and $G$ is the Newtonian constant of gravity. This is a very
steep function of \mwd\ because of the mass-radius relationship for white
dwarfs\footnote{Throughout this review, I use the convenient formula due
to Webbink as reported in \cite{PW1975} for numerical calculations. While
this is not the most up-to-date mass-radius relationship for white dwarfs,
and the core temperature and composition do change the relationship
somewhat, the Webbink formula is sufficiently accurate for most current
needs in regard to \awds.}. The strong shock condition dictates that the
post-shock flow initially travels at $1/4 v_{ff}$,
and the difference is converted to thermal energy, leading to
$$	kT_{s,ff} = \frac{3}{8} \frac{GM_{wd} \mu m_{\rm H}}{R_{wd}}	$$
where $m_{\rm H}$ is the proton mass and $\mu$ is the mean molecular
weight of the accreting gas (here assumed to be 0.615), and $k$ is
Boltzman's constant. The maximum shock temperatures are 22, 36, 57,
and 92 keV, respectively, for 0.6, 0.8, 1.0, and 1.2 \msun\ white dwarfs
(see also Figure\,\ref{tshock}).

The Bremsstrahlung cooling time scale $t_c$ of a single temperature,
X-ray emitting plasma is
$$	t_c = 6.8 \times 10^{14} \frac{\sqrt{kT}}{n} s	$$
where the temperature \kt\ is in units of keV, the number density
$n$ is in units of cm$^{-3}$.  Therefore, a 10 keV plasma with
a density of 10$^{15}$ cm$^{-3}$ will cool in roughly 2 seconds,
numbers not atypical of those encountered in \awds.

\begin{figure}[t]
\plotone{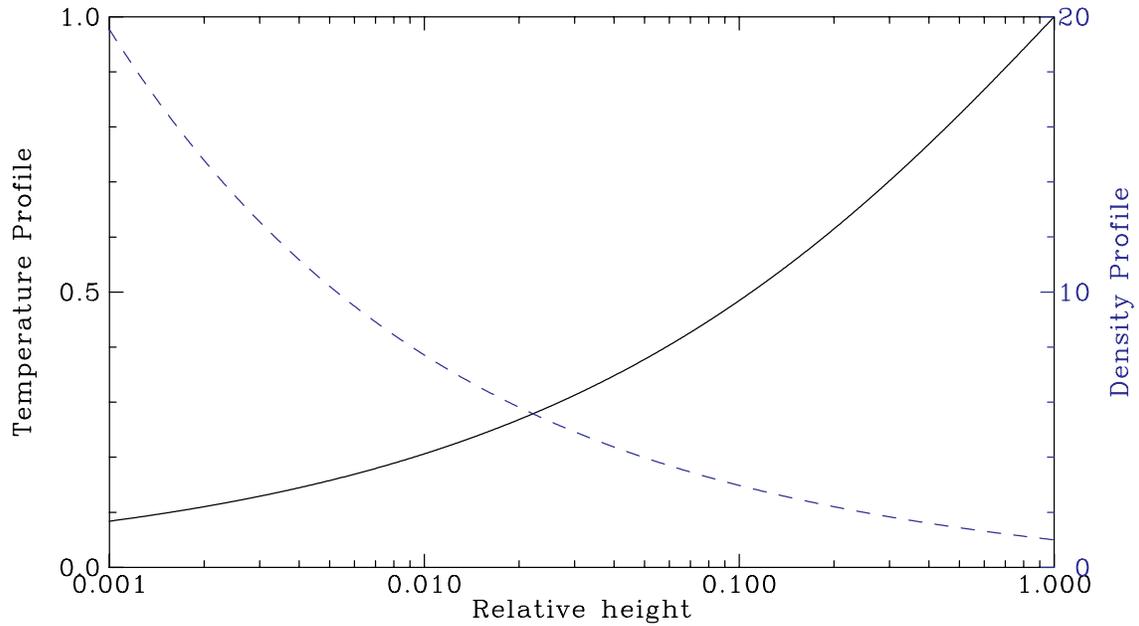}
\caption{Relative temperature (solid black, left axis) and density (dashed
         blue, right axis) of the post-shock region, in an Aizu-type
	 accretion column, normalized to the respective values at the shock.
	 \label{aizuprof}}
\end{figure}

As first investigated by \cite{Aizu1973}, the resulting shock height
is such that the time for the shocked plasma to reach the white
dwarf surface equals the total cooling time, which has to be calculated
along with the density and the temperature profile of the post-shock region.
A sample profile is shown in Figure\,\ref{aizuprof}. As can been seen,
the density rises quickly towards the white dwarf surface. As a result,
most of the cooling takes place near the bottom of the post-shock region,
which is the origin of the lowest temperature emission, while the origin
of the highest energy photons is more uniform.

A simple expectation for non-magnetic \cvs\ is that half the available
potential energy is radiated away in the Keplerian accretion disk.
The disk around \awds\ is never hot enough to emit X-rays; instead,
they are sources of IR, optical, and UV photons.  If the white dwarf
is spinning slowly compared to the break-up rate, then the other half
is available in the boundary layer between the disk proper (i.e., the
Keplerian part) and the white dwarf. If the boundary layer remains
optically thin, it should emit hard and medium energy X-rays broadly
similar to those seen in magnetic \cvs\ \citep{PR1985a}.  However,
the theory of the boundary layer is much more complex than that of
accretion column in magnetic \cvs: while a one-dimensional theory
is sufficient to establish the overall characteristics of an accretion
column, the theory of boundary layer must be at least two dimensional.
The most pressing question is whether the shock can be strong. If it is,
a scaled (by a factor of 2 in available energy) version of the Aizu
model might provide a good initial approximation. On the other hand,
if there may be a series of weak shocks, the maximum temperature of the
emitting plasma is lower still.

Optically-thin, thermal X-ray emission is also seen during nova
eruptions. When the nova occurs in \syss, the ejecta are embedded
in the dense wind of the mass donor. In this case, the ejecta promptly
encounters the red giant wind and strongly shocked. In contrast,
when the nova occurs in \cvs, any shocks present must be internal,
in that collisions are between multiple components of the ejecta
with different velocities. The expected density is a strong function
of time, as the ejecta expand. Since a 1 keV, $n = 10^6$ cm$^{-3}$
plasma has a cooling time of $\sim$20 yrs, X-ray emission observed
several weeks after eruption may represent a small fraction of the
thermal energy of the shocked plasma; the predominant cooling
mechanism in this case is likely to be adiabatic expansion.
Significant radiative cooling requires a high density: either
the very early phase of an eruption when the ejecta occupies a
small volume, or near the red giant mass donor in a relatively
tight orbit around the white dwarf.

It is also possible for the accretion disk wind \citep{DP2000}
to collide with something else, get shocked, and emit medium
energy X-rays. In both novae and in accretion disk wind shocks,
non-equilibrium ionization (NEI) effect may need to be considered.
In NEI plasma, the ionization state of elements have not caught
up to the current temperature, a situation that is expected when
the ionization time $nt$ is less than 10$^{12}$ cm$^{-3}$\,s \citep{SH2010}.
This effect is not of significant concern for the accretion powered
X-rays, which is in a steady-state situation where the injection of
freshly shocked material is balanced by radiative cooling in \awds.
By definition, steady-state requires a radiative shock, in which
case $n t_c > 10^{14}$ cm$^{-3}$\,s, hence the X-ray emitting plasma
is in ionization equilibrium.

%
%
%

Soft X-ray and EUV emissions in \awds\ can be from the white dwarf surface
itself, or from an optically thick boundary layer
\citep{PR1985b}.  The former is inevitable at some level, even when the
primary energy release mechanism is the optically thin X-ray emission,
because it is expected to occur near the white dwarf surface. A substantial
fraction (up to a half) must therefore be intercepted by the white dwarf
photosphere, much of it absorbed and reprocessed (see below for a discussion of
``reflection''). This is the baseline mechanism leading to the soft component
in magnetic \cvs\ \citep{LM1979}.  Other mechanisms include accretion of
dense blobs, which liberate the kinetic energy deep within the white dwarf
photosphere \citep{KP1982}, the accretion energy liberated within the
optically-thick boundary layer \citep{PR1985b}, and nuclear burning seen
in \ssss\ and in novae \citep{KvdH1997}.

The combination of the typical emitting area area ($A$; some fraction of
the total surface area of the white dwarf) and the typical luminosity
($L$) often found in the optically-thick component in \awds\ results
in effective temperatures ($T_{eff}$) in the range $\sim$100,000--1,000,000K
(\kt\ of $\sim$9--90 eV). Note, however, that the lower limit of this
observed range of temperatures may well be determined by the combination
of interstellar absorption and instrument sensitivity, rather than the
intrinsic properties of \awds. The upper limit, on the other hand, is
constrained by radiation pressure. Although this is often considered
in terms of the Eddington luminosity, the local version of the argument is also
valid: accretion rate per unit area cannot be so high that the resulting
radiation pressure exceeds the local gravity. This translates to an upper
limit to the local accretion rate, and since $L/A \propto T_{eff}^4$, an
upper limit to the temperature. The Eddington limit of white dwarf temperature
is shown in, e.g., Figure 5 of \cite{WKB1987}. This figure also illustrates
the atmosphere limit, which is based on a higher opacity than pure electron
scattering.  Since matter is unlikely to be fully ionized in \awds, this
is a more realistic case and results in lower maximum temperatures than
the Eddington limit.

The partially ionized nature of matter also means that the blackbody
model is only an approximation. As photons emerge from deep inside
the optically thick region through the $\tau \sim 1$ zone, it
leaves an imprint on the outgoing spectrum in the form of absorption
lines and edges; the temperature, the pressure and the bulk motion of this
region determines what features are present, and how broad they are.
The detailed models of such spectra are called stellar atmosphere
models, but the situation in \awds\ is more complex than that typically
found in stars. Composition is a key issue, determined by the
competition between fresh accretion and gravitational settling, and
the exact mixture of pristine gas vs. nuclear burning products. Another
is the temperature structure.  Irradiation by X-rays results in a
temperature inversion above the photosphere and a flatter temperature
profile with optical depth; the edges and lines will be weak if the
temperature profile is flat where these features are formed \citep{WKB1987}.

The X-ray photons that are emitted by the \awds\ often interact with
matter inside the binary, circum-binary matter, and the interstellar medium
(ISM). For X-rays below 10 keV, photoelectric absorption is the dominant
process. The strong energy dependence of photoelectric absorption is encoded
the very terminology ``hard'' and ``soft'' X-rays. For reference,
equivalent hydrogen column density (\nh; assuming solar composition) at
which the transmission is $1/e$ is 1.2$\times 10^{20}$ cm$^{-2}$,
1.4$\times 10^{21}$ cm$^{-2}$, 2.3$\times 10^{22}$ cm$^{-2}$ and
2.8$\times 10^{23}$ cm$^{-2}$, respectively for photon energies of
0.2, 0.5, 2.0, and 5.0 keV. For an assumed ISM density of 0.1 cm$^{-3}$,
the resulting \nh\ is 1.2$\times 10^{20}$ cm$^{-2}$ at a distance of $\sim$400 pc.

However, absorption significantly in excess of the expected ISM
absorption is often seen in the X-ray spectra of \awds. The measured
X-ray \nh\ is often significantly greater than that expected from
optical or UV extinction (using, e.g., the calibration
by \citealt{PS1995} of \nh\ = 1.79$\times 10^{21} \times$ A$_V$ cm$^{-2}$).
Moreover, spectral fits often indicates the presence of a partial
covering absorber: a fraction of the photons is observed directly,
the rest through an intervening absorber. This suggests that the size
of the absorber is of the same order as the X-ray emission region,
most likely indicating a location near the white dwarf, or at least
somewhere within the binary itself.

Above 10 keV, Compton scattering, rather than photoelectric absorption,
is the dominant form of interaction between X-rays and matter.
Compton-scattering optical depth reaches 1 for \nh\ $\sim 1.5 \times 10^{24}$
cm$^{-2}$. While only a small fraction of intrinsic absorbers observed
in \awds\ have a column density high enough for Compton scattering
not to be negligible, the white dwarf itself is always, and the
accretion disk is sometimes, Compton thick. So, when the X-rays
strike the white dwarf surface, soft and medium energy X-rays are
largely absorbed and reprocessed, while hard X-rays are ``reflected''
\citep{Retal1981}. At the same time, X-rays above 7 keV have a high
probability of interacting with K-shell electrons in iron atoms.
A vacancy in the K-shell of a neutral (or mildly ionized) iron
atom often leads to the emission of fluorescent line at 6.4 keV.

It is customary to measure the X-ray flux of \awds, and also estimate
the ``unabsorbed'' flux, by setting all the absorbers to zero after
spectral fitting. One then calculates the X-ray luminosity by multiplying
the unabsorbed flux by $4\pi d^2$, where $d$ is the estimated distance
to the source. However, there is some ambiguity in the definition of
X-ray luminosity: how to account for the X-rays that are reprocessed by
the white dwarf surface, or those that are absorbed and reprocessed by
intrinsic absorbers. There probably is no single solution that is
applicable to all \awds\ for all situations. For now, the best option
may be to stick with the simple procedure described above, while paying
attention to its implications in individual situations.

\section{Observations and Analysis Methods}

\subsection{X-ray Surveys}

\awds\ are discovered using a variety of methods, including
optical variability, optical colors and emission lines, and
X-ray surveys. Over the last several decades, X-ray surveys
have been instrumental in establishing magnetic CVs as an
important class, and in providing valuable samples with which
to study the space densities of various types of \awds.

\begin{figure}[t]
\plotone{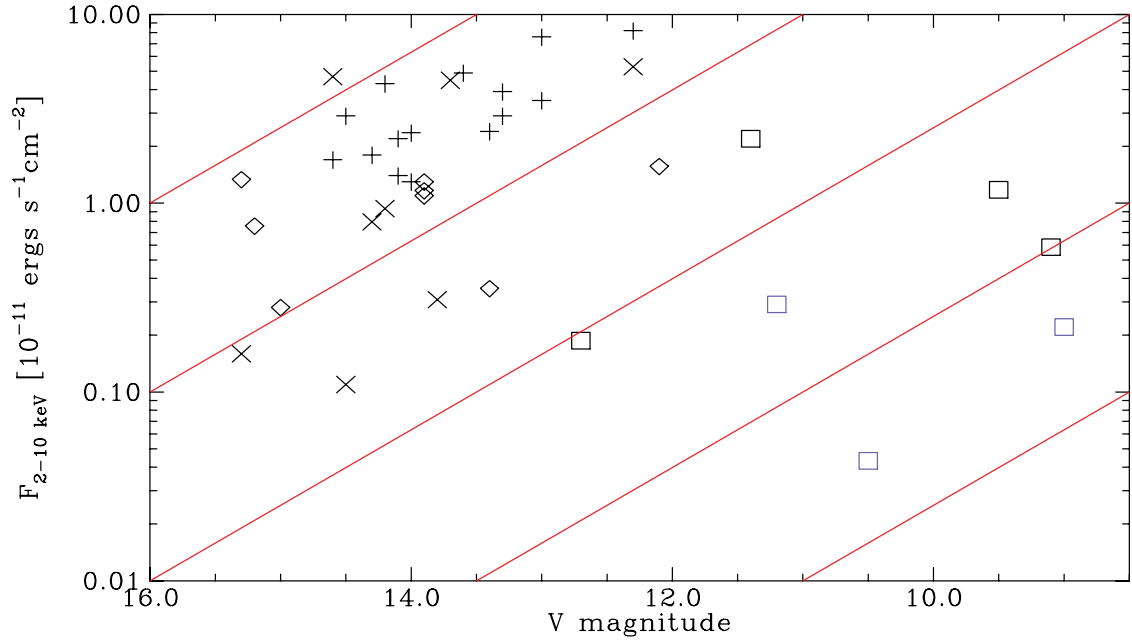}
\caption{Selected \cvs\ of various subtypes in the F$_X$/F$_{opt}$ plane.
         Pluses are IPs, crosses are polars, diamonds are dwarf novae in
         quiescence, and open squares are non-magnetic nova-like systems
         (black) and dwarf novae in outburst (blue). The diagonal red
         lines are for equal F$_X$/F$_{opt}$ ratios, separated by factors
         of 10. \label{fxfopt}}
\end{figure}

Several all-sky surveys in the medium energy X-rays were conducted
during the 1970s, including those with {\sl Uhuru\/} and {\sl Ariel V\/},
all with non-imaging instrument. \heao\ (1977-1979) was the last to
carry out such an all-sky survey, which also localized point source
positions using a scanning modulation collimator \citep{Getal1978},
Of the 660 sources with modulation collimator positions, \cite{S1992}
considered 43 to be \cvs, although some of these proposed identifications
have not been confirmed using imaging X-ray observations.  \cite{S1992}
further concluded that roughly half the \heao\ sample of \cvs\ were
magnetic, including both IPs and polars.  Subsequent studies of
individual magnetic \cvs\ established most IPs and some polars
as luminous, hard X-ray sources (see, e.g., \citealt{I1991}).
To illustrate the differences among various subclasses of \cvs\ in
terms of the importance of the X-ray band, I show a sample of
objects in the 2--10 keV X-ray flux vs. V-band optical magnitude plane
in Figure\,\ref{fxfopt}.

The \rosat\ mission carried an imaging soft X-ray telescope
and was operated in a scanning mode during the first 6 months
of the mission to conduct an all-sky survey. In addition, even
during the rest of the mission when it was operated in a pointed
mode, its wide filed-of-view led to the serendipitous discovery
of many additional sources. The most significant impact of
\rosat\ mission on \awds\  is the large increase in the number
of known polars \citep{B1999}.

\startlongtable
\begin{deluxetable}{lrlrrr}
\tablewidth{0pt}
\tablecaption{\awds\ in the BAT 70 month catalog\label{bat70tab}}
\tablehead{
\colhead{No.\tablenotemark{a}}  & \colhead{BAT designation} &
\colhead{Common Name}  & \colhead{Type\tablenotemark{b}} &
\colhead{flux\tablenotemark{c}}  & \colhead{22 mo.\tablenotemark{d}}
}
\startdata
1010 & Swift J1855.0$-$3110 & V1223 Sgr               & IP    & 117.71 &  97.1 \\
 782 & Swift J1548.0$-$4529 &    NY Lup               & IP    &  91.75 & 101.4 \\
 625 & Swift J1234.7$-$6433 &    RT Cru               & S     &  81.41 &  71.9 \\
 181 & Swift J0331.1+4355   &    GK Per               & IP    &  77.52 &  48.3 \\
  15 & Swift J0028.9+5917   &  V709 Cas               & IP    &  74.72 &  89.1 \\
 887 & Swift J1730.4$-$0558 & V2731 Oph               & IP    &  69.46 &  67.8 \\
 790 & Swift J1559.6+2554   &     T CrB               & S,RN  &  63.88 &  64.0 \\
 281 & Swift J0529.2$-$3247 &    TV Col               & IP    &  59.99 &  51.1 \\
1123 & Swift J2142.7+4337   &    SS Cyg               & DN    &  54.45 &  45.7 \\
1115 & Swift J2133.6+5105   & 1RXS J213344.1+510725   & IP    &  52.64 &  52.3 \\
1147 & Swift J2217.5$-$0812 &    FO Aqr               & IP    &  52.40 &  58.5 \\
 873 & Swift J1712.7$-$2412 & V2400 Oph               & IP    &  50.28 &  52.7 \\
1044 & Swift J1940.3$-$1028 & V1432 Aql               & AP    &  49.28 &  54.0 \\
 877 & Swift J1719.6$-$4102 & IGR J17195$-$4100       & IP    &  40.36 &  35.9 \\
 158 & Swift J0256.2+1925   &    XY Ari               & IP    &  36.45 &  31.2 \\
 962 & Swift J1816.1+4951   &    AM Her               & P     &  36.17 & \\
 312 & Swift J0558.0+5352   &  V405 Aur               & IP    &  34.93 &  31.7 \\
1175 & Swift J2255.4$-$0309 &    AO Psc               & IP    &  31.72 &  29.2 \\
 393 & Swift J0750.9+1439   &    PQ Gem               & IP    &  31.32 &  30.3 \\
 299 & Swift J0542.6+6051   &    BY Cam               & AP    &  29.95 &  29.1 \\
 422 & Swift J0838.0+4839   &    EI UMa               & IP    &  29.23 &  32.1 \\
 375 & Swift J0732.5$-$1331 &  V667 Pup               & IP    &  27.48 &  29.7 \\
 758 & Swift J1509.4$-$6649 & IGR J15094$-$6649       & IP    &  25.81 & \\
 835 & Swift J1649.9$-$3307 & IGR J16500$-$3307       & IP    &  25.65 & \\
 253 & Swift J0502.4+2446   & V1062 Tau               & IP    &  25.27 & \\
 642 & Swift J1252.3$-$2916 &    EX Hya               & LLIP  &  24.34 &  25.4 \\
 845 & Swift J1654.7$-$1917 & IGR J16547$-$1916       & IP    &  23.48 & \\
 248 & Swift J0457.1+4528   & IGR J04571+4527         & IP    &  23.07 &  \\
 374 & Swift J0731.5+0957   &    BG CMi               & IP    &  22.82 &  21.5 \\
 491 & Swift J1010.1$-$5747 &  V648 Car               & S     &  22.18 &  22.5 \\
  40 & Swift J0055.4+4612   &  V515 And               & IP    &  20.62 &  19.7 \\
 798 & Swift J1617.5$-$4958 & IGR J16167$-$4957       &       &  20.56 &  20.8 \\
 890 & Swift J1731.9$-$1915 & V2487 Oph               & RN    &  20.03 &  \\
 278 & Swift J0525.6+2416   & 1RXS J052523.2+241331   & IP    &  19.43 & \\
1071 & Swift J2015.9+3715   & RX J2015.6+3711\tablenotemark{e}   & P?    &  19.23 & \\
 563 & Swift J1142.7+7149   & YY (DO) Dra             & LLIP  &  18.65 &  17.4 \\
1107 & Swift J2123.5+4217   & V2069 Cyg               & IP    &  17.64 &  12.1 \\
 749 & Swift J1453.4$-$5524 & IGR J14536$-$5522       & P     &  17.07 &  22.8 \\
 424 & Swift J0838.8$-$4832 & IGR J08390$-$4833       & IP    &  16.83 & \\
 332 & Swift J0625.1+7336   &    MU Cam               & IP    &  16.58 & \\
 853 & Swift J1701.3$-$4304 & IGR J17014$-$4306       &       &  16.15 &  \\
1055 & Swift J1958.3+3233   & V2306 Cyg               & IP    &  15.61 & \\
 288 & Swift J0535.1$-$5801 &    TW Pic               &       &  15.56 & \\
 300 & Swift J0543.2$-$4104 &    TX Col               & IP    &  15.54 &  18.0 \\
 386 & Swift J0746.3$-$1608 & 1RXS J074616.8$-$161127 &       &  15.49 &  \\
 963 & Swift J1817.4$-$2510 & IGR J18173$-$2509       & LLIP: &  14.46 & \\
 943 & Swift J1800.5+0808   & V2301 Oph               & P     &  14.45 & \\
 415 & Swift J0820.6$-$2805 & 1RXS J082033.6$-$280457 & IP:   &  12.86 & \\
 951 & Swift J1807.9+0549   &  V426 Oph               & DN    &  12.83 & \\
 341 & Swift J0636.6+3536   &  V647 Aur               & IP    &  12.38 & \\
 598 & Swift J1212.3$-$5806 & IGR J12123$-$5802       &       &  12.31 & \\
1035 & Swift J1924.5+5014   &    CH Cyg               & S     &  11.85 &  22.6 \\
1022 & Swift J1907.3$-$2050 & V1082 Sgr               &       &  11.69 &  27.5 \\
1144 & Swift J2214.0+1243   &    RU Peg               & DN    &  11.34 &  \\
 724 & Swift J1424.8$-$6122 & IGR J14257$-$6117       &       &  11.32 & \\
 324 & Swift J0614.0+1709   &                         & IP    &  11.04 & \\
 355 & Swift J0704.4+2625   &  V418 Gem               & IP    &  10.94 & \\
 356 & Swift J0706.8+0325   &                         & P     &  10.14 & \\
1190 & Swift J2319.4+2619   &                         & P     &   9.86 &   2.8 \\
 462 & Swift J0927.7$-$6945 &                         & IP    &   9.76 & \\
 479 & Swift J0958.0$-$4208 & 1RXS J095750.4$-$420801 & IP    &   9.16 & \\
 331 & Swift J0623.9$-$0939 &                         &       &   9.11 &  \\
1103 & Swift J2116.0$-$5840 &    CD Ind               & AP    &   9.04 & \\
 627 & Swift J1238.1$-$3842 & V1025 Cen               & LLIP  &   8.99 & \\
 275 & Swift J0524.9+4246   & Paloma                  & LLIP  &   8.97 & \\
1109 & Swift J2124.6+0500   &                         &       &   8.93 & \\
1004 & Swift J1848.4+0040   &  V603 Aql               & N     &   8.74 & \\
 322 & Swift J0610.6$-$8151 &    AH Men               &       &   8.59 & \\
  11 & Swift J0023.2+6142   & V1033 Cas               & IP    &   8.43 & \\
1101 & Swift J2113.5+5422   & 1RXS J211336.1+542226   &       &   8.40 & \\
 709 & Swift J1409.2$-$4515 &  V834 Cen               & P     &   8.31 & \\
 255 & Swift J0503.7$-$2819 & 1WGA J0503.8$-$2823     & IP    &   8.23 & \\
 393 & Swift J0749.7$-$3218 &                         &       &   8.10 & \\
 365 & Swift J0717.8$-$2156 & 1RXS J071748.9$-$215306 &       &   7.86 & \\
 708 & Swift J1408.2$-$6113 & IGR J14091$-$6108       & IP:   &   7.82 & \\
 417 & Swift J0826.2$-$7033 & 1RXS J082623.5$-$703142 &       &   7.72 &  20.8 \\
 508 & Swift J1039.8$-$0509 &    YY Sex               & P     &   7.56 & \\
1148 & Swift J2218.4+1925   & 1RXS J221832.8+192527   & P     &   7.40 & \\
1201 & Swift J2341.0+7645   & 1RXS J234015.8+64207    & P:    &   7.19 & \\
 219 & Swift J0426.1$-$1945 &    IW Eri               & P     &   6.69 & \\
 469 & Swift J0939.7$-$3224 & 1RXS J093949.2$-$322620 & IP:   &   5.97 & \\
\enddata
\tablenotetext{a}{Entry number in the BAT 70 month catalog \citep{BAT70}}
\tablenotetext{b}{Abbreviated types: IP for intermediate polars, LLIP for
                  low luminosity IPs, P for polars, AP for asynchronous polars,
                  N for (old) novae, RN for recurrent novae, DN for dwarf
		  novae, and S for symbiotic stars. A blank in this column
                  indicates objects of unknown subtype, while a colon indicates
		  proposed but unconfirmed type. For references of IP nature,
                  see http://asd.gsfc.nasa.gov/Koji.Mukai/iphome/iphome.html.}
\tablenotetext{c}{The BAT-band flux in the 70 month catalog.}
\tablenotetext{d}{The BAT-band flux in the 22 month catalog \citep{BAT22},
                  for those objects that were also detected.}
\tablenotetext{e}{This object is one of possible counterparts of the BAT source.}
\end{deluxetable}

\begin{figure}[t]
\plotone{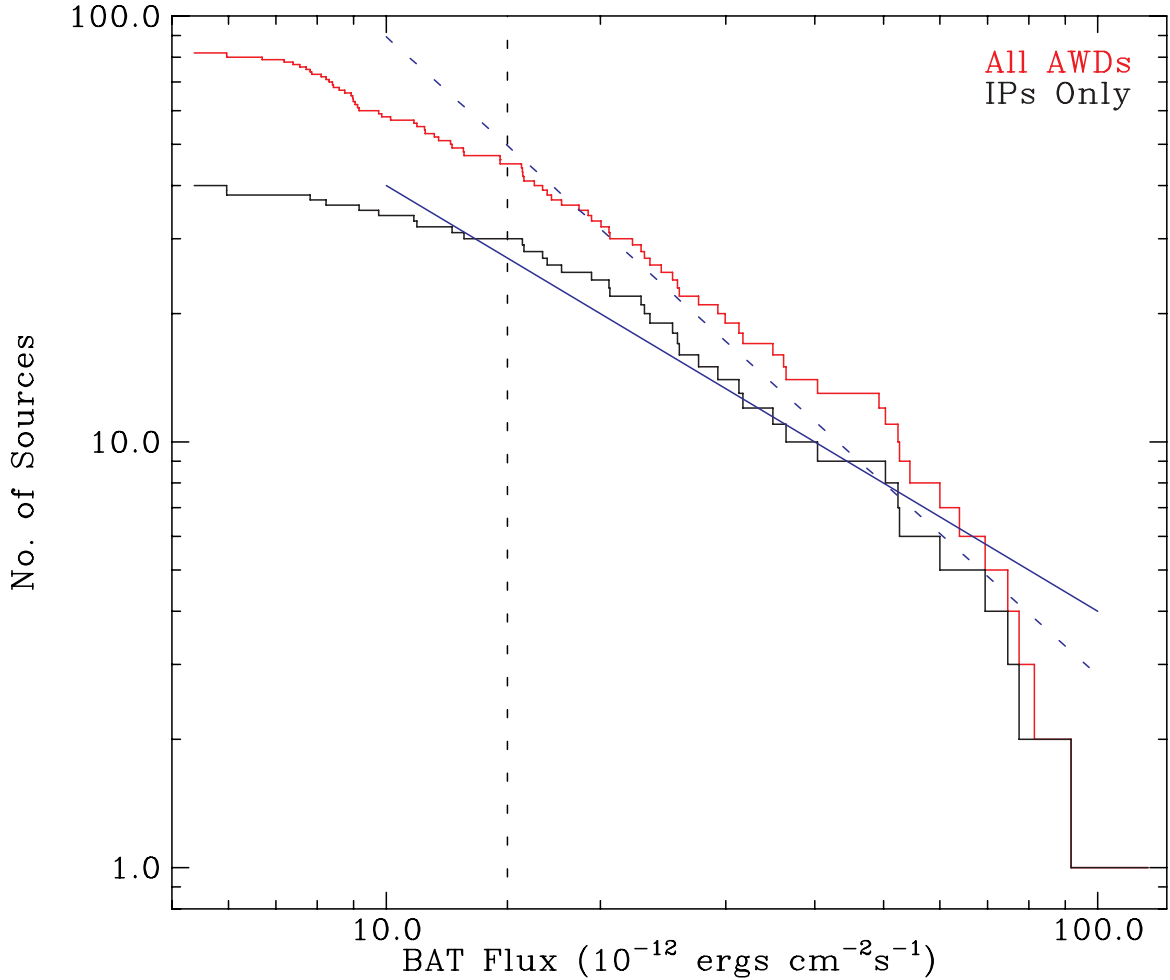}
\caption{$log$ N-$log$ S diagram for BAT detected \awds. The red histogram
         shows the cumulative number of all \awds\ in Table\,\ref{bat70tab}
         above given flux, while black histogram is for confirmed IPs only.
         The dashed vertical line is the approximate completeness limit of
         the BAT 70 month survey. Blue solid line is the expected slope
         for objects with a 2-dimensional spatial distribution, while
	 dashed line is for those with a 3-dimensional distribution.
         \label{lognlogs}}
\end{figure}

The \integral\ mission has detected a large number of previously unknown
hard X-ray sources, as a by-product of pointed observations using its
wide-angle, coded-aperture mask instruments. \cite{EJBetal2006} was the
first to carry out a systematic search for \cvs\ among \integral\ sources.
Of the 15 \cvs\ thus identified, the majority were confirmed or candidate
IPs. Since many \integral\ observations are aimed at studies of Galactic
sources, the \integral\ survey is particularly deep along the
Galactic plane, and this has led to the discoveries of many \cvs.
This is both an advantage and a disadvantage. Here I focus instead
on the hard X-ray survey with \swift\ BAT, particularly the 70-month
catalog \citep{BAT70}, whose sky coverage is relatively uniform
(see their Figure 1) because \swift\ is primarily a gamma-ray burst mission.
In Table\,\ref{bat70tab}, I list 81 \awds\ in the literature as of 2016,
with confirmed, probable, or proposed classification, sorted by the BAT-band
flux.  Roughly half are confirmed or proposed to be IPs.
This fraction is sufficiently high that it makes sense to search for IP
signatures in all newly discovered hard X-ray emitting \cvs; it is not
so high that all such systems should be assumed to be IPs. The remaining
objects include polars, non-magnetic CVs, and four \syss\ \citep{Ketal2009}.
I show a $log$N -- $log$S plot of all BAT detected \awds, as well as
that of a subset of them that have been confirmed as IPs, in
Figure\,\ref{lognlogs}. In the BAT 22-month catalog \citep{BAT22},
35\footnote{Of these, 34 are listed in Table\,\ref{bat70tab}; V2491~Cyg
was also listed in the 22-month catalog, although the association of
the BAT source with this nova is not 100\% certain \citep{v2491pre},
and it does not appear in any subsequent versions of the BAT source
catalog.} out of 479 sources are \awds; in the BAT 77-month catalog
\citep{BAT70}, the corresponding numbers 81 out of 1210, subject
to some uncertainties in the source classification.

There are also deeper surveys concentrating on smaller areas of the
sky. For example, \cite{Petal2007} took advantage of the fact that
the \rosat\ all-sky survey was deeper near the north ecliptic pole
to infer the space density of \cvs. Another example is the survey
of the Galactic center region: \cite{Metal2009} cataloged 9017
X-ray sources in the \chandra\ observations of a
$2^\circ \times 0.8^\circ$ field, of which 6760 have high absorption
consistent with sources lying at or beyond the Galactic center distance,
and argued that a large fraction of the latter were probably magnetic
\cvs. Such focused surveys, while useful, must deal with either
small number statistics (if the field is at a high Galactic latitude,
and a survey of local population) or faint counterparts (if detectable
at all) in crowded regions of the sky (Galactic plane and center).

Finally, many stellar clusters have been the subject of X-ray surveys,
mainly to probe dynamical formation of close binaries. For example,
\cite{Getal2012} discovered a candidate cataclysmic variable in their
survey of the intermediate-age rich open cluster, NGC 6819, using
\xmm. Globular clusters have been more frequent targets of X-ray
surveys, usually with \chandra; for example, of the X-ray sources
in 47 Tuc \citep{Getal2001}, about 30 are considered likely \cvs\
\citep{Eetal2003a,Eetal2003b}.

\subsection{Imaging Analysis}

The X-ray emission region on a white dwarf is far too small
to be resolved by direct imaging for the foreseeable future.
However, nova ejecta, after several years, can become large
enough to be resolved with current technology. First such
detection was made by \cite{BO1999} in their \rosat\ observation
of GK~Per (Nova Persei 1901). This is by far the best observed
X-ray shell around a nova, likely due to the unusual circum-binary
environment \citep{Betal1987}. Most recently, \cite{Tetal2015} compared
the \chandra\ images taken in 2000 and 2013 and detected the
expansion of the X-ray nebulosity at a rate of 0$"$.14 yr$^{-1}$.
The remnant has also significantly faded in X-rays by 30--40\%
during these 13 years, which they interpret as due to adiabatic
expansion, since radiative cooling time is far too long to account
for this.

In the case of GK~Per, extended X-ray emission is obvious to the
eye in the \chandra\ data without sophisticated analysis. Even so,
a careful analysis is needed to quantify the spatial extension:
the central source, the binary GK~Per itself, is a bright enough
source that one must worry about readout streak, and the source
is piled up at the center. For all other \awds, the spatial extent
of the detectable X-ray nebulosity is of order 1$''$, only slightly
larger than the point-spread function (PSF) of \chandra, so specialized
processing and analysis are essential. This involves sub-pixel
event repositioning, ray-tracing simulation to model the PSF
appropriate for the spectral shape and detector position of the
central source, and maximum likelihood de-convolution (as was done
for RS Oph: \citealt{Letal2009}). The resulting X-ray images, in
conjunction with images taken in other wavelengths, have the potential
to provide valuable insights into the energetics and spatial distribution
of the nova ejecta as well as the circum-binary environment.

In addition, extended X-ray emissions have been detected in a few \syss, 
and are interpreted as due to thermal emission from non-relativistic
jets (see, e.g., \citealt{Netal2007,Ketal2010}). In the case of Mira AB
system, there is a faint X-ray source linking the two stars \citep{Ketal2005}.
However, such analysis is technically challenging and positive detection
of extended X-ray emission is rare.

\subsection{Timing Analysis}

The X-ray emission from \awds\ is variable on a variety of time scales,
from seconds to $>$centuries. Periodic signals can be seen on the orbital
and the spin periods, \edit2{sometimes also on the beat periods of the two,
and their harmonics \citep{NBT1996}}; quasi-periodic oscillations (QPOs) have
been detected in some cases; and aperiodic variability also contain some
valuable information, at least in principle.

\subsubsection{Sampling}

Pointed observations with observatories in high Earth orbits
(\chandra\ and \xmm, as well as \exosat, which operated during
1983--1986: \citealt{EXOSATref}) can observe a target continuously
for up to several days. More often, X-ray observatories are
placed in low Earth orbits, with spacecraft orbital periods near
95 min. Some of these observatories minimize the number of slews by
pointing towards a single target for hours to days. In such cases,
good on-source data are interrupted by Earth occultations, and
passage of the satellite through high particle background regions
such as the South Atlantic Anomaly (SAA). Some low Earth orbit
observatories (e.g., \rosat, \xte, and \swift) achieve high overall
observing efficiencies by interleaving observations of multiple
targets during each spacecraft orbit. Data taken with such missions
have even longer gaps and span a longer calendar time to achieve the
same good on-source exposure time.

The presence of frequent data gaps of significant durations (comparable
to the periods of interest) makes the use of the fast Fourier transform
(FFT) algorithm rather problematic. Discrete Fourier transform (DFT)
is far more suitable, has well understood statistical properties
\citep{Scargle1982}, and there is an implementation that is computationally
fast enough to tackle large data sets \citep{PR1989}.

Regardless of how the light curves are analyzed, one unfortunate fact
remains: whenever one uses data taken with low Earth orbit satellites,
one encounters difficulties when searching for periods near $\sim$95
min, $\sim$190 min, $\sim$285 min etc., where quite a few \cvs\ are
found. This situation is analogous to the problem of 1-day aliases
and the difficulties detecting period near 1 day that plague ground-based
observations.

\subsubsection{Background Subtraction}

The need for background subtraction is obvious in non-imaging
observations. For example, in the \xte\ PCA observation of the
IP, V709~Cas, reported by \cite{dMetal2001}, the raw count rate
averaged 110 \cps\ in all five proportional counter units
of the PCA. After background subtraction, the net
rate was 35 \cps.  Note that V709~Cas is among the X-ray brightest
\cvs\ on the sky. Useful results have been obtained for \awds\ more
than an order of magnitude fainter than V709~Cas using \xte. While
the accuracy background subtraction (rather than the counting
statistics of the source photons) is the limiting factor in the
resulting light curves in such cases, there is a clear recommendation
from the instrument team on how to do this, and the PCA background
model has proved to be sufficiently accurate for most purposes.

During a \suzaku\ observation reported by \cite{Yetal2010},
the count rate for V709~Cas with one of the units of XIS was
$\sim$1.8 \cps, of which $\sim$0.14 \cps\ is likely to be
due to background\footnote{Note that Table 1 of \cite{Yetal2010}
reported the net XIS rate in the 3--12 keV range; the numbers above
are for the entire XIS bandpass.}. In this case, some timing analysis
can be performed without any background subtraction, but that is
far from ideal. Background subtraction is clearly essential for
fainter \awds.

Both for non-imaging and imaging instruments, the background is
a mixture of X-ray and particle events. The former includes contributions
of faint, unresolved, background AGN and hot gas in our Galaxy.
The particle event rate is strongly variable as a function of the
location of the satellite in the geomagnetic field, hence of time.
To account for both in data taken with imaging X-ray telescopes, one
generally extracts the background from a source-free region on the
same detector during the same observation. An ideal background region
must be large enough to introduce little additional statistical errors,
close enough to the source extraction region so that it can be considered
to have the same X-ray and particle background levels as the source extraction
region, yet far enough from the object of interest so that it contains
no source photons.

In practice, one often assumes both X-ray and particle background to
be flat near the center of the FOV, so that a single scaling factor
(the ratio of detector areas covered by the background and the
source extraction regions) can be used. Since this is the default for
spectral analysis when using {\tt xspec}, spectral and timing analyses
can serve as a useful cross-check. If the light curve is modulated
on the spacecraft orbital period, this may indicate a background
subtraction problem. If the net spectrum shows features known to be
present in the background spectrum (such as instrumental lines),
there probably is a problem with background subtraction. When neither
the light curve nor the spectrum shows suspicious features, it is
a likely indication that any systematic errors due to background
subtraction are smaller than statistical errors.

The other ideals of background regions are difficult to realize for
missions with large PSFs, particularly those with large extended
wings in the PSFs, such as \asca\ and \suzaku. Some compromises
are inevitable; one should always keep the potential consequences
of those compromises in mind. Also, background light curves can
be averaged over multiple time bins to smooth out statistical
fluctuations, as long as they add up to an interval much smaller
than the spacecraft orbit. However, if the background region also
contains a fraction of the source count due to the wings of the PSF
(true for \asca\ and \suzaku), and if the source varies suddenly (e.g.,
during a total eclipse), this procedure can introduce artifacts
in the net light curve.

\subsubsection{Warner's Warning}

\cvs\ are variable across many time scales. As \cite{Warning} warned
in the context of optical observations, random processes can lead to
the impression of a periodic signal where only aperiodic variability
exists, when a short segment of a light curve is inspected. This is
also true of periodograms: there is always a highest peak in any
periodogram, so finding a peak is not the same thing as establishing
the presence of a periodic signal. If the underlying aperiodic
variability is frequency independent (``white noise''), then DFT
as modified by \cite{Scargle1982} provides a statistical test, in the form
of ``false alarm probability.'' However, the variability of \awds\ is
often frequency dependent. Ideally, more sophisticated methods
should be used for \awds, such as the autoregressive technique of
\cite{Hetal2004}.

Another simple, yet highly effective, check is to see if the same
period is detected in multiple data sets. This also allows to test
the coherence of the signal. On the other hand, if a period is detected
with apparently high statistical significance but not seen in another
data set of comparable or higher quality, it is quite possible that
the assumption behind the statistical test did not match the reality
of the object.

\subsubsection{Eclipse Analysis}

The procedure to analyze eclipse light curves of CVs is well established
for optical data (see, e.g., \citealt{P1981b}) and can also be used for X-ray
light curves if a total eclipse is observed. First one measures the
times of the four contacts (the beginning and the end of ingress and
the beginning and the end of egress in order). One then measures the
eclipse duration, defined as the length between the mid-ingress
to the mid-egress, as well as the ingress/egress durations (the two should
be the same). This procedure differs from that used for high mass X-ray
binaries (HMXBs; see, e.g., \citealt{CCK2015}), for which the eclipse
duration is defined as the time interval between the second and the third
contacts.

In both cases, the eclipse duration reflects the size of the part of the
mass donor that passes in front of the X-ray emission region divided by the
relative velocity of the two stars in the plane of the sky. In HMXBs, the X-ray
emission region has the size of the same order as that of the neutron
star, and is negligible, so the duration of totality can be used. For
\cvs, the diameter of the white dwarf (of order 15,000 km) is a significant
fraction of the size of the secondary, and therefore one must use the
eclipse duration defined as mid-ingress to mid-egress: these are when
the edge of the secondary passes through the center of the eclipsed object.
The size of the eclipsed object can be inferred from the ingress/egress
durations in \awds, because the edge of the secondary is sharp. This
is not the case for HMXBs, in which the ingress and egress durations
are often seen to be variable, due to the variable absorption by the
stellar wind.

Note also that the ratio of the projected size of the secondary to the
diameter of the white dwarf is of order 10 for \cvs\ below the period
gap. Therefore, it is inevitable that a fraction (of order 1/10th) of
\cvs\ in which the white dwarf is eclipsed should show a partial,
rather than a total, eclipse of the white dwarf. Prominent examples
are the low luminosity IP EX~Hya \citep{Retal1991} and the dwarf nova
V893~Sco \citep{MZS2009}.

\subsection{Spectral Analysis}

Once source and background spectra are extracted, and the response
file(s) downloaded or generated, one can proceed to spectral fitting.
Here, I provide some guidance on the usage of some common models
available in the X-ray fitting package, {\tt xspec}.

\subsubsection{Optically Thin Emission}

When the nature of the object is unclear, it is often customary
to fit the the medium energy X-ray spectrum with simple analytical
models, such as a blackbody, a power law, and a Bremsstrahlung.
Of the three, we do not expect a blackbody fit to be successful
for \awds\ above 1 keV, and if it is, it should indicate an unphysically
high temperature (see the Eddington limit argument in \S 2). Medium
energy X-rays from \awds\ most likely originate from optically thin,
thermal plasma, and the Bremsstrahlung model is likely to provide
a good description of the continuum. However, power law can also
often provide an adequate description, regardless of the underlying
emission mechanism, within a limited energy range. Moreover, an
optically thin thermal component seen through a complex absorber
can give the appearance of being flatter than a Bremsstrahlung
continuum (see \S 5.4 below).

The first discrete spectral feature to be noticed in the X-ray spectrum
of \awds\ is the Fe K complex near 6.6 keV, initially fit as a single
Gaussian. An early example includes the \edit2{\ginga} spectrum of SS~Cyg
\citep{YIO1992} who derived a Bremsstrahlung temperature of 17.5 keV
and a line centered at 6.7 keV. The line energy suggests that the
line is dominated by the He-like Fe, confirming the thin thermal
nature of the emission. At this point, it makes sense to fit X-ray
spectra of \cvs\ using modern models of thermal emission from collisionally
excited thermal plasma, including both the Bremsstrahlung continuum
and discrete lines\footnote{See the home pages of the two major efforts
currently active in this field, ATOMDB (http://atomdb.org/) and SPEX
(http://www.sron.nl/spex). In addition to ever increasing, but still
imperfect, knowledge of atomic physics, the parameter regime occupied
by \awds\ is often outside the main focus of both these groups, sometimes
resulting in the lack of suitable models (e.g., for plasma of sufficiently
high temperatures).}.

However, in a pure \kt=17.5 keV plasma, Fe should be completely ionized.
Therefore, \cite{YIO1992} inferred that the emission was multi-temperature
in nature, one of the first observational demonstration of this fact in
any non-magnetic \cvs. High S/N observations with CCD detectors can resolve
the Fe complex into H-like (6.97 keV), He-like, and fluorescent (6.4 keV)
components, and the presence of K-shell lines from medium-Z elements clearly
demonstrate the multi-temperature, optically-thin, thermal nature of the
emission (see, e.g., the \suzaku\ observation of SS~Cyg; Figure 3 of
\citealt{Ietal2009}).

\begin{figure}[t]
\plotone{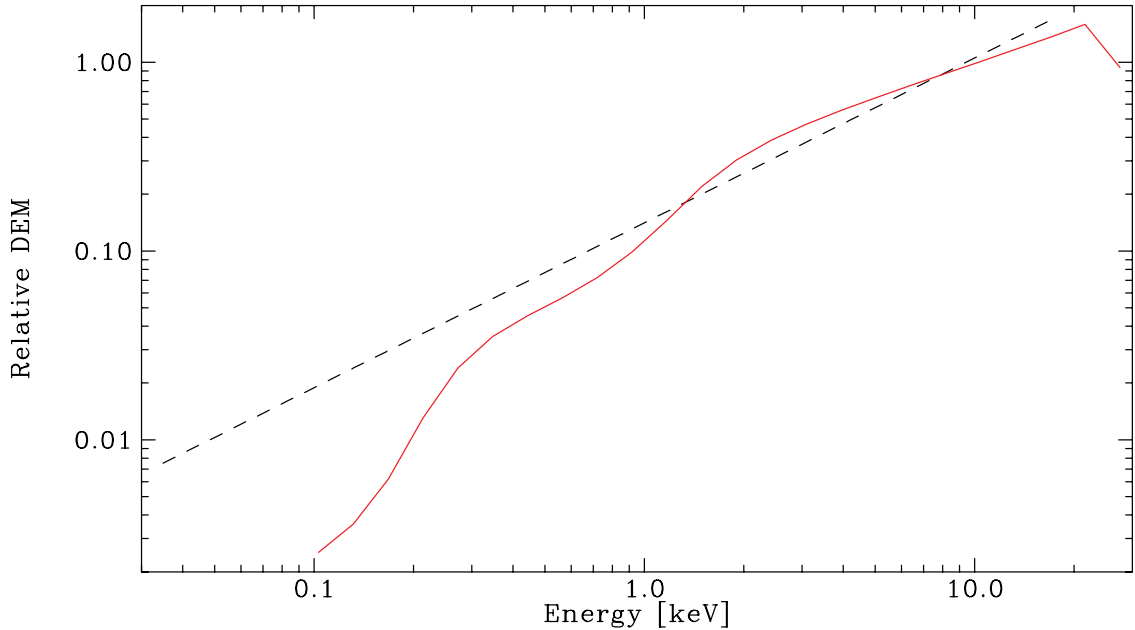}
\caption{A comparison of a power-law DEM distribution with power law index
         1.0, appropriate for pure Bremsstrahlung cooling with no additional
	 energy source or cooling term, and that of the cooling flow model,
	 {\tt mkcflow}. \label{demcomp}}
\end{figure}

Finally, the plasma pressure is nearly constant in an accretion column.
Therefore, the isobaric cooling flow model ({\tt mkcflow} in XSPEC;
\citealt{MS1988}) provides a good approximation for many \awds.
Note that, when using the {\tt mkcflow} model in XSPEC, a small but non-zero
redshift ($z$) must be specified. This parameter is used to redshift the model
and, in combination with the cosmological parameters, as a proxy for distance.
One can reset the Hubble constant to 50 km\,s$^{-1}$Mpc$^{-1}$
and the cosmological constant to 0.0 for ease of conversion between
$z$ and distance (e.g., $z=5\times 10^{-8}$ corresponds to a distance
of 300 pc using these cosmological parameters). One important advantage
of this model is that the cumulative contribution of line cooling is
properly taken into account in calculating the cooling time at each
temperature, which is often not the case for models specifically created
for \cvs\ (including the original Aizu model). The expected emission
measure distribution (DEM) is a pure power law with index of 1.0 for
Bremsstrahlung cooling,
while the DEM for a cooling flow is more complex (Figure\,\ref{demcomp}).
Note that, if one calculates the temperature and density structure assuming
that only Bremsstrahlung cooling operates, and then integrates the
appropriate plasma (continuum+line) emission based on that profile,
the strength of low-temperature lines will be overestimated.
On the other hand, an important limitation of {\tt mkcflow} model
is that it does not account for the pressure increase toward the
bottom of the accretion column. This is a small but real effect
in magnetic \cvs, as the post-shock plasma near the top has a
non-negligible ram pressure, and retain 1/16th of the kinetic
energy of the pre-shock flow.

\subsubsection{Optically Thick Emission}

The first model to be applied to the soft components in \awds\ is
the blackbody. Although this is an idealized model and one expects
that real astrophysical objects do not radiate as a perfect blackbody,
it provides a useful first approximation. Since the total luminosity
is proportional to the emitting area and effective temperature ($T_{eff}$)
to the fourth power ($L \propto A \times T_{eff}^4$), it can be used to
estimate the emitting area $A$ (\S 5.2). However, one must be cautious
because we often do not observe the peak of the blackbody curve. This
can lead to the best-fit $kT$ and \nh\ values to be strongly anti-correlated,
and hence a large uncertainty in inferred luminosity and area,
even within the blackbody approximation.

The problem is even more severe if one considers a more physically
realistic model: stellar atmosphere. As I already discussed in
\S 2, the physical structure of the emission region is an active
research topic and there are a large number of free parameters
in the case of novae. I discuss the case of the soft component in
magnetic \cvs\ in depth in \S 5.6.

\subsubsection{Absorption Models}

Currently, the best model of X-ray absorption by the ISM is {\tt tbabs}
and its successor, {\tt tbnew} \citep{WAC2000}, which accounts for the
fact that a significant fraction of ISM exists in the form of molecules
and dusts. However, intrinsic absorbers within \cvs\ and \syss\ probably
exist largely as atoms. In modeling them, it is in fact a drawback that
{\tt tbabs} and {\tt tbnew} account for the presence of molecules
and dusts. The older absorption model, {\tt phabs}, may well be the
best option, in combination with the {\tt XSPEC} convolution model
{\tt partcov} which creates a partial covering absorber model based
on any absorber model. Note that {\tt pcfabs} is a partial covering
model that is hard-wired to use the even older absorption model,
{\tt wabs}, and so no longer recommended.
Instead of molecules and dusts, one effect that should be considered
is the partial ionization of intrinsic absorber, a topic we will return
to in \S 5.4.

\section{Identifying \awds\ and Determination of Magnetic Nature}

As this review makes clear, \awds\ display a wide range of characteristics
and so there is no hard and fast rules that can be applied mechanically
to identify all of them when an unknown X-ray source is examined. For
non-experts' convenience, I list typical properties of X-ray emissions
from \awds\ in Table\,\ref{typical}. These properties, along with
information obtained through optical observations, including the X-ray
to optical flux ratios (Figure\,\ref{fxfopt}), provide a useful guide
to identify likely \awds\ among poorly studied X-ray sources.

\begin{deluxetable}{llrlcc}
\tablewidth{0pt}
\tablecaption{Typical X-ray Characteristics of \cvs\ and \syss\ \label{typical}}
\tablehead{
\colhead{Energy Source}  & \colhead{Energy Range} &
\colhead{Size\tablenotemark{a}}  & \colhead{Time Scale\tablenotemark{b}} &
\colhead{Luminosity (\eps)}  & \colhead{Temperature (keV)}
}
\startdata
Accretion       & Hard   & WD     & Short & 10$^{29}$--10$^{33}$ & 5--50 keV \\
                & Soft   & WD     & Short & 10$^{31}$--10$^{34}$ & 20--50 eV \\
Nuclear Burning & Hard   & Large  & Days  & 10$^{33}$--10$^{35}$ & 1--30 keV \\
                & Soft   & WD     & Short & 10$^{35}$--10$^{38}$ & 30--80 eV \\
Colliding Winds & Medium & Binary & Days? & 10$^{32}$ ?          & 1 keV \\
\enddata
\tablenotetext{a}{X-ray emission region size, compared to the white dwarf, the binary, or larger. Only the hard X-rays from novae (under``nuclear burning'') can be spatially resolved using the current and near-future technology.}
\tablenotetext{b}{Shortest variability time scale, which can be as short as $<$10 s.}
\end{deluxetable}

Objects showing a distinct soft component can be considered candidate polars,
since many polars are known to be soft X-ray bright \citep{KW1987}.
See, however, \cite{HM1995} for IPs with a soft component.  In the X-rays,
polars exhibit a strong modulation on a single period (\pspin = \porb) or
two similar periods (if asynchronous). However, true verification of the
polar nature requires optical observation, such as detection of circular
polarization or cyclotron humps.

The presence of the HeII $\lambda$4686 is often used to judge if a CV
is magnetic or not. While this is a useful diagnostic, the mere presence
of this line is far from conclusive. Polars generally have a soft component
with strong ionizing continuum shortward of the HeII 228\AA\ edge, and IPs
probably have a similar soft component. Note, however, that the hot white
dwarf photosphere of a recent nova, or the optically thick boundary layer
of non-magnetic CVs, are also capable of ionizing HeII. Another important
factor is the accretion column or curtain of magnetic \cvs, which produce
strong emission lines but little continuum in the optical. In polars,
there is no accretion disk to dilute the stream emission; in IPs,
the truncated disk provides some dilution. This explains the empirical
rule-of-thumb proposed by \cite{S1992}: if the equivalent width of the
H$\beta$ line is greater than 20\AA, and the HeII $\lambda$4686/H$\beta$
ratio is greater than 0.4, then the object is likely to be a magnetic CV.
In \cvs\ studied by \cite{S1992}, however, many IPs have weaker H$\beta$
lines and cannot easily be \edit2{separated} from non-magnetic systems using
this diagnostic.

A large fraction of \awds\ detected in hard X-ray surveys are IPs.
Most IPs have a flat X-ray continuum that cannot be explained
as due to simply absorbed thermal emission; these same systems
have strong 6.4 keV fluorescent lines (see, e.g., \citealt{EI1999}).
Non-magnetic \cvs, on the other hand, tend to be less absorbed
and have weaker 6.4 keV lines. However, there are exceptions to
both these trends, so they cannot be used for a definitive
classification. A stronger statement can only be made if the X-ray light
curve shows a coherent and persistent period, and if this is not the
orbital period: it likely \edit2{represents} the detection of the spin period.
In many well-known IPs, there is an obvious and dominant peak in the
power spectrum at the spin period, with amplitudes in the 10s of percent
range (see, e.g., \citealt{NW1989}), and the exact details of how the
data are analyzed do not matter.

However, there are also IPs with relatively weak spin modulations,
and there are objects which are presumed non-magnetic even though they share
some characteristics with magnetic objects. In searching for periodicities,
one can emphasize sensitivity and accept certain number of false
positives, or one can emphasize reliability and accept a lower
sensitivity (see \S 3.3 above). Note, however, that most studies
estimate false alarm possibility assuming a white (i.e., frequency
independent) \edit2{noise} using, e.g., the Lomb-Scargle periodogram
\cite{Scargle1982}. In reality, many \awds\ exhibit a red noise,
so a more sophisticated methodology (see, e.g., \citealt{Retal2008})
is required to assess the significance of an apparent peak in
the periodogram.  \cite{Retal2008} considered three objects to
be candidate IPs based on the complex absorber in their X-ray
spectra. Of the three, one (EI~UMa) was later recognized as an IP
\citep{TWRetal2008} while the other two (V426~Oph and LS~Peg)
still lack a definite spin signature. They may well be non-magnetic
systems with some resemblance to IPs, or they may be IPs whose
spin signature is almost perfectly hidden.

The true hallmark of the white dwarf spin period is the stability
of the clock \citep{P1981a}, so it is a good idea to treat any
proposed spin period as merely a candidate until the same period
is detected in another observation, to within the expected limits
of white dwarf spin up/down. This also presupposes that the spin
period is often, if not always, detectable given observations of
a comparable quality.

Another useful check is the energy dependence of the spin modulation.
In IPs, the spin modulation is generally more prominent at lower
energies, although not as strongly as would be expected if the
spin modulation was purely due to a single photoelectric absorber
\citep{NW1989}. However, this is not generally the case for polars,
or for some IPs.

When a late type giant is found to be the likely optical candidate
for an unknown X-ray source, it may be a symbiotic star with a white
dwarf accretor. While they exhibit a wide variety of X-ray spectra
(see \S 6.5), all are thermal emissions\footnote{A late type giant associated
with a non-thermal X-ray source probably is a symbiotic X-ray binary
with a neutron star accretor (see, e.g., \citealt{Cetal2008}).}. This
should be checked by searching for emission lines (see, e.g.,
\citealt{LC2005}) or additional blue and variable continuum, as per
the original definition of \syss. Note, however, that the optical
emission lines in some X-ray bright \syss\ are relatively weak.
For example, the recurrent nova and BAT source, T~CrB, can become very
weak at times \citep{MDC2016}. Recently, \cite{Metal2016} identified
a newly discovered BAT source with a poorly studied red giant, SU~Lyn,
whose emission lines are undetectable in low resolution optical
spectroscopy (they are clearly present in high resolution spectra).
Thus, if a candidate optical counterpart of a X-ray source is a red
giant without obvious emission lines in low resolution data,
it should not be immediately discounted without first taking high
resolution optical spectra or UV data.

In a comprehensive study of X-ray sources in M31 and M33 detected
with \rosat, \xmm, and \chandra, \cite{Petal2005} found that a large
fraction of \ssss\ in M31 are novae. This demonstrate that X-ray
observations are capable of discovering novae in their supersoft
phase out to local group galaxies and beyond, even though M31 itself is
well monitored optically for novae so few are likely to be discovered
through their X-ray emissions first. Such independent discoveries
are harder in the Galaxy, because of the larger solid angle involved
and due to the ISM on the Galactic plane. For example, assuming an
average \edit2{ISM} density of 0.1 cm$^{-3}$, the column density is
2.0$\times 10^{21}$ cm$^{-2}$ at 6.5 kpc. This is high enough to
significantly reduce the count rate of any supersoft component.
Nevertheless, one object, V1375~Cen (=XMMU~J115113.3$-$623730),
originally discovered in the \xmm\ slew survey as a soft source,
is thought to have been a nova, based on post X-ray discovery
follow-up observations \citep{Getal2010}.

The hard X-rays from shocked nova ejecta are less well observed,
but they appear to be widespread and have peak luminosities in
the 10$^{33}$--10$^{35}$ \eps\ range \citep{MOD2008}. They predicted
that some faint X-ray transients in the Galactic Center region
may turn out to be associated with novae, which may have been
confirmed by the pre-discovery \chandra\ detection of the IR-discovered
nova, VV-NOV-13 \citep{OMD2016}.

\section{X-ray Emission from Magnetic \cvs}

Next, I highlight selected results on accretion onto magnetic white
dwarfs. Because Z~And is the only confirmed magnetic symbiotic system
\citep{SB1999}, and there are no high quality X-ray data on this object
(the \xmm\ and \chandra\ observations obtained during an outburst
are of low statistical quality, and probably tell us more about the
outburst than about the accretion processes; \citealt{Setal2006}),
this section will entirely focus on magnetic \cvs. However, 
the apparent lack of magnetic \syss\ may simply reflect the difficulty
in identifying magnetic systems while shell burning is in progress.
When more \syss\ driven purely by accretion are discovered, a fraction
may well turn out to contain a magnetic white dwarf.

There are relatively few in-depth observations of polars with
\chandra\ and \xmm. One reason for this is that polars are frequently
found in low state (16 of 37 in an \xmm\ survey of polars;
\citealt{Retal2004}). This makes observations risky: proposals
for regularly scheduled observations of polars are often rejected
because the success is not guaranteed, and proposals for 
target-of-opportunity (TOO), or triggered, proposals are often
rejected because competition for such constrained time is stronger.
Regardless of the reasons, there is no option but to rely heavily
on older X-ray observations for several key aspects of polars.

For magnetic \cvs, timing and spectral analyses must be considered
together to establish a coherent picture. Not only can such studies
reveal the detailed physics of the X-ray emission regions, but also
allow us to estimate the white dwarf mass (\mwd) and gain some insight into
how the magnetic field interacts with the accretion stream.

\subsection{X-ray Modulation of Polars}

In discussing magnetic \cvs, it is useful to establish a terminology
to distinguish the two poles of a magnetic dipole. In this review,
I will call the pole closer to Earth (i.e., on the same side of the
orbital plane as Earth) the ``upper pole,'' and the other the ``lower
pole.'' I will largely avoid the terms ``1-pole'' and ``2-pole'' due
to possible confusion, as these can refer to the number of poles that
are visible from the Earth, or the number of poles that are accreting.

Polars accrete without an intervening accretion disk. The matter
lost from the inner Lagrangian (L$_1$) point of the secondary
first follow a ballistic trajectory, and subsequently captured
by the magnetic field and follow the field lines to the white
dwarf surface. The locations of the interaction region, and hence
the accretion footprint, are determined by the complex interaction
between the accretion flow and the magnetic field. To first order,
the interaction region is where the ram pressure of the ballistic
stream and the magnetic pressure are equal. If the accretion
is predominantly to the upper pole, then the magnetic stream should
also be curved above the orbital plane; lower pole, below.

Some polars, such as ST~LMi (=CW~1103+254) and VV Pup, exhibit simple
X-ray light curves consisting of a faint phase, during which they are
hardly detected, and a bright phase (see Fig.~2 of \citealt{Mason1985}).
The faint phase is indicative of self occultation: the accretion region being
hidden by the solid body of the white dwarf. In these systems, the faint
phase lasts for over half a cycle; one therefore infers that accretion occurs
predominantly onto the lower pole\footnote{Therefore, ST LMi is an
example of a ``1-pole'' system in terms of physical amount of accretion,
which is, at the same time, a ``2-pole'' system if one counts the number
of observable poles.}. Since the hard X-rays are emitted above the surface
in optically thin region, the hard X-ray light curves are shaped
predominantly by self occultation. On the other hand, the soft X-rays
are emitted from the surface that is optically thick; projection and
limb-darkening effects are also important. The hard and soft X-ray
light curves of ST~LMi are qualitatively consistent with such simple
expectations, with perhaps a modest spatial extension of the emission region.

The prototype polar, AM~Her typically show a faint phase lasting
$\sim$0.2 cycle (see, e.g., \citealt{Metal2000}), suggesting that
accretion mainly occurs on the upper pole that suffers a short self
occultation. However, using archival {\sl SAS-3\/} data, \cite{PMH1987}
reported the disappearance of the faint phase in 1976 November in both
soft and hard X-rays. In contrast, during the 1998 August high state,
modulation was absent in soft X-rays but persisted in hard X-rays.
More puzzlingly, 1983 \exosat\ observation caught it in a so-called
reverse soft X-ray mode, during which the hard X-ray light curve was
similar to that during normal mode, but soft X-rays were brighter for
$\sim$half a cycle centered on what is normally the faint phase
\citep{Hetal1985}.

The soft X-ray light curves of EF~Eri, AN~UMa, V834~Cen (=E1405$-$451),
and QQ~Vul (=E2003+225) are even more complex (Fig.~1 of
\citealt{Mason1985}). The hard X-ray light curve of EF~Eri can be
described as sinusoidal, and that of V834~Cen may be considered similar.
The lack of faint phase indicates either that the active pole is always
visible (so it must be the upper pole) or that both poles accrete. One
particular feature is a dip lasting about 0.1 cycle, which is thought
to be due to occultation by the magnetic stream toward the upper pole
\citep{KW1985}. There is a narrower dip $\sim$0.1 cycle later in AN~UMa
and V834~Cen \citep{Mason1985}, which these papers do not explain. Neither
is there a full explanation for the general complex shape of the soft X-ray
curves of these polars.

One potential scheme to interpret such complex light curves is
treat each peak as representing a distinctive emission region,
possibly invoking a complex magnetic field geometry beyond a simple
dipole. However, it is worth emphasizing that a presence of multiple
emission regions, even if confirmed, does not necessarily imply a complex
magnetic field geometry.  The presence of other factor is obvious
because both QQ~Vul and V834~Cen, seen to have complex soft X-ray light
curves in their first \exosat\ observations \citep{Mason1985},
were seen to have drastically different light curves in later
observations \citep{Oetal1987,Setal1994}. In the first two
\exosat\ observations of QQ~Vul, there was a broad peak at
phase 0.8--1.0, and a much narrower peak at 0.45--0.55.
During the third observation, there were two roughly equal peaks
at phases 0.25 and 0.65. Since the magnetic field configuration
cannot conceivably change within a few years, the changes, hence
an important part of the reason for complex light curves,
must lie in factors that can change from epoch to epoch.

\begin{figure}[t]
\plotone{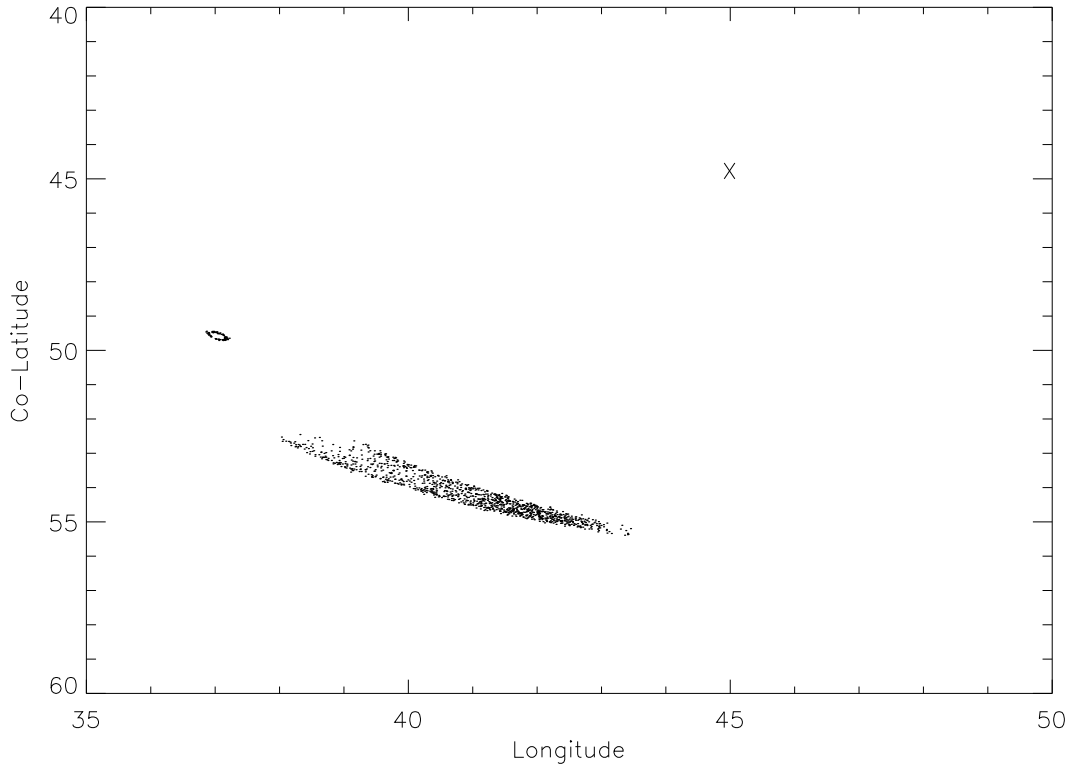}
\caption{An example of the elongated spot shape expected in polars,
           in the simplest magnetic field geometry of a centered dipole.
	   See text for details. This is Figure 4.16 of \cite{CashThesis},
           reproduced with permission. \label{CashSpot}}
\end{figure}

The interaction of the magnetic field and the ballistic stream
almost certainly leads to an elongated accretion spot, or spots,
even if the white dwarf has a simple, centered, dipole field
\citep{M1988,CashThesis}. The key ingredients of these studies
are (1) a realistic density profile (higher density core and
lower density halo) of the stream that leaves the L$_1$ point
\citep{LS1975}; and (2) the assumption that each element of the stream
is captured by the magnetic field according to the local balance
of ram and magnetic pressures.  Thus, different parts of the
stream couple to magnetic fields at different locations along
the ballistic trajectory.  A sample map of accretion region
from \cite{CashThesis} is reproduced in Figure\,\ref{CashSpot},
in which a cross marks the location of the magnetic pole.
As can be seen, the main accretion region is significantly
offset from the magnetic pole. Moreover, there is an additional
spot (upper left of the figure), which result from low density
stream coupling directly to the magnetic field at or very near
the L$_1$ point, where the stream velocity and hence the ram
pressure is very low.  

Such double spot structure, with an elongated main spot, has
the potential to generate complex soft X-ray light curves,
which is yet to be fully studied. Moreover, by changing the
mass transfer rate, the spot geometry and the light curve
morphology will change. Finally, the double stream implied
by the double spot structure may explain the double dips seen
in AN~UMa and V834~Cen.

\edit1{Asynchronous polars provide a unique opportunity to study
the coupling of accretion stream to the magnetic field, as the
orientation of the magnetic field relative to the secondary
changes slowly over days. An example of this type of study
is \cite{CLetal2015}, who analyzed optical data on V1432~Aql.}

\subsection{Size of the Accretion Region(s)}

In magnetic \cvs, the accretion rate per unite area (\smdot),
also called the specific accretion rate, is more important in
determining the physical characteristics of the outgoing X-rays
than the total accretion rate ($\dot M$). In recent years,
therefore, X-ray emission models are usually provided as a function
of $\dot m$, typically expressed in the units of g\,s$^{-1}$cm$^{-2}$.
In the early literature, however, it this was more often expressed as
$\dot M$/$f$, where $f$ is the size of the accretion region expressed
as the fraction of the white dwarf surface, allowing a direct comparison
with models of spherically symmetric, radial accretion onto non-magnetic
white dwarfs (see, e.g., \citealt{LM1979}). Note that, for a wide range
of \mwd, \smdot\ of 1 g\,s$^{-1}$cm$^{-2}$
corresponds to $L/f$ of 8.0--9.5$\times 10^{35}$ \eps, because the larger
surface area of the lower mass white dwarf compensates for the smaller
amount of potential energy released per unit mass.

The soft X-ray spectrum contains information about the emitting area,
modulo the distance to the system. However, there are practical
difficulties. First, one does not know the true spectral shape, and
a blackbody fit may not be reliable. Second, in blackbody fits,
there often are correlated errors on \kt\ and \nh, resulting in
large uncertainties in the emitting area and the luminosity determined
from X-ray spectral analysis (see, e.g., Fig. 10 of \citealt{Petal1984}).
Note also that the precise meaning of the accretion spot size, whether
expressed as area or as $f$, can be murky, particularly when blobby
(inhomogeneous) accretion is considered \citep{King2000}.

The most robust method to estimate $f$ is via the measurement of
eclipse light curves. In the X-ray eclipsing polar, HU~Aqr,
\cite{Setal2001} found that ingress is compromised because of
the stream dips of the kind discussed above. However, the egress
is clean and lasts 1.3 s (or 4$^\circ$ or 450 km). Such rapid
transitions can be measured only with the highest quality X-ray
light curves. For example, \cite{Petal2002} could not resolve the
egress duration for DP~Leo or WW~Hor. Note also that eclipse
light curves provides a measurement in one dimension (in longitude),
and hence provides incomplete, if important, information when
the accretion region has a complex shape.

Multiple emission regions have the potential to complicate
eclipse analysis further. In the hard X-ray bright eclipsing
polar, V2301~Oph, \cite{RC2007} found that the ingress consisted
of two steps of rapid ($<$5 s) decline separated by a `standstill'
lasting 26$\pm$4 s. The egress lasted 23$\pm$4 s in a single
step, without a standstill.

Among IPs, only XY~Ari is X-ray bright and deeply eclipsing,
and hence the only object for which emission region size has
been estimated using the eclipse method. There is an obvious
complication, however: the location of the emission region,
relative to the secondary, changes as a function of the spin
phase. Nonetheless, \cite{Hellier1997} found that, by aligning
the individual eclipses, the composite egress (their Fig. 8)
showed 2/3rd of the flux emerging in $<$2 s, with the rest
emerging over a $\sim$20 s period. However, the timing changes
not strictly as a function of spin phase but has jitters
(their Figure 6). Therefore, it appears that, at any \edit2{given} moment,
the majority of the flux is emitted in a small ($f < 0.002$) area,
but the accretion foot points wanders around on a larger
($f<0.01$) area on the white dwarf.

While these measurements can inform our understanding of the
X-ray light curves and spectra, they are not precise enough
to lead to definite, quantitative conclusions. For example,
while the $L/f$ values, or \smdot, of HU~Aqr
and XY~Ari are constrained, it does not pinpoint the vertical
location of these systems in the magnetic field -- $L/f$ diagram
in which X-ray dominated and cyclotron dominated shocks can be
separated (see, e.g., Figure 2 of \citealt{LM1979}).

\subsection{X-ray Spin Modulation of IPs}

\begin{figure}[t]
\plotone{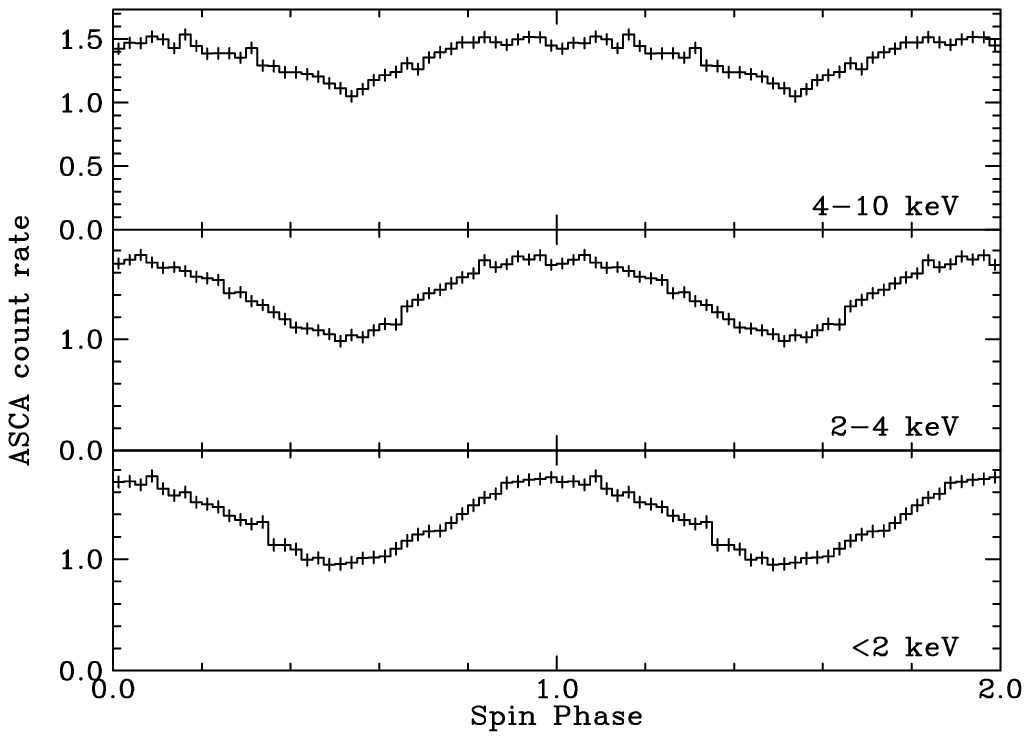}
\caption{An example of energy-dependent X-ray spin modulation of IPs, V1223~Sgr
           observed with \asca. \edit1{Each panel shows the sum of count rates
           in the four telescopes on-board \asca\ in the indicated energy
	   range, folded on the spin period.} \label{v1223spin}}
\end{figure}

As mentioned previously, the amplitude of X-ray spin modulation of IPs
is usually a function of the photon energy, in that the amplitude
is larger at lower energies. However, \cite{NW1989} found that
the energy dependence of spin modulation seen in the \exosat\ data
was not as strong as expected if it was caused by a single photoelectric
absorber: I show an example in Figure\,\ref{v1223spin}. A simple absorber
with a column density  high enough to modulate the 4--10 keV count rate
(\nh $\sim 10^{23}$ cm$^{-2}$) should completely absorb all photons
below 2 keV, yet roughly one half of soft photons are unabsorbed.
\cite{NW1989} therefore introduced the concept of partial-covering
absorber to IPs, which improved the spectral fit for both the
phase-averaged and phase-resolved spectra. Changes in the column
density and/or the covering fraction of the partial covering absorber
can reproduce the energy dependence of spin modulation observed. Since then,
it has become routine for spectral fit to IPs to use partial covering
absorbers ({\tt pcfabs} or any absorber model convolved with {\tt partcov}
in {\tt xspec}), sometimes more than one depending on the
data quality.

\begin{figure}[t]
\plotone{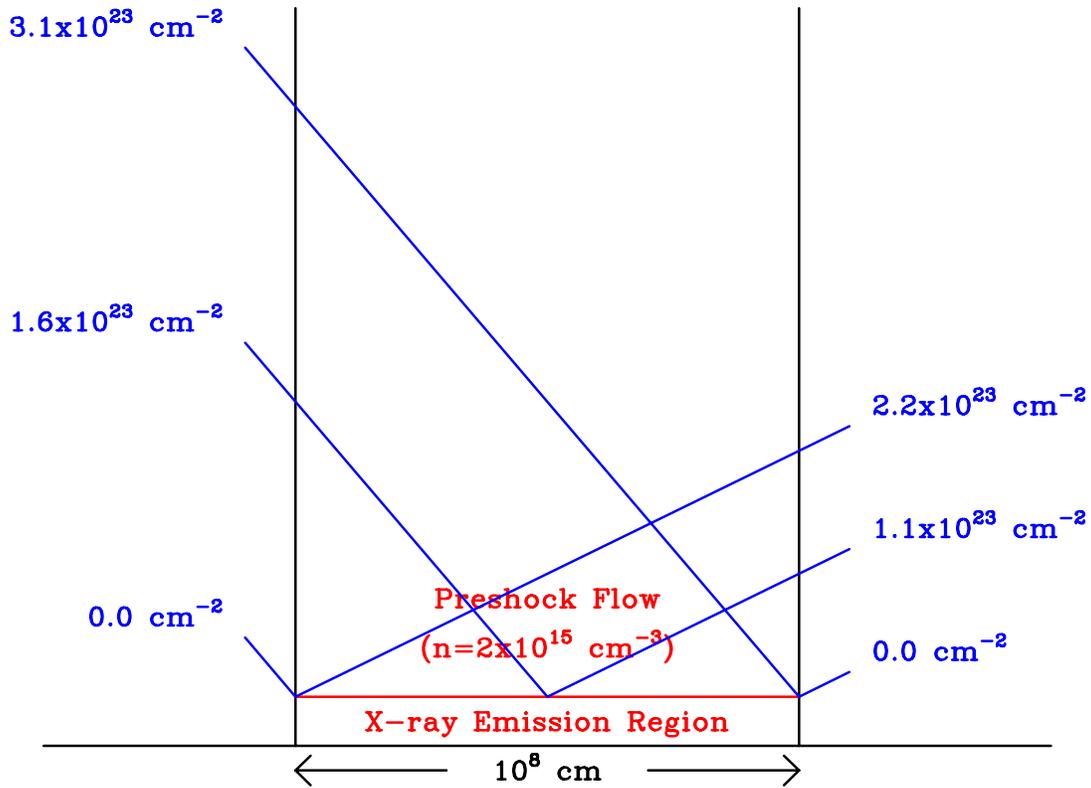}
\caption{A schematic diagram illustrating the pre-shock flow
acting as a complex absorber. \label{pwabschema}}
\end{figure}

\begin{figure}[t]
\plotone{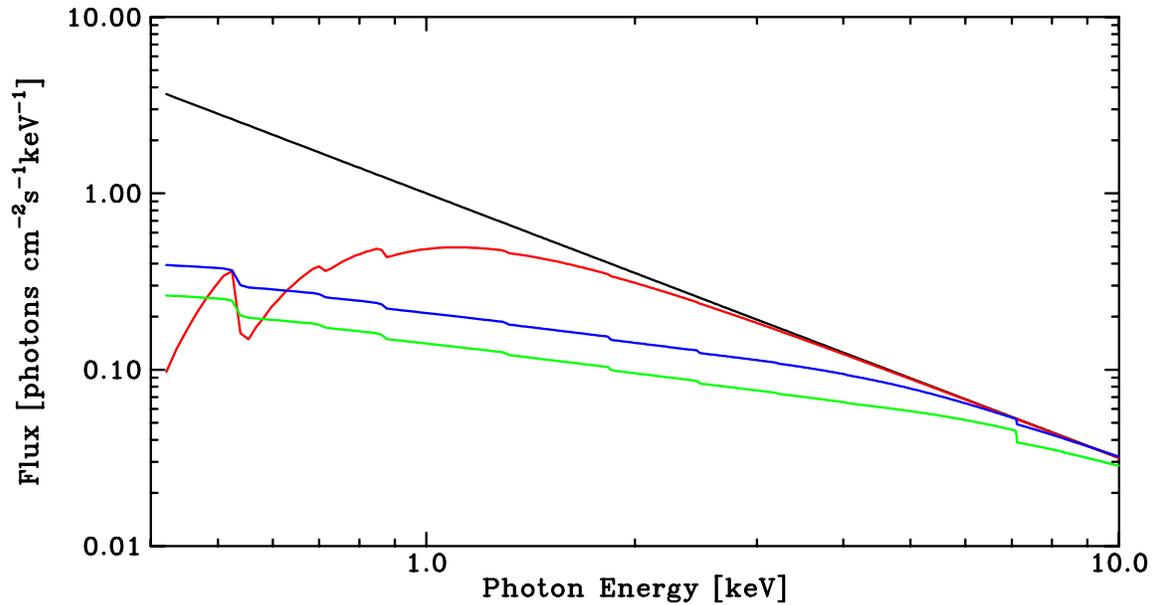}
\caption{The energy dependence of the complex absorber as encoded in
the {\tt pwab} model. Unlike a simple photoelectric absorber, which
results in an exponential cutoff at low energies (red curve),
this absorber is power-law like (blue and green lines, shown for
different maximum values of \nh). \label{pwabresult}}
\end{figure}

The partial covering nature of the spin modulation of IPs is in
clear contrast to the stream dips in polars, which is caused by
a distant part of the stream. The logical candidate location
for the absorber is the immediate pre-shock flow: since the
emission region (the post-shock region) and the absorption
region are of similar size and are adjacent, it is hard for
the absorption to be anything but partial in this case.
In fact, a deeper examination of the geometry leads to the
conclusion that the traditional partial covering models are
not complex enough \citep{DM1998}. As the lines of sights to
different parts of the X-ray emission region go through differing
amounts of matter (Figure\,\ref{pwabschema}), the absorption cannot
be modeled as an absorber with a unique \nh\ or as partial covering
absorbers with several discrete values of \nh, each covering a
fraction of the X-ray source. The geometry of absorption by
the pre-shock flow requires a distribution of covering fraction
as a function of \nh, which \cite{DM1998} approximated as a power
law, and implemented in {\tt xspec} as {\tt pwab}. While a simple
photometric absorber results in an exponential cut-off at low
energies, this complex absorber has an approximately power-law
dependence on photon energy (Figure\,\ref{pwabresult}). In the context
of this model, the spin modulation of IPs can be explained as
changes in the parameters of such a complex absorber, most
likely the maximum \nh.

However, this model may not be applicable to all IPs. In EX~Hya,
there is little evidence for absorption (simple or complex) either
in the phase-averaged spectrum (see next section) or in the
energy-resolved spin light curves. \cite{Aetal1998} therefore
advocated the possibility that the spin modulation in this unusual
IP is due to self-occultation (i.e., by the body of the white dwarf).
\cite{Mukai1999} introduced the concept of ``horizon angle,'' the angle
between the line connecting the emission region to the white dwarf 
and the line of sight. An emission region infinitesimially above the
surface will disappear behind the limb of the white dwarf when the
horizon angle is 90 degrees. However, for a realistic range of
\smdot, the shock height can easily be a few percent
of the white dwarf radius (see their Figure 1), so it would not be
surprising if the horizon angles in some IPs were in the 100--120$^\circ$
range. This means that, assuming these IPs accrete at both poles of
a centered dipole, there is a wide range of viewing angles at which
emissions from both poles are simultaneously observable. This is the
kind of effect envisioned by \cite{Aetal1998}. The same model may also
apply to HT~Cam, another IP with little evidence of absorption
and energy-independent spin amplitude \citep{dMetal2005}.
Note that both EX~Hya and HT~Cam are short period IPs below the
period gap, and hence expected to have low accretion rates. This
naturally results in low \smdot\ and hence tall shocks, which makes
pre-shock absorption geometrically less important, likely making
the visibility of two tall post-shock regions the main mechanism
for spin modulation. Such low luminosity IPs (LLIPs) will be discussed
again in \S 8.

\subsection{Hard X-ray Spectra of Magnetic \cvs}

\begin{figure}[t]
\plotone{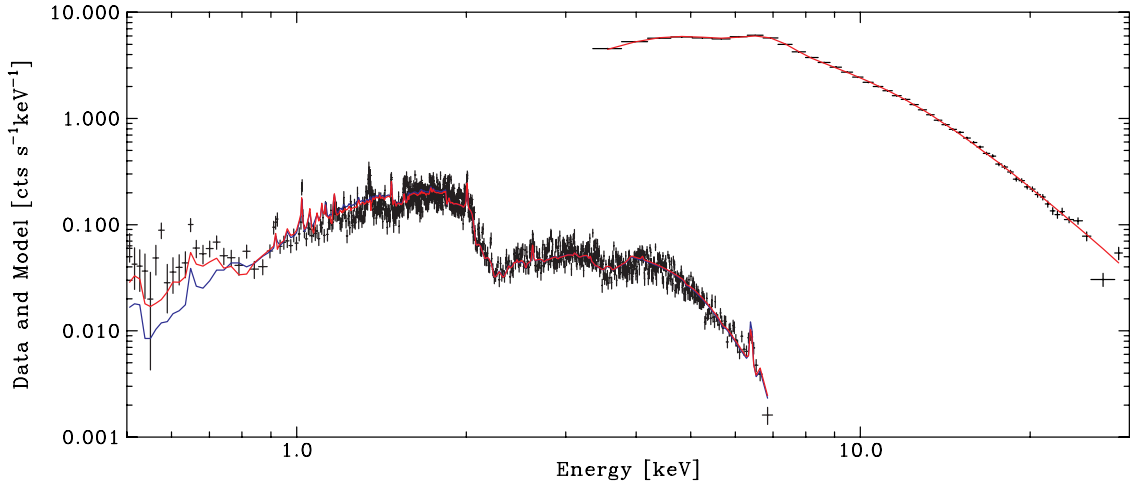}
\caption{Preliminary joint spectral fit to simultaneous  \xte\ PCA and
\chandra\ HETG observations of V1223~Sgr.  A single model (but allowing for
a cross normalization factor between \xte\ and \chandra\ data) was fit to MEG,
HEG, and PCA data (HEG spectrum is omitted from the figure for simplicity).
The blue model is the {\tt mkcflow} cooling flow model modified by the
complex absorber model, {\tt pwab}; in the red model, {\tt pwab} was
replaced with an ionized absorber version of {\tt pwab} currently
in development. \label{v1223spec}}
\end{figure}

\cite{Metal2003} presented early HETG spectra of 7 \cvs\ then
available in the \chandra\ archive, and showed that they can be
divided into two types, ``cooling-flow'' and ``photoionized.''
Non-magnetic \cvs\ SS Cyg, U Gem, and V603 Aql, as well as the
LLIP, EX~Hya, belong to the first type.
The second group is the photoionized type, comprised of 3 IPs,
V1223~Sgr, AO~Psc, and GK~Per.  Their X-ray spectra are characterized
by a hard, power law-like continuum in the 0.5--8 keV range that is too flat
to be reproduced by the cooling flow model modified by simple absorber.
Also, their emission line ratios do not follow the prediction of the
cooling flow model. \cite{Metal2003} therefore fitted the \chandra\
HETG-band spectra of these systems using an {\sl ad hoc\/} power law
with a model of photoionized emission.  However, such a model is clearly
unphysical: the strong shock is the primary energy
source in IPs, and their X-ray spectra above 10 keV are well fit by
cooling-flow type models.  The solution is likely to lie in the
effect of the complex absorber: \cite{Metal2015} showed that the
joint \nustar -\xmm\ spectra of IPs above 1--2 keV can be fit well with
an {\tt mkcflow} model modified by {\tt pwab}. In Figure\,\ref{v1223spec},
I show an example of fitting simultaneous \chandra\ HETG and \xte\ PCA
spectra of V1223~Sgr, again showing that this model can reproduce
the gross features of the X-ray spectra of IPs.

As already discussed in \S 2, reflection plays an important role in
the observed X-ray spectra of \cvs\ and \syss, and one can borrow from the
studies of reflection off the accretion disk in active galactic nuclei
(AGN) and in X-ray binaries \citep{GF1991}.  In magnetic \cvs, the 6.4 keV
line is usually present often with equivalent widths of order 150 eV.
Although this line is also expected from the intrinsic absorber,
statistical analysis of the \asca\ sample led \cite{EI1999} to
conclude that some of the 6.4 keV line is indeed from reflection
off the white dwarf surface. The Compton reflection hump was first
detected in AM~Her by \cite{Retal1981}, and a few other reports
followed sporadically (see, e.g., \citealt{Betal1995,Detal1995}).
Recently, the analysis of joint \edit2{\nustar\ and} \xmm\ spectra of V709~Cas,
NY~Lup and V1223~Sgr by \cite{Metal2015} led to a direct and robust
detection of the Compton hump in these systems.

Not only did \cite{Metal2015} securely detect reflection, they
measured the reflection amplitude. The amplitude is defined as
1.0 when the reflector subtends 2$\pi$ steradian of sky as seen
from the X-ray emitter. If the emission region is just above
the white dwarf surface in magnetic \cvs, we expect a
reflection amplitude close to 1.0. However, at least in V709~Cas,
the reflection amplitude was estimated to be significantly lower
($<$0.60 in the cooling flow fit to the joint \nustar -\xmm\ data).
This can be accomodated within the standard framework if the shock
height is not negligible. For example, an X-ray emission region 0.2
\rwd\ above the surface would result in a reflection amplitude of 0.45.
Such an emission region will also have a horizon angle of 123.5$^\circ$,
so both poles should remain visible for a substantial fraction of the
spin period. Indeed, spin modulation above 10 keV was already seen in
V709~Cas by \cite{dMetal2001} and even more clearly in the \nustar\ data 
by \cite{Metal2015}. This is the most likely mechanism for
hard ($>$10 keV) X-ray spin modulation of IPs, where the
photoelectric absorption is negligible, although viewing geometry
dependence of reflection hump and Compton scattering in the
accretion flow \citep{Rosen1992} also needs to be considered.

Additional complication is the warm absorber edge reported in the
\chandra\ HETG spectrum of V1223~Sgr \citep{Metal2001} and the \xmm\ 
RGS spectrum of V2731~Oph \citep{dMetal2008}. Mukai et al. (in preparation)
have began experimenting with an ionized version of the {\tt pwab}
model, with at least qualitative success (Figure\,\ref{v1223spec}).
This is to be expected, since the immediate pre-shock flow in IPs,
the site of the complex absorber, is situated adjacent to the post-shock
region emitting X-rays at a rate of several times 10$^{33}$ \eps.
\edit2{One} expects warm absorber features when directly viewing the post-shock
region through the pre-shock flow, and photoionized emission features
from parts of the pre-shock flow not in front of the post-shock region,
consistent with the findings of \cite{Metal2003}.

\subsection{The White Dwarf Mass}

The potential utility of X-ray spectra of magnetic \cvs\ for estimating
\mwd\ has been known since at least \cite{Retal1981}.
Several groups have applied this basic idea using a variety of X-ray
data and several different spectral models.

\cite{CRW1998} applied their post-shock region model to the \edit2{\ginga}
2--20 keV spectra of 14 magnetic \cvs\ (polars and IPs) and often
found high values for the \mwd. However, when \cite{Retal1998}
compared \mwd\ estimates for XY~Ari derived from X-ray spectra
(using \edit2{\ginga}, \asca, and \xte\ data) using the same model,
against that derived from X-ray eclipse timings, the former were
found to be systematically too high. One plausible
interpretation is that the continuum shape below 20 keV is too heavily
influenced by complex absorbers and the reflection hump to allow
a reliable estimate of \mwd.

\cite{EI1999} took a different approach, instead focusing on the
intensity ratio of He-like and H-like Fe K lines. In an isothermal
plasma, the ratio can be interpreted as ionization temperature; in
a multi-temperature plasma, it is the emission measure weighted
average of the ionization temperature within the post-shock region,
which can be related to the shock temperature, hence \mwd.
Although this method is potentially subject to systematic
uncertainties of its own due to the accuracy of continuum fit (particularly
the depth of the Fe K edge) and detailed physics of the lines
(see \S 9.5), it does provide a useful check of the broadband fitting
method.

\cite{SRR2005} applied their own structured post-shock region
model to \xte\ PCA and HEXTE data of magnetic \cvs, covering the
3--100 keV range, and obtained lower \mwd\ estimates than those
previously obtained from X-ray spectral fitting. The key is the use of
the broadband data: this approach has the advantage that the data
include the energy range where the exponential cut-off of the
multi-temperature Bremsstrahlung continuum is expected.
Although their spectral model does not include
line emission, and the Fe K triplet is modeled as a single Gaussian,
this is not an issue as long as the high energy continuum shape is
well constrained by the data. The same basic approach was followed by
\cite{Yetal2010} using \suzaku\ XIS and HXD data and their own model
of the post-shock region, which does include the lines.  The disadvantage
of this approach is the relatively poor quality of the HEXTE and HXD
data for most objects, which, combined with the additional complication of
reflection, might introduce systematic errors. For the \suzaku\ work,
an additional weakness is that these IPs are near the faint limit
for HXD, where the systematics of background subtraction is an issue,
coupled with the uncertainties in the XIS-HXD cross calibration.
Nevertheless, these methods have potential to make routine determination
of \mwd\ possible given X-ray data of sufficient quality.

\subsection{The Soft Component of Magnetic \cvs}

In the simplest picture, one might expect that the hard and the soft
luminosities of magnetic \cvs\ are roughly equal \citep{LM1979}. In
contrast, observations through the early 1990s indicated a soft
X-ray excess in polars and a lack of detectable soft components in IPs.
Before going into more recent observations that modified these
views, it is important to consider the reason why polars can have
soft excess: at low \smdot\ and high magnetic field,
cyclotron cooling of the post-shock region becomes important.
Furthermore, it may enter the non-hydrodynamic regime \citep{LM1979}
and a stream of fast ions may directly heat the white dwarf atmosphere
(``cyclotron-dominated low-$\dot m$ bombardment solution''; \citealt{FB2001}
and references therein). In addition, the shock height is lowered
compared to the Aizu values for the same \mwd\ and \smdot,
and the electron temperature is lower than in corresponding Aizu case.

The differences in the physics of post-shock region will also impact
the spectrum of the soft component. As discussed in Section 2, stellar
atmosphere models are necessary to describe the outgoing radiation
of a white dwarf whose energy source is deep inside the atmosphere,
while irradiation from above alters the temperature gradient and
the outgoing spectrum is closer to a blackbody \citep{WKB1987}.
This was studied by \cite{Mauche1999} using \euve\ grating spectra of
polars, who found the blackbody to provide superior fits to the data
compared to pure-H or solar abundance stellar atmosphere models
suggesting that irradiation from above plays a significant role.
However, there are departures from the blackbody, so what is needed
is an irradiated stellar atmosphere model. Note also that the effective
temperatures derived from the blackbody fits are in the 15--25 eV range
for the 9 polars, lower than typically found using lower resolution
X-ray data. Also note that the bolometric fluxes inferred from the
fits depend strongly on the choice of models. Any inferences we can
currently draw from X-ray data should be treated with caution, until
a full set of of irradiated stellar atmosphere models are developed
and high quality data that can test them are obtained.

Two further complications exist in fitting the soft component. First,
it is likely that the heated area of the white dwarf cannot be characterized
using a single temperature \citep{BBR2012}. This might result from
intrinsic density gradients in the accretion flow from the L$_1$ point
\citep{LS1975} that survives beyond the interaction region \citep{CashThesis}.
Even a uniform density stream can lead to a distribution of temperatures,
if the soft component is primarily due to irradiation and the shock
is well above the white dwarf surface. Only the center of the accretion
foot point is fully illuminated and reaches the highest temperature, which
is surrounded by a halo of partially illuminated region with lower
temperatures, the outer parts of which may primarily emit in the UV
\citep{Getal1998}.

The other complication is whether the soft component can escape going
through the same complex absorbers that hard X-rays do. Traditional view
is that the soft component is unaffected. In general, the soft X-ray
component of polars appears unabsorbed, except by the ISM. Moreover,
there are no obvious spectral changes as a function of spin phase, even
as intensity changes by a significant factor (see, e.g., the
\exosat\ grating observation of QQ~Vul; \citealt{Oetal1986}). These
arguments do exclude the model in which the soft component is affected
by a simple absorber or a partial covering absorber. However, as noted
above, complex absorbers of the {\tt pwab} type do not manifest
themselves as exponential cut-off of low energy photons
(Figure\,\ref{pwabresult}), so the possibility of a {\tt pwab}-type
absorber is not ruled out by the data. Geometrically speaking, while the
outer halo may effectively escape the shadow of the pre-shock flow, the
soft X-rays are likely to be from the smaller core and is probably subject
to absorption in the pre-shock flow.

\cite{RC2004} published the most up-to-date summary of the soft excess
to date, using a snapshot survey of polars using \xmm. They find
large soft excesses (more than a factor of 5) in only 13\% of their sample,
significantly down from earlier studies based on \rosat\ data (58\% according
to \citealt{Retal1994}, but 24\% in the re-analysis using updated calibrations
by \citealt{RC2004}). They also found several polars without a detectable
soft component. The \xmm\ data are far better suited for the accurate
characterization of the hard component, and \cite{RC2004} used their
multi-temperature shock emission model with complex absorbers, as necessary.
However, they used a blackbody model with simple absorber for the soft
component, which clearly is a limitation of this study, or rather, that of our current
understanding of the soft component. Moreover, the blackbody temperatures
determined using the \euve\ grating data \citep{Mauche1999} are systematically
lower than those determined from \rosat\ data \citep{Retal1994} ---
unfortunately, \cite{RC2004} did not tabulate their best-fit temperatures
--- possibly indicating that both may have been influenced by the choice
of an imperfect spectral model. Also, these recent paper concentrated
solely on the luminosity ratios, and not the actual luminosity values.
A direct comparison of hard component luminosity against evolutionary
expectations \citep{KW1987} would have provided an additional check of
the soft excess.

On the IP side, PQ~Gem and V405~Aur were discovered to have a distinct
soft component, by \cite{HM1995}, and were called
soft IPs. \cite{EH2007} analyzed the \xmm\ data on 12 IPs and found
that the fit improved with the addition of a blackbody component for
8 of them. Note, however, that this is not necessarily the same as proving
the existence of a soft component. In some cases, the conclusion
that there is a separate soft component appears inescapable, because
the data \edit2{points} are plainly above any reasonable extrapolation of models
that fit above 2 keV. In other cases, however, the blackbody
may be fitting the residuals created by the fact that the spectral
model for the hard component is imperfect (emission model with incorrect
DEM, the use of double partial covering absorber model etc.). Discoveries of
additional IPs with apparent soft components have continued since
then (see, e.g., \citealt{Aetal2008}). Taken at face value, these
fits indicate that the soft component in IPs tend to be hotter
(many above 50 eV and as high as 120 eV; Figure 11 of \citealt{Aetal2008})
than in polars, arise from smaller spots, and represent smaller
fractions of the total luminosity of these systems. The inferred
temperature are close to the Eddington limit of very high mass
white dwarfs, and are higher than the atmosphere limit for
\citep{WKB1987}, however.

\subsection{The Magnetospheric Boundary of IPs}

Theories of interaction between a partial accretion disk and the
magnetic field of a compact object start with the simplest geometry,
that of an aligned rotator (magnetic pole=rotational pole), and often stop
there. In an aligned rotator with a centered dipole \edit2{field}, there is
azimuthal symmetry. The accretion flow must makes a 90$^\circ$ change
in direction of motion at the magnetospheric boundary.
The inner radius of the disk is determined by the balance between the
kinetic energy density of the disk and the magnetic energy density.
The equilibrium spin period of the white dwarf is determined by the
balance between the material torque and the magnetic torque. In
equilibrium, the inner disk radius is near the co-rotation radius
\citep{GL1979}, although the exact location (their fastness parameter)
is subject of debate. Other things being equal, a weaker magnetic
field leads to a smaller inner disk radius, hence a faster equilibrium
spin period. Long-term campaign of optical photometry can reveal
spin period changes over several years, and this can be used to
infer the magnetic moment and other properties of the IPs \citep{Joe1994}.
Finally, if the inner edge of the disk is sufficiently close to
the white dwarf surface, then one of the basic assumptions of
\cite{Aizu1973} breaks down: the flow velocity is determined by
the free-fall condition starting from the magnetospheric boundary,
and not the free-fall velocity from infinity \edit1{\citep{SRR2005}}. 

If the magnetospheric radius is inside the co-rotation radius, the
Keplerian flow must first decelerate; outside, the flow must accelerate
to match the motion, if it is able to accrete at all. However, IPs by
definition are oblique rotators, or spin modulations will not be observable.
There is no well-developed theory of how accretion disk matter couples to
the magnetic field lines in this situation. Observers sometimes
vaguely describe two accretion curtains, each 180$^\circ$ wide,
centered on the sub-polar point of the inner edge of the disk.
How wide are the curtains, and what is the azimuthal dependence of
accretion rate?

X-ray observations of two IPs may provide a clue \citep{Hellier2014}.
Both FO~Aqr and PQ~Gem show sharp absorption notches in the spin profiles
in addition to the general, smoother, spin modulation. The latter is
due to the complex absorption by the immediate pre-shock flow, while the
former is thought to be the equivalent of the stream dips in polars,
and similarly thought to be caused by the distant part of the accretion
curtain. Based on the relative phasing of the two, accretion in FO~Aqr
appears to occur predominantly from points behind the poles
\citep{Eetal2004} while that in PQ~Gem is predominantly ahead
\citep{EHR2006}. Since the white dwarf in PQ ~Gem is spinning down
\citep{EHR2006}, it is reasonable to assume that the inner edge of
the disk is outside the co-rotation radius. The X-ray data can then
be explained if the more rapidly moving field lines sweeps up the
disk material from behind in regions where they are sufficiently
tilted and only small-angle deflections are necessary. Similarly,
in FO~Aqr, which has been spinning up in recent years \citep{Williams2003},
disk material experiences a small-angle deflection to start following
the tilted field line which is moving more slowly than the local
Keplerian velocity.

\subsection{The Partial Disk and Related Phenomena}

\begin{figure}[t]
\plotone{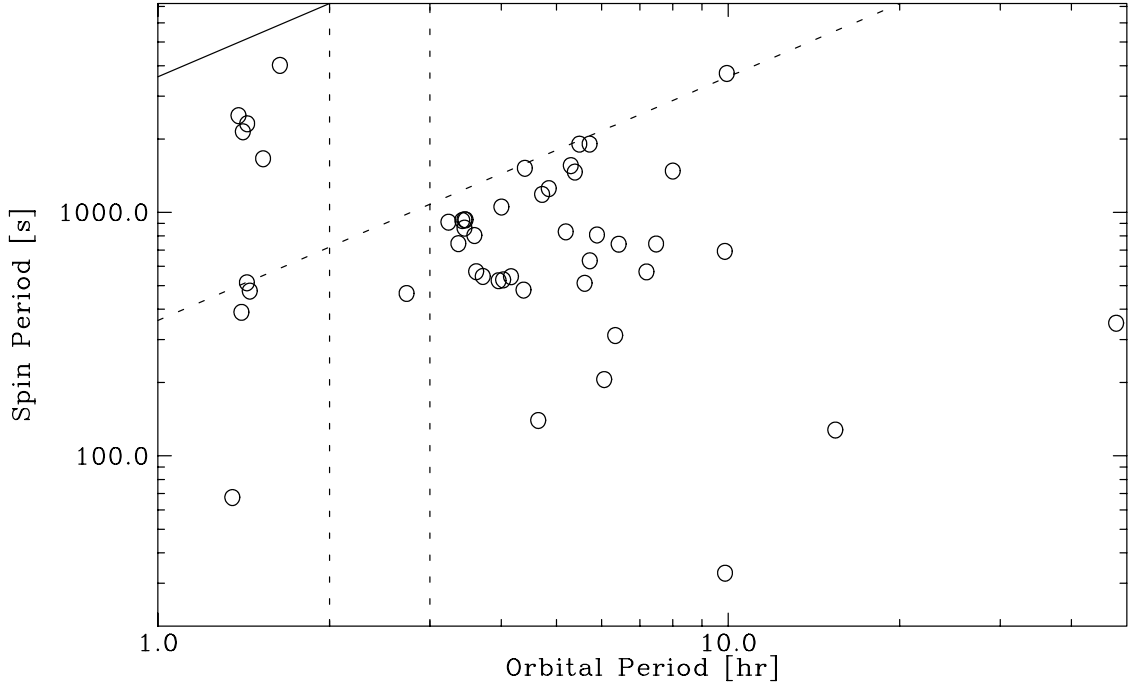}
\caption{Locations of 47 confirmed IPs with well-established orbital
and spin periods. Vertical dashed lines show the approximate location
of the CV period gap (2--3 hrs), while the diagonal lines are shown
for \pspin=\porb\ (solid; i.e., polars) and for \pspin=0.1 $\times$ \porb.
\label{ipspinorbit}}
\end{figure}

Observers often take it for granted that most IPs have a partial
accretion disk, but this issue needs a careful examination.
\cite{KL1991} reviewed the conditions for the existence and
the formation of partial accretion disks. When accretion
initially starts, the stream from the L$_1$ point follows a
ballistic trajectory. If the minimum distance that this ballistic
trajectory takes, $R_{\rm min}$, is less than \rwd, no disk can
form. Similarly, if the magnetospheric radius $R_{\rm mag}$ is
greater than $R_{\rm min}$, then no disk can form, and the accretion
proceeds directly to the magnetic pole(s): such a system should be
a polar. Once a partial disk is well established, the accretion
stream can be absorbed into the disk at its outer radius, but the
conservation of angular momentum dictates that the inner and outer
radii of the disk ($R_{\rm in}$ and $R_{\rm out}$) must encompass
$R_{\rm circ}$, the circularization radius where the specific angular
momentum of the Keplerian motion equals that of matter at L$_1$ point.
If one further assumes spin equilibrium, then the spin period is a proxy
for $R_{\rm in}$=$R_{\rm mag}$. \cite{KL1991} noted that IPs with
\pspin=0.1$\times$\porb\ (see Figure\,\ref{ipspinorbit} for an
updated plot) were near the limit of where partial disks could
exist, depending on the mass ratio.

There are now multiple examples of IPs below the period gap in which
\pspin\ significantly exceeds 0.1 $\times$ \porb, including EX~Hya.
They may be accreting via a standard, partial, accretion disk, with
$R_{\rm mag} < R_{\rm in}$, in which case they are severely out of
spin equilibrium and one expects them to be spinning up much faster
than those with \pspin=0.1$\times$\porb. On the other hand, if they
are at or near spin equilibrium, their accretion flow is unlike the
standard accretion disk in non-magnetic CVs \citep{Netal2008}. If
the X-ray spectrum of EX~Hya indeed indicates that $R_{\rm in}$
is small \citep{Letal2015}, this strongly favors the first interpretation.

Although most IPs with \pspin\ $<$ 0.1 $\times$ \porb\ are believed to
accrete via a partial accretion disk, V2400~Oph (=RX~J1712.6$-$2414) is
an exception. Its X-rays are modulated at the 1003 s period, not the 927 s
period seen in optical polarimetry \citep{Betal1997}, which is
almost certainly the true spin period.  The X-ray modulation
is at the beat (or synodic) period:
\pbeat\ =(\pspin$^{-1}$-\porb$^{-1}$)$^{-1}$. This is the expected
behavior for a diskless IP seen at a low inclination angle \citep{WK1992}
due to pole-switching. That is, one pole is favored to capture the majority
of the matter in the accretion stream from the secondary during $\sim$half
the beat cycle. While beat period modulation in the optical is often seen
and is likely due to reprocessing of X-rays in structures fixed
in the binary frame (the secondary and/or the bright spot region of the
disk), any X-ray beat modulation almost certainly requires pole-switching.

While the X-ray modulation in FO~Aqr is spin-dominated, some modulation
is also observed at the beat period. To explain this, \cite{Hellier1993}
proposed disk overflow accretion as the model: at the outer edge, some of
the accretion stream from the secondary is incorporated into the disk,
but the rest continues on the ballistic trajectory above the orbital
plane, and can accrete directly onto the magnetic poles. The relative
importance of disk and overflow accretion appears to change from epoch
to epoch in FO~Aqr \citep{Betal1998} and in TX~Col \citep{Netal1997}.

For disk overflow accretion to take place, the magnetic field of the
white dwarf must reach the ballistic overflow. Since a well-developed
disk can in principle shield its outer region from the field, it is
likely that the ballistic stream needs to reach or approach the inner
edge of the accretion disk. Both FO~Aqr and TX~Col has \pspin\ close
to 0.1 $\times$ \porb, so this condition is likely to be satisfied.
In addition, \cite{Hetal2000} discovered an X-ray beat modulation in
EX~Hya during the 1998 outburst (itself a likely sign of partial
accretion disk; \citealt{AV1989}), suggesting disk overflow accretion.
If so, $R_{\rm in} \sim R_{\rm circ}$ in EX~Hya even in outburst,
and $R_{\rm in}$ is expected to decrease during outburst (cf. the
case of XY~Ari: \citealt{Hetal1997}), contrary to the inference from
X-ray spectroscopy mentioned above. Thus, the true nature of accretion
flow in EX~Hya remains a mystery.

\edit1{Finally, the difference between the orbital and spin periods
of the object ``Paloma'' (156 min and 130 min, respectively) are
too large for it to be a short-lived deviation. Since the well-established
asynchronous polars are thought to synchronize within several centuries
\citep{CLetal2015}, this transitional object may be best classified
as a low luminosity diskless IP, with the highest ratio of \pspin\ to \porb.
Note also that Paloma is located inside the period gap.}

\subsection{AE Aqr: a Probable Propeller}

AE~Aqr is a highly unusual IP with a rapidly rotating (\pspin=33 s)
white dwarf. Optical spectroscopy failed to show the signatures of
an accretion disk, with the H$\alpha$ line likely originating in
matter being ejected by a propeller mechanism \citep{WHG1998}.
At the same time, AE~Aqr is a well-known X-ray source, if atypically
soft for IPs \citep{EI1999}. If it is a propeller, what is the origin
of these soft X-ray emission? \cite{Ietal2006} analyzed the \xmm\ RGS
data, and found that the intensity ratios of N and O helium-like triplets
implied a modest plasma density of $\sim 10^{11}$ cm$^{-3}$. This,
combined with the emission measure, led them to conclude that the
emission region is much larger than the white dwarf. \cite{Mauche2009}
analyzed the \chandra\ HETG data, and found that Ne, Mg and Si lines
favored progressively higher densities (or else subject to higher
degree of photoexcitation). Considering this and also the velocity
information, \cite{Mauche2009} proposed that the X-rays may be
produced by accretion onto the white dwarf in the bombardment regime. 
\cite{Ketal2014} observed AE~Aqr with \nustar\ and detected pulsed
thermal emission up to $\sim$20 keV. They also considered it likely
that they originated in accretion onto the white dwarf surface, although
their interpretation differed from that of \cite{Mauche2009} in details.
That is, despite the propeller effect well established through optical
spectroscopy, some of the matter appears to accrete onto the white
dwarf successfully.

The origin of X-rays in AE~Aqr can be questioned because the white
dwarf is spinning down, with sufficient spin-down power to explain
the luminosity in the X-rays and at other wavelengths in principle
\citep{Mauche2006}. In this respect, AE~Aqr remained unique among all
magnetic CVs, until it was realized that an previously obscure
object, AR~Sco, was an M star-white dwarf binary that was also
spin-down powered \citep{Marshetal2016}. The actual emission mechanisms
in AR~Sco appear completely different from those in AE~Aqr, however.

\section{X-ray Emission from Non-Magnetic \cvs\ and Symbiotic Stars}

In non-magnetic \cvs\ and \syss, the starting point for interpretation
is the boundary layer model, in which the Keplerian accretion disk
extends all the way down to the white dwarf surface. The disk converts
half the potential energy to kinetic energy and the rest is radiated away
in the IR, optical, and UV. The boundary layer, in which the Keplerian
disk connects with the white dwarf surface, should emit equal luminosity
as the disk, if the white dwarf is a slow rotator. At high accretion rates,
the boundary layer should become optically thick and emit in the soft
X-ray/EUV range, yet optically thin, medium energy X-ray component is
usually observed \citep{PR1985b}. At low accretion rates, the boundary
layer should be optically thin and emit hard X-rays \citep{PR1985a};
however, one should note that models in which the accretion disk develops
a central hole have also been proposed.

\subsection{Observational Overview}

Although X-rays from non-magnetic \cvs\ had been detected using
non-imaging instruments, it was the imaging capability of
\einstein\ that established them as a class of X-ray sources
(see \citealt{CM1984} and references therein). More recent compilation
papers include those of \asca\ data by \cite{BWO2005}, and \xmm\ data
by \cite{DPetal2005}. These and other works firmly established
non-magnetic CVs in general, and quiescent dwarf novae in particular,
as an important class of medium-energy X-ray sources, at luminosities
in the 10$^{30}$ -- 10$^{32}$ \eps\ range. They also established that
their X-rays originated in an optically thin, multi-temperature region.
Furthermore, \cite{Metal2003} showed that the cooling flow model
provided a reasonable description of their X-ray spectrum. At high
accretion rates, a few nearby dwarf novae in outburst have also been
detected as soft X-ray/EUV sources. While the soft component is almost
certainly from the optically thick boundary layer, the optically thin emission
never completely disappear, as was already noted by \cite{PR1985b}.

The presence of exhibit X-rays from the accretion disk boundary layer
is now well established. For example, \cite{Letal2013} conducted a
\swift\ XRT snapshot survey of known \syss, and found a number of
them to emit optically thin, thermal X-rays in the medium energy band,
which they named the $\delta$-type X-ray emission. Some $\delta$-type
emissions are hard enough and luminous enough for five \syss\ to have
been detected in the \swift\ BAT survey \citep{Ketal2009,Metal2016}.
The $\delta$ component is often heavily absorbed. Some \syss\ show a
$kT \sim$ 1 keV, optically thin, thermal ($\beta$-type; \citealt{MWJ1997})
component, which is likely due to colliding winds. Several systems show
a combination of a relatively unabsorbed $\beta$ component and a heavily
absorbed $\delta$ component \citep{EIM1998}. Shell-burning \syss, on the other
hand, have been detected as supersoft ($\alpha$-type; \citealt{MWJ1997})
X-ray sources, in the few cases where absorption is relatively
low\footnote{\cite{MWJ1997} used the term, $\gamma$-type,
to describe the spectrum of the symbiotic X-ray binary, GX 1+4, which
has a neutron star accretor.}.  These, as well as the spatially extended,
optically thin, thermal X-rays (see \S 3.2) are discussed further in \S 6.5.

\subsection{Quiescent Dwarf Novae}

The outburst behavior of dwarf novae is widely interpreted in
the framework of the disk instability model (DIM; see \citealt{DIM}
and references therein)\footnote{Note, however, this is not a
universally accepted model for all dwarf novae (see, e.g.,
\citealt{BBO2016}).}. According to DIM, quiescence is when the
disk is in a low temperature, low surface density state that is
incapable of transporting all the incoming matter onto the white
dwarf surface; the disk mass increases gradually, until a critical
value is reached and the disk transitions to a hot state, allowing
a large amount of matter to accrete onto the white dwarf in a relatively
short span of time. DIM has been successful in reproducing the
observed properties of dwarf nova outbursts (see, e.g.,
\citealt{Cetal2012}), less so those of quiescence (see below).
Regardless of the details, the accretion rate onto the white dwarf
($\dot M_{acc}$) in quiescence must always be smaller than the
average mass transfer rate from the secondary ($\dot M_{tr}$,
usually taken to be constant) for DIM to work. The latter can
be estimated by integrating the luminosity of a dwarf novae
through outburst cycles, while X-ray observations in quiescence
offers a simple and plausible method for estimating the former.

The X-rays from quiescent dwarf novae are
optically thin, thermal emissions from multi-temperature plasma.
While such models are required to achieve a good fit to high quality
data (spectral resolution of a CCD detector combined with good
signal-to-noise ratio), current data are insufficient to derive
DEM distribution purely from data. Thus, \cite{BWO2005} used
a purely phenomenological power-law DEM distribution in their
analysis of \asca\ data, while \cite{Metal2003} used the cooling
flow model (see Figure\,\ref{demcomp}) in the analysis of \chandra\ HETG
data.  A hybrid approach was adopted by \cite{DPetal2005} in their
analysis of \xmm\ data: they used their own, modified version of
{\tt mkcflow} model which allowed the DEM to be tilted relative
to that of the standard cooling flow by multiplying with a power law
(see their equation 2). They often found statistically significant
deviation of this power law index from 0.0, which should be investigated
further in the future, even though they remark that the isobaric
cooling flow model provides a good description of the X-ray spectra.

An important parameter of such fits is the maximum temperature
of the plasma. \cite{DPetal2005} started their comparison with
the Virial temperature ($kT_{vir}$) and concluded that the maximum
temperature is $\sim 0.6 kT_{vir}$. This is close to the temperature
expected of a strong shock for Keplerian flow
(9/16 $kT_{vir}$=0.5625$kT_{vir}$) that \cite{Betal2010} compared
their data to (see also Figure\,\ref{tshock}). Note that both these
papers took the white dwarf mass estimates from the literature,
with varying precision and accuracy. In particular, unless the
system is eclipsing, the inclination angle is often uncertain,
which \edit2{translates} to a large uncertainty in the mass estimate.
Having said that, currently available data are consistent with
the idea that the X-ray emission from quiescent dwarf novae are
powered by a strong shock in the boundary layer between the white
dwarf surface and the Keplerian disk that extends all the way down to it.

This picture is consistent with the fact that the X-ray emission
region in quiescent dwarf novae is compact, as shown by sharp
transitions in the X-ray light curves of deeply eclipsing dwarf novae
HT~Cas, OY~Car, and Z~Cha \citep{Metal1997,Retal2001,Netal2009,Netal2011}.
For OY~Car, \cite{WW2003} found that the eclipse duration
(mid-ingress to mid-egress) and the ingress/egress durations (see \S 3.3.4)
were both shorter in X-rays than in the optical, but in such a way that
the second and the third contacts were aligned in these two different
wavelength regimes to within measurement errors. This implied an effective
vertical displacement of X-ray emission region. \cite{WW2003} interpreted
as implying that X-rays were emitted at the polar regions, but the lower
pole was occulted, presumably by the accretion disk. The absorption
(and hence the apparent displacement) is less prominent for X-rays above 2 keV
than those below 2 keV in Z~Cha, the latter dominating the total
\xmm\ EPIC light curve \citep{Hetal2009}.

Absorption in the disk had already been invoked by \cite{vTBV1996} to
explain the inclination dependence of the soft X-ray fluxes among
non-magnetic CVs studied with \rosat. The \xmm\ spectrum of the deeply
eclipsing system, HT~Cas had a measured absorbing column of 1.6$\times
10^{21}$ cm$^{-2}$, far in excess of what would be expected from the ISM
for an object at a distance of 130 pc \citep{TLS2008}. The \suzaku\ spectrum
of the partially eclipsing system, V893~Sco required a partial covering
absorber with $\sim$50\% covering fraction at \nh\ $\sim 2 \times 10^{22}$
cm$^{-2}$ \citep{MZS2009}, again suggestive of absorption by the inner disk.
However, if it is a disk, it is quantitatively very different from
the standard optically thick, geometrically thin, steady state disk,
which predict a surface density of order 1 g\,cm$^{-2}$ near the
inner edge of the disk (see, e.g., \citealt{Hetal1998}), hence should
result in a much higher \nh.

Once the modest absorption (\nh\ of order 10$^{21}$ to 10$^{22}$ cm$^{-2}$)
in the disk plane is considered, the eclipse results on OY~Car \citep{WW2003}
does not necessarily imply the displacement of the emission region.
It only proves that a large fraction of photons detected with \xmm\ EPIC
instruments are from above the orbital plane, which is still qualitatively
consistent with a boundary layer emission.

These and other observational results pose a severe challenge to
the basic version of DIM. It predicts that matter simply accumulates
in the outer disk with very low accretion rate onto the white dwarf
during quiescence. \cite{Betal2010} used a sample with parallax
based distances\footnote{Even so, the distance to SS~Cyg has
since been revised \citep{MJea2013}.}, and derived a luminosity
in the 10$^{30}$--10$^{32}$ \eps\ range, which requires an accretion
rate that is many times higher than can be achieved through a cold
disk \citep{DIM}. DIM also predicts that the accretion rate should
increase during quiescence from the end of one outburst to the
beginning of the next, as the surface density of the disk increases.
This has never been observed at any wavelengths: In particular, in
X-rays, dwarf novae appear to remain constant
or decline slowly in luminosity during the inter-outburst intervals (e.g.,
\citealt{MPT2004,CW2010,Fetal2011}). Moreover, the WZ~Sge type stars,
i.e., dwarf novae with extremely long quiescent intervals, cannot be
explained with DIM with $\alpha \sim 0.01$ that works for shorter
recurrence period dwarf novae. Finally, \cite{Betal2015} discovered
a correlation between the quiescent X-ray luminosity and the duty
cycle of dwarf novae. While this does not immediately contradict DIM,
there does not appear to be a ready explanation for it, either.

For the unexpectedly high quiescent X-ray luminosity, relative
to the basic version of DIM, an often cited fix is the truncated
disk, perhaps via the coronal siphon flow \citep{MMH1994}. For
the exceptionally long inter-outburst of WZ~Sge, a proposed solution
is also a truncated disk \citep{LKC1999}. It is implausible for
the truncated disk to be the solution for both problems, however:
If quiescent dwarf novae with normal inter-outburst intervals
(weeks to a few months) needs a truncated disk, WZ~Sge with 30-year
inter-outburst interval would need an even bigger hole, leaving little
room for a disk to exist. A possible alternative is the evaporative
instability of the accretion disk \citep{SW1986} combined with the
irradiation from the boundary layer \citep{dKW1999}. Unfortunately,
none of these models have provided, to date, an explanation of
how quiescent dwarf novae maintain a constant or slowly declining
X-ray luminosity during quiescence\footnote{This is in contrast
to the situation for the constant optical/UV luminosity, where
an increasing disk luminosity can in principle be hidden
by a decreasing luminosity of a cooling white dwarf.}.

The question of the possible central hole has been explored by
several groups by studying the aperiodic variability of X-rays
in quiescent dwarf novae. They concluded that these disks are
truncated \citep{BR2012,DMN2014}, by interpreting the data in
the framework of propagating fluctuation model. However, this
is a model-dependent statement, and these studies did not explain
how this model applies to a cold disk in which little matter is
transported (as required for DIM to work at all). A quantitative
study is necessary to assess if a cold disk can nevertheless propagate
fluctuations sufficiently effectively for this interpretation
to hold. In addition, proponents of central holes must also
explain the lack of obvious central holes in the eclipse maps
of quiescent dwarf novae such as V4140~Sgr, both in total optical
emission and its fluctuations \citep{BBO2016}.

While observers have used cooling flow and other analytical
models, there are theoretical efforts to model the X-ray
emission from quiescent dwarf novae from first principles
(see, e.g., \citealt{Letal2008} and references therein).
Several such studies have started with the assumption that
the optically thin boundary layer is radiatively inefficient,
hence spatially extended, and advection plays an important
role, based on a study by \cite{NP1993}. However, it is puzzling that
this paper claimed the optically thin boundary layer to be radiatively
inefficient, considering that the Bremsstrahlung cooling time for a 10$^8$ K,
10$^{-9}$ g\,cm$^{-3}$ (appropriate for the 10$^{-10.5}$
M$_\odot$\,yr$^{-1}$ case; their Figure 2) plasma is $\sim$ 2 s
(0.2 s for the  10$^{-9.5}$ M$_\odot$\,yr$^{-1}$ case). Note that \cite{NP1993}
based their calculations on the accretion disk structure of \cite{H1990},
in which disk is represented by a set of mutually non-interacting annuli.
As \cite{H1994} noted, this approach is inapplicable at both the inner
and the outer boundaries of the disk. A study of the boundary layer
structure that fully incorporates the interaction among annuli is
therefore highly desirable.

\subsection{The Optically Thick, Soft X-ray/EUV Component}

At high accretion rates (dwarf novae in outburst and non-magnetic
nova-like systems), the boundary layer is expected to become
optically thick \citep{PR1985b,PN1995}. Indeed, luminous soft X-ray/EUV
emission has been seen in some nearby dwarf novae in outburst
\citep{WMM2003}. At the same time, the optically thin emission never
disappears. However, before the X-ray data can be used to infer the
properties of the boundary layer, it is necessary to investigate
if the observed X-rays are indeed from the boundary layer.

It is clear that there is a soft X-ray component that is not from the
boundary layer. This was first demonstrated by \cite{Netal1988}, who
reported that the soft X-ray emission observed with \exosat\ in OY~Car
during its 1985 May superoutburst was never eclipsed (their Figure 13).
Note that OY~Car is a well known high inclination system in which both
the white dwarf photosphere (in optical/UV) and the optically thin X-ray
component are eclipsed in quiescence \citep{WW2003}. Thus, the soft X-rays
observed during the superoutburst had an extended origin. Using an
\euve\ observation of OY Car during its 1997 March superoutburst,
\cite{MR2000} confirmed the lack of eclipse, and showed that its
EUV spectrum consisted of prominent emission lines on a weak continuum. 
However, this should not be taken as the evidence for a lack of
the expected optically thick boundary layer emission: In fact,
\cite{MR2000} successfully modeled the observed emission as due
to a warm accretion disk wind photoionized by a T$\sim$90,000--130,000 K
boundary layer emission.

\cite{Petal2004} observed the deeply eclipsing nova-like system,
UX~UMa, with \xmm, and discovered that there were two X-ray components.
The component that dominates in the 3--10 keV range is heavily absorbed
(\nh $\sim 8 \times 10^{22}$ cm$^{-2}$) and is deeply eclipsed. The component
that dominates below 2 keV is also optically thin, does not suffer the
heavy absorption (observed \nh $\sim 8 \times 10^{19}$ cm$^{-2}$), and
is UN-eclipsed. This is an intermediate-temperature component whose
origin remains unclear (for the photoionized wind explanation
for the softer component in OY~Car to apply here, the temperature and the
luminosity of the boundary layer component need to be much higher). The fact
that we do not detect the expected soft, blackbody-like component from the
optically thick boundary layer cannot be taken as evidence for the
lack of such a component, given the hard component goes through a column
sufficiently high to hide it from our view. In fact, all eclipsing,
high accretion rate CVs may have sufficiently high column density
to hide the soft component from the optically thick boundary layer. 

The soft component is best studied \edit2{in} systems such as SS~Cyg, which
is nearby (d=114 pc: \citealt{MJea2013}) and seen at a low inclination
angle ($i \sim 50^\circ$: \citealt{BRB2007}). The best data to date have
been obtained with \chandra\ LETG and show a strong, blackbody-like
continuum with a host of absorption features \citep{M2004}. In fact,
the theory of optically thick boundary layer was developed to explain
an earlier detection of this component in SS~Cyg \citep{P1977}. However,
this soft X-ray component as predicted by \cite{P1977} is rarely detected,
leading to the idea of the missing boundary layer \citep{Fetal1982}.
In a compilation of \euve\ spectra of dwarf novae in outburst
(SS~Cyg, U~Gem, VW~Hyi and OY~Car), \cite{M2002} found that,
with the exception of SS~Cyg, they did not have any significant
flux shortward of $\lambda \sim 80 \AA$ (E$>$0.15 keV), with
estimated blackbody temperatures of order kT$\sim$10 eV, much less
than theoretically expected, presumably because the boundary layer
occupies a much larger area than the theory assumed\footnote{That is,
the specific accretion rate, $\dot m$, is lower than assumed, using
the terminology frequently used for magnetic \cvs; see \S 5.2}. This
makes it impossible to detect this component unless the interstellar
\nh\ was exceptionally low. That is, the missing boundary layer
``problem'' is really a mismatch between the expected and observed
properties of the optically thick boundary layer.

Nevertheless, the optically thick boundary layer of SS~Cyg appears
less luminous than the disk (L$_{bl}$/L$_{disk} \sim 0.06^{+0.18}_{-0.03}$),
according to the blackbody fits \citep{M2004}. So a version of the
``missing boundary layer problem'' may still be present, if the blackbody
fit results are reliable. For OY~Car, on the other hand, for which the
properties of boundary layer emission are inferred indirectly through
the observation of photoionized wind, \cite{MR2000} found a plausible solution
in which L$_{bl}$/L$_{disk} \sim 1.0$. Interestingly, \cite{M2004}
inferred a small fractional area of the boundary layer for SS~Cyg
(best fit value of 5.6$\times 10^{-3}$ and an upper limit of
6.6$\times 10^{-2}$, both for d=160 pc) while \cite{MR2000}
inferred a much larger ($>$0.1) value for OY~Car. Perhaps the
most alarming aspect of the boundary layer emission of SS~Cyg
is its homologous spectral evolution as the luminosity changed
by two orders of magnitude \citep{Metal1995}, implying that its
temperature remained roughly constant. In contrast, if the
fractional area remained constant, the temperature should
have changed by a factor of $\sim$3, which is well within our
capability to detect. This suggests caution in proceeding
further with our interpretation of the spectra of the optically
thick boundary layer.

Dwarf nova oscillations (DNOs), the short period ($<$1 min) oscillations
in the optical light of high accretion rate, non-magnetic CVs with
high coherence ($Q \sim 10^4$--10$^6$), are also seen in their soft
X-ray/EUV component. \cite{MR2001} discovered that EUV and optical
DNOs of SS~Cyg had the same period and phase, and also observed
a frequency doubling from P=6.59 s to 2.91 s. If the period is associated
with the Keplerian motion around the white dwarf, then the minimum
period of 5.6 s (observed as 2.8 s oscillations after period doubling)
requires a white dwarf mass of at least 1.1 M$_\odot$. This is also
consistent with the maximum temperature inferred for the cooling flow
of kT$\sim$42 keV \citep{Betal2010}, which implies a 1.15 M$_\odot$
white dwarf\footnote{While the most recent optical determination of the
white dwarf mass in SS~Cyg is 0.81$\pm$0.19 M$_\odot$ \citep{BRB2007},
this is based on the inclination angle they infer ($45^\circ < i < 56^\circ$).
This, in turn, was determined using an unstated assumption that
the fractional contribution of the accretion disk light remained
the same between two epochs, one at which they measured the average
flux from the secondary, and the other when they measured the amplitude of
the ellipsoidal variation.}.

\subsection{The Residual Hard X-rays}

One would not expect any optically thin, hard X-ray emission from
the boundary layer if it is completely optically thick. However,
nova-likes and dwarf novae in outbursts do emit optically thin
X-rays in the 0.5--10 keV range (see, e.g., \citealt{Wetal2017} and
references therein). 

\begin{figure}[t]
\plotone{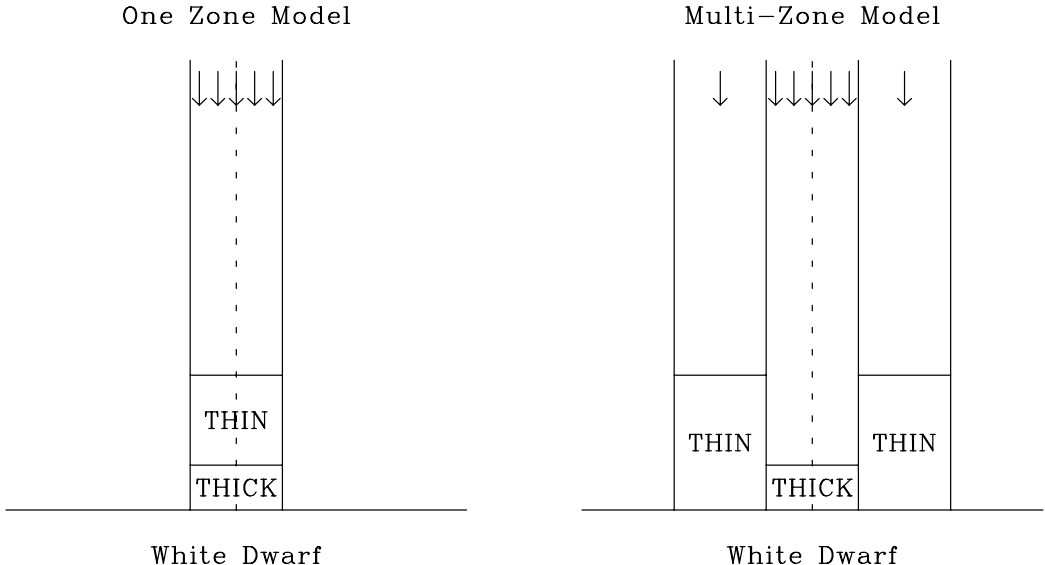}
\caption{These schematics show two extremes of how optically thick and
thin regions might co-exist in the boundary layer. In both, the white
dwarf surface is shown as a horizontal line, and the orbital plane is
indicated by dashed vertical line. (Left) In a one-zone scheme, the
boundary layer may consist of an optically thick layer near the stellar
surface and an optically thin layer on top. (Right) In a multi-zone
scheme, of which this shows a simplest version, there could be an
optically thick region near the orbital plane where the accretion rate
is high; since the density in the disk falls off exponentially with
height above the mid-plane, this may well be sandwiched between optically
thin regions. \label{blsketch}}
\end{figure}

When a single dwarf \edit2{nova has} been observed both in quiescence
and in outburst, preferably using identical set of instruments,
it is often the case that the hard X-ray component is more luminous
and harder in quiescence (see, e.g., SS~Cyg: \citealt{Ietal2009}).
Known exceptions to this general rule are GW~Lib, which is so
X-ray faint in quiescent that this may not be surprising
\citep{Betal2009}, and U~Gem, whose outburst X-ray spectrum is
too complex to allow a straightforward measurement of the
temperature change \citep{Getal2006}. The picture
of \cite{PR1985b}, in which an optically thin region exists
in the outskirts of an optically thick boundary would
not provide a prediction for the luminosity. In principle,
the optically thick and thin regions could co-exist in different
configurations (Figure\,\ref{blsketch}). In either geometry,
one might expect the maximum temperature to be set by the
gravitational potential of the white dwarf, and therefore
remain constant. This is often taken as a severe difficulty
of the ``optically thin skin'' interpretation for the residual
hard X-ray emission.

It is clear that there is \edit2{an} intermediate temperature \edit2{(kT} $\sim 1$ keV),
optically thin component that is not from the boundary layer in
at least a subset of high state objects. I have already noted the case
of UX~UMa in \S6.3. The UN-eclipsed X-ray component seen in OY~Car in
superoutburst with the \rosat\ HRI \citep{Petal1999}, which was assumed
to be a component distinct from that seen with EUVE \citep{MR2000}, may
be the same. \cite{Wetal2017} found it necessary to add an optically thin
component of similar temperatures to obtain satisfactory fits to their
\suzaku\ data on dwarf novae in outburst. The evidence may favor a
non-boundary layer origin of this intermediate temperature, optically
thin component, but this is a separate question from the origin of the
residual X-rays seen in the medium energy band.

One possibility is that these residual, medium-energy X-rays are not
from the boundary layer. This is the option preferred by \cite{WMM2003},
citing the observed temperature drop compared to the quiescence as
key evidence. \cite{Ietal2009} proposed a corona above the accretion
disk to be the origin of the residual X-rays\footnote{Such models
probably are motivated, in part, by analogy with the corona invoked
for the power-law component in black hole X-ray binaries and in AGN.
However, recent observations appear to favor compact ``corona'' of
very limited spatial extent \citep{Fetal2015}, unlike that envisioned
for SS~Cyg by \citep{Ietal2009}.}. They based this model partly on the
reflection fraction they measured in quiescence and during outburst
using \suzaku\ XIS and HXD. Since the HXD signal in outburst was weak,
this is subject to considerable systematic uncertainty due to our limited
ability to model the HXD background, so this needs to be confirmed
with higher quality data with \nustar.  They also argued that the
line broadening observed in SS~Cyg in outburst was consistent with
an accretion disk corona. While undoubtedly true, there may well
be other explanations for the increased line width. Additional
consideration should be given to the smooth recovery of the
hard component temperature during outburst decay of SS~Cyg
(Figure 5 of \citealt{MPT2004}) which might be easier to interpret
if the hard X-rays in quiescence and in outburst shared a single
origin\footnote{The temperature does change suddenly during
the rise.}.

The other possible explanation for the lower temperature is that
the residual X-rays are indeed from the optically thin part of the
largely optically thick boundary layer, but there is a second cooling
mechanism that competes with hard X-ray emission. In particular, 
\cite{Fetal2011} proposed that the optically thick part could provide
sufficient number of seed photons to Compton-cool the optically thin part
of the boundary layer. The same model was used to explain the relatively
faint, very soft emission from RS~Oph observed in 2007 and 2008, 537 and
744 days since its 2006 nova eruption \citep{TNetal2011}. In this case,
the still hot photosphere of the white dwarf contributes a significant
number of UV photons that can serve as seed photons. Finally, this scenario
also explains the lack of detection of optically thin emission in \ssss,
even though accretion is necessary to fuel the continued nuclear burning.
The weakness of this scenario \edit2{is} that it is feasible if and only if the
density of the plasma in the optically thin part of the boundary layer
is relatively low, which has not been established yet.

Another possibility is suggested by the UV detection of a rapidly
spinning accretion belt in several dwarf novae \citep{Sionetal1996,Cetal1997}.
It is possible that the temperature drops in outburst because the boundary
layer is between the Keplerian disk and the accretion belt, and
the kinetic energy of the belt is radiated later.

If the quiescent X-rays are from the boundary layer, which transitions
to an optically thick state when the accretion rate exceeds a critical
value, one would expect a hard X-ray brightening at the onset of a
dwarf nova outburst, before the sudden appearance of the soft component,
and the reverse is expected during decline \citep{SHL2003}. This is
indeed seen in SS~Cyg \citep{WMM2003} and has been taken by many as the
paradigm applicable to all dwarf novae. However, as \cite{Fetal2011}
pointed out, such behavior has only ever been seen in SS~Cyg; in other
dwarf novae, the optically thin X-rays simply increase during outburst
(U~Gem, GW~Lib) or is suppressed during outburst without a temporary
increase. There is no published explanation for why the SS~Cyg like
behavior is atypical. Moreover, \cite{Fetal2011} attempted to derive
the critical accretion rate empirically, and found values that are much
lower than the theoretical values \citep{PN1995}.

Nova-like systems are thought to be systems in which the
mass transfer rate is high enough to maintain the disk in the
high (dwarf nova outburst-like) state. If so, the boundary layer
properties of nova-like systems should be similar to those of
dwarf novae in outburst, perhaps with a lower temperature,
lower luminosity (than in quiescent dwarf novae) optically
thin component and an optically thick, blackbody-like, soft
X-ray component. Unfortunately, the number of high quality
X-ray observations of such systems is limited, and the results
are often confusing. In particular, \cite{Zetal2014}
studied VY~Scl type systems in high and low states, likely caused
by variations in the mass transfer rate, and found no clear pattern
of correlated or anti-correlated changes of optical and X-ray
brightness. 

It is quite possible that the residual hard X-rays in dwarf novae
in outburst has the same origin as the hard X-ray component in UX~UMa,
almost certainly the boundary layer since it is deeply eclipsed
\citep{Petal2004}. On the other hand, the lack of a consistent
pattern for nova-likes leaves open the possibility of multiple
distinct origins.

\subsection{X-ray Emission from Symbiotic Stars}

Symbiotic stars can be powered by accretion, shell burning,
or a combination of both. In principle, shell burning produces
$\alpha$ type, supersoft, emission. However, the observability
of this component depends on the combination
of the effective temperature and absorption. For the latter,
one must consider both the interstellar absorption and that
due to the M giant wind, following a procedure similar to that
provided by \cite{vdBetal2006}\footnote{However, one should
not merely copy the specific \nh\ values they estimated.
This is because the final \nh\ values depend both on the
mass loss rate and the velocity of the M giant wind. In particular,
\cite{vdBetal2006} assumed a 230 \kms\ wind for their specific
case, which is much higher than typically seen in normal red
giant. For wind velocities in the 10--20 \kms\ range, wind can
easily provide \nh\ of several times 10$^{22}$ cm$^{-2}$, which
is sufficient to hide a supersoft ($\alpha$-type) emission.}.

Accretion-powered \syss\ are often detected as $\delta$-type
X-ray sources, indicating optically thin emission from the
accretion disk boundary layer. This is the case with the four
symbiotic stars detected in the BAT 22-month survey \citep{Ketal2009}.
While one of them, CH~Cyg, has a relatively soft spectrum for
a BAT source, the other three are very hard, suggesting that they harbor
massive white dwarfs. Their medium energy spectra require a partial-covering
absorber, which is seen to be variable from epoch to epoch
on time scales as short as several days. This suggests that
the absorber is located close to the white dwarf. In contrast,
the slow M giant wind, whose origin is located of order 1AU
away, cannot change significantly over time scales of days.

While the BAT-detected \syss\ have high X-ray fluxes, high X-ray
temperatures, and high absorbing column, other $\delta$-type \syss\ are
less extreme \citep{Netal2016}.  Since both the $\beta$ and $\delta$-type
emissions are optically thin thermal X-rays, it is not always trivial to
assign an unambiguous classification. The $\beta$-type emission appears
to peak around 1 keV, and is consistent with $kT \sim 1$ keV.
This requires a shock velocity of order 1000 \kms, commonly seen
in the optical and UV spectra of winds in \syss\ (see, e.g.,
\citealt{CEMcC2008}). Accretion can produce $kT$ in excess of
10 keV, depending on the white dwarf mass. When the observed
$kT$ is 3 keV, for example, this probably indicates a $\delta$-type
origin, but other factors should be considered to make a definitive
classification.

As with the optically thin X-rays observed in non-magnetic \cvs, the
$\delta$ type component of \syss\ can be fit successfully with {\tt mkcflow}
and minute time-scale X-ray variability is observed (e.g., RT~Cru:
\citealt{LS2007}). \cite{Letal2013} found that the $\delta$-type
\syss\ exhibit significant UV flickering, which is absent
in shell-burning systems. Although the distances to many \syss\ are
poorly known, the X-ray luminosity appears to range from that similar
to non-magnetic\cvs\ to as high as 10$^{34}$ \eps.

The $\beta$-type emission is interpreted as due to colliding winds.
One is from the red giant, but what about the other? It is clear that
shell burning can result in a fast wind from the white dwarf
or its vicinity; this appears to be the case in, e.g., AG~Peg
\citep{Retal2016}. CH~Cyg \citep{EIM1998,Metal2007} is the prototype
of \syss\ that show composite $\beta$/$\delta$ type X-ray spectra
(see also UV~Aur, ZZ~CMi, NQ~Gem and V347~Nor: \citealt{Letal2013}).
Such systems prove that the $\beta$-type emission can exist in
purely accretion-powered \syss. These system may produce accretion disk
winds that are strong enough to result in $\beta$-type emission.

As noted in \S 3.2 \edit1{(see also \citealt{Setal2017})},
spatially resolved X-ray emission has been detected
from several \syss\ including R~Aqr (\citealt{Netal2007} and references
therein) and CH~Cyg \citep{GS2004}, and are interpreted as thermal
emissions from a jet. Interestingly, jets, which are typically detected
at radio wavelengths, are produced by \syss\ with a wide range of X-ray
spectral types, e.g., AG~Dra ($\alpha$-type), Hen~3-1341 and V1329~Cyg
($\beta$ type), CH~Cyg ($\beta$/$\delta$ type), and Mira AB ($\delta$
type; \citealt{SB2010}).

\subsection{X-rays from AM~CVn Systems}

For the purposes of studying their X-ray emission, AM~CVn systems
can be divided into two classes: diskless systems HM~Cnc and V407~Vul,
and the rest \citep{AMCVN}. The former have extremely short periods
(5.4 min and 9.5 min, respectively, for HM~Cnc and V407~Vul). As
a consequence, the ballistic trajectory from the L$_1$ point
encounters the white dwarf photosphere before it can circularize
to form a disk. In other words, they are direct impact accretors
\citep{MS2002}. They were originally discovered using \rosat\ as
soft X-ray sources, and their spectra can be approximated with
a blackbody with emission features (HM~Cnc: \citealt{S2008}) or
with absorption component with non-solar abundances, in particular
with enhanced with neon (V407~Vul: \citealt{R2008}). Other than the
possible emission lines in HM~Cnc, no optically thin component has
been detected. There does not appear to be a quantitative study of
whether the lack of optically thin X-rays is to be expected in the
direct impact shock regions, given the likely white dwarf mass and
the specific accretion rate \citep{DWS2008}. The phasing
of the optical and X-ray light curves, and the long-term period history
derived from repeated X-ray observations (see, e.g., \citealt{Retal2005})
of HM~Cnc and V407~Vul are consistent with the direct impact accretor
model.

Longer orbital period AM~CVn systems all appear to accrete via a disk,
and their X-ray emission \edit2{is} similar to \edit2{that from} normal hydrogen-rich \cvs,
including the dichotomy between the optically faint and X-ray bright
state and the optically bright and X-ray faint state dichotomy (see
\citealt{BWO2005}, which included AM~CVn stars GP~Com and CR~Boo along
with normal \cvs).  The major difference is in elemental composition.
While the abundances of H or He cannot be inferred from X-ray spectra,
those of elements from O to Fe can be directly studied in the optically
thin X-ray emission.
Using the \xmm\ RGS data, \cite{S2004} found overabundance of N and Ne in
GP~Com. \cite{Retal2006} also found N overabundance in V396~Hya and
SDSS~J124058.03$-$015919.2, although one must caution that the fit was
done using a plasma model with hydrogen abundance set to 0.0. Since
these models usually assume that the bulk of electrons, responsible
for the Bremsstrahlung continuum, are provided by hydrogen atoms,
they may or may not work properly under these conditions. Different
evolutionary scenarios for AM~CVn systems predict different elemental
abundances \citep{AMCVN}, which can be tested using future X-ray
observations in principle.

\section{X-ray Emission from Novae}

Two types of X-ray emission are commonly observed from novae in
eruption. One is the optically thin emission from an extended region,
and the other is the optically thick emission from the white dwarf
surface.

\subsection{Shock Emission}

When nova eruptions happen in \syss, the ejecta are embedded in the
dense M giant wind. In symbiotic recurrent novae, the ejecta velocities
are high and the binary separations are often of order 1 AU
(\porb\ $\sim$ 1 yr). Since 3000 \kms\ ejecta can travel 1 AU in
$\sim$half a day, strong shock and subsequent X-ray emission is
expected almost immediately.
This has indeed been found to be the case, most recently during the 2014
eruption of V745~Sco, which was strongly detected with \swift\ XRT 3.7
hr after the optical discovery \citep{Petal2015}. During its 2006 eruption,
RS~Oph was an even brighter X-ray source as observed with \xte\ PCA and
\swift\ XRT, although the coverage was not quite as prompt or as densely
sampled \citep{SLMK2006,Betal2006}. Note, in particular, that it was luminous
enough and hard enough during the first few days to have been detected above
10 keV using \xte\ PCA and \swift\ BAT. \cite{SLMK2006} found the flux and
temperature evolution was consistent with a low-mass ejecta that swept
up its own mass in red giant wind within the first few days. \cite{Betal2006}
found the observed \nh\ evolution was consistent with that expected
from the unshocked red giant wind ahead of the shock. \cite{VOBB2007}
further compared the observations with a series of 1-dimensional
hydrodynamic models and found that the shock was radiatively efficient.
Moreover, the manner in which matter was ejected (e.g., the duration of
strong mass loss) was found to have a strong influence on the evolution
of the shocks. For example, the end of plateau in the shock velocity
directly  corresponds to the end of the fast wind in the context of their
model.

Not all embedded novae immediately become X-ray luminous.
In the case of V407~Cyg, a fast nova in a symbiotic system, there was only
faint X-ray emission for the first $\sim$20 days, whose origin remains
unclear \citep{Netal2012}. The X-ray luminosity rapidly increased thereafter.
They interpreted this as the time it took for the blast wave to reach
the immediate vicinity of the Mira type mass donor in a wide (20--25 AU
separation) binary. In the \suzaku\ XIS spectrum of V407~Cyg 30 days
since optical discovery, \cite{Netal2012} found strong He-like likes of
Si and S, inconsistent with the measured electron temperature of 2.8 keV
if the emission was from a single temperature plasma in collisional
ionization equilibrium (CIE). They interpreted this as due to some of
the X-rays being from non-CIE plasma, which is understandable since
some ejecta are running into low density wind away from the M giant.

Shock X-ray emission is a common, if not universal, feature of
novae, whether embedded or not \citep{MOD2008}. The very presence
of such X-rays proves that mass ejection in novae is a complex
process. Presumably a fraction of mass is ejected at low velocity
first, and a faster system catches up to it. One of the best-observed
cases is that of V382~Vel \citep{MI2001}. It was undetected at first,
then detected as an intrinsically hard and highly absorbed source,
becoming softer and less absorbed in subsequent observations.
\cite{MI2001} interpreted
these features as due to a collision of massive, slow shell that was
ejected first, and a fast wind inside it. The slow expansion of outer
shell provides the required \nh\ decline, without invoking a significant
amount of ionization (that is, it is consistent with being so massive as
to remain largely unshocked). In the peculiar recurrent nova, T~Pyx,
the combined X-ray and radio data suggest that the initial ejecta was
far less massive than the later, faster system \citep{Netal2014,Cetal2014},
but it may well be an exception in this regard. \edit1{A recent review by
\cite{Setal2017} contains a summary of multi-wavelength observations
that have revealed the complexity of the mass ejection processes in
novae.}

One of the most exciting recent findings in the field of nova research
is that classical novae produce GeV $\gamma$-rays
\citep{FermiNovaI,FermiNovaII}.
The detection of X-ray emission from shocks contemporaneous with radio
synchrotron emission \citep{Ketal2011,JWetal2016a,JWetal2016b}
suggests that X-rays could play a vital role in diagnosing the shocks
that accelerate particles and produce $\gamma$-rays.

\subsection{Supersoft Emission}

TOO observations with \chandra\ and \xmm\ gratings have provided a
number of high quality, high spectral resolution spectra of the
supersoft emission from novae (see, e.g.,
\citealt{JUNetal2003,TNetal2008,JUNetal2011}). In addition, the
flexible observing capability of the \swift\ mission has enabled
monitoring observations of a large number of novae \citep{Setal2011},
and has become a pre-requisite for triggering \chandra\ or \xmm\ TOOs.
Currently, the data are far ahead of our ability to model them,
which must consider non-LTE effects, non-stationary effects,
and non-solar abundances. One potential use of the supersoft emission
is to measure the effective temperature and \edit2{then} convert it to white
dwarf mass by assuming, e.g., the Eddington limit \citep{WKB1987}.
Even if the absolute values are not trustworthy, novae can certainly
be placed in mass order when comparable data are available.

Combined X-ray and optical monitoring of novae in M31 has resulted
in a large sample of novae with known distances. This allowed
Henze and co-workers to derive correlation between the blackbody
temperature, turn-on times, and turn-off times of the supersoft
emission against each other, and against expansion velocities
measured through optical spectroscopy \citep{Hetal2014}. The
supersoft emission is powered by continued shell burning, and the
turn-off time is believed to indicate when the residual nuclear fuel
is exhausted and shell burning turns off. The turn-on time, on the
other hand, is purely observational: shell burning is thought to
continue from the time of the thermonuclear runaway, but the supersoft
emission hidden by the ejected shell. Thus, the turn-on time is
set by the combination of the total ejecta mass and its velocity.
The standard understanding of the white dwarf mass as the key parameter
is confirmed by this study: more massive white dwarf requires smaller
mass for nova eruption, eject matter more violently, and its supersoft
emission reaches a higher temperature.

Variability has emerged as a major feature of supersoft emission
in novae. The \chandra\ LETG data on V4743~Sgr showed a $\sim$22 min
oscillation followed by a sudden drop in count rate \citep{JUNetal2003}.
High amplitude variabilities on hours to days timescales (oscillations
seen from one \swift\ pointing to the next, or state changes during
a single \chandra\ or \xmm\ observation) have been discovered in RS~Oph
\citep{Oetal2011} and several other novae. In addition, QPOs on 30--70 s
time scale have been detected in 4 novae as well as persistent supersoft
source, CAL~83 \citep{JUNetal2015}. A possible explanation is that it is
a non-radial oscillation that is excited by the nuclear burning.

While the spectra of supersoft emission are blackbody-like when
observed at CCD resolution, grating spectroscopy has revealed
the presence of absorption or emission lines. \cite{JUNetal2013}
conducted a systematic study of this and compared the properties
of those with emission lines and those with absorption lines. The
former appear to be less luminous than the latter, and appear to
be found predominantly in high inclination systems
(again, with the exception of the atypical recurrent nova, T~Pyx).
Among them are several eclipsing systems in which the smooth
orbital modulation of the supersoft emission is seen, including
the persistent supersoft source, CAL~87. For this object,
\cite{Eetal2001a} argued
that the central source was permanently hidden from our view
and only scattered emission is seen, which is essentially the
same explanation as was applied to the boundary layer
emission of OY~Car in superoutburst \citep{MR2000}. Among the
novae, the same explanation clearly holds for U~Sco \citep{Oetal2013}
and V959~Mon \citep{Petal2013}. The degree of orbital modulation
allows us to infer the size of the scattering to be the same order
as the secondary in these systems. This suggest that the accretion disk
was present in these novae during the supersoft phase to hide the
direct emission. As to the absorption lines in the low inclination
systems, two possible origins had been considered: they may be intrinsic
feature of the white dwarf atmosphere, or caused by discrete shells ejected
by the novae \citep{Petal2012}. In addition, it may be worth exploring
the scenario in which a single scattering region is seen in emission
in high inclination systems and in absorption in low inclination systems.

\section{Collective properties}

\subsection{Population and Luminosity Function}

\cvs\ are discovered through X-ray surveys (\S 3.1) and through optical
surveys. Because the discovery efficiency varies greatly for different
method--subclass combinations, the aggregate catalog of all known \cvs\ is
highly heterogeneous. It is therefore meaningless simply to list the fraction
of magnetic \cvs, for example, in published catalogs: the results are as
likely to tell us about the selection effects as about the underlying
populations.

It is far more meaningful to derive the space densities of various
subclasses. For this purpose, volume-limited samples would be ideal,
but this is impractical for the time being. Flux-limited samples are
therefore the method of choice, assuming that one can estimate the
distances to individual objects to the required degree of accuracy.
An early, perhaps the first, application of this method to X-ray
discovered \cvs\ was that by \cite{Hetal1990}. While they derived
a local space density of \cvs\ of at least a few $\times 10^{-5}$ pc$^{-3}$,
there are reasons to be wary of this number. One is the perennial
difficulties in estimating distances to \cvs; the other, specific to
this work, is that they did not consider a realistic Galactic distribution
of \cvs.

The most up-to-date work on this subject is the series of papers by
Pretorius and collaborators using \rosat\ and \swift\ BAT flux-limited
samples \citep{PK2012,PKS2013,PM2014}. Summarizing the highlights of
these papers: the local space density of non-magnetic CVs detected with
\rosat\ is 4$^{+6}_{-2} \times 10^{-6}$ pc$^{-3}$, but this number may
double when the X-ray faint population (L$_x < 5 \times 10^{29}$ \eps)
is also considered.  Polars, if their duty cycle is 0.5, have a total
space density of 9.8$^{+5.4}_{-3.1} \times 10^{-7}$ pc$^{-3}$, of which
roughly three quarters are below the period gap. Long-period IPs have
an estimated space density of 1.0$^{+1.0}_{-0.5} \times 10^{-7}$ pc$^{-3}$.
These are systems with BAT band luminosities in the 10$^{32}$--10$^{34}$
\eps\ range; there appears to be a separate population of LLIPs
(see also \S 5.3), with a few systems around L$_{BAT} \sim 10^{31}$ \eps,
and they could be more numerous than the normal, X-ray luminous, IPs.

Pretorius and colleagues found the intrinsic fraction of magnetic
\cvs\ (polars and IPs) to be $\sim$16\% to within a factor of 2.
This value is smaller than, but consistent within errors with,
the fraction of magnetic \cvs\ found in catalogs. This value is
also consistent with the figure of $\sim$10\% for the fraction
of magnetic white dwarfs among non-interacting samples \citep{LBH2003}.
Note, however, even among field white dwarfs, the incident of
magnetism is an active research area (see, e.g., \citealt{Vetal2014}).

Several caveats should be kept in mind. Current studies rely on our
ability to measure distances to a sample of \cvs\ using indirect methods.
In addition, the scale height of Galactic distribution of \cvs\ is an
additional source of uncertainty. Pretorius adopted a scale height of
120 pc for long-period (and hence relatively young) \cvs, and 260 pc
for short-period (old) ones. While these are reasonable assumptions,
these space density numbers should be reviewed once \gaia\ distances
become available for a large number of \cvs.

In these studies, X-ray luminosity function (XLF) and space density are
strongly linked. Both luminous but extremely rare type of objects
and numerous but faint class can be missed in a flux-limited survey.
This is the reason for the uncertainty in the true population of LLIPs.
Similarly, there
are reasons to believe that the current X-ray surveys are missing
the bulk of non-magnetic \cvs. Early X-ray studies of non-magnetic
\cvs\ focused on well-known objects already detected in X-rays, many
of which had X-ray luminosities above 10$^{31}$ \eps\ \citep{MS1993}.
\cite{Betal2010} studied optically selected sample, those with
parallax-based distance measurements, and found that their XLF peaked
in the 10$^{30}$--10$^{31}$ \eps\ range. However, even this sample
is biased towards optically bright \cvs. Non-magnetic CV population
probably is dominated by the very faint population discovered with Sloan
Digital Sky Survey (SDSS), which are thought mostly to be WZ~Sge
type dwarf novae with extremely long recurrence types \citep{Getal2009}.
\cite{Retal2013} observed 20 optically faint non-magnetic \cvs\ (16
of which were discovered in SDSS) with \swift, and found all of
them to have 0.5--10 keV luminosity less than $\sim 2 \times 10^{30}$
\eps\ (and the majority less than $\sim 1 \times 10^{30}$).

As noted earlier, \cite{Betal2015} found a correlation between the
X-ray luminosity and the duty cycle of local dwarf novae. They applied
this relationship to the optical variability study of \chandra\ Galactic
bulge survey sources, and derived the space density of dwarf novae, which
was consistent with those derived for local, long period dwarf novae
\citep{PK2012}. The \cite{Betal2015} measurement is based on the distant
CV population, it is consistent with the local sample of \cite{PK2012}
only when the Galactic distribution of dwarf novae is taken into account.

Finally, the space density of \syss\ is highly uncertain, as discussed
by \cite{Metal2016}. They considered it likely that there \edit2{are} up to several
dozen \syss\ within 1 kpc, which implies a lower space density than for
\cvs.

\subsection{X-rays from the Galactic Ridge, Bulge, and Center Regions}

There is an apparently diffuse X-ray emission throughout the Galaxy,
around the Galactic Center, in the Galactic Bulge, and in the disk
(the last is usually referee to as the Galactic ridge X-ray emission,
or GRXE; see, e.g., \citealt{Wetal1985}). Considerations of fluctuations
and the lack of \einstein\ point source led \cite{Wetal1985} to
conclude that, if the GRXE was made up of discrete sources, those
with luminosities in the 10$^{33.5}$ to 10$^{36}$ \eps\ were not
major contributors. \cite{MS1993} proposed dwarf novae as a potential
major contributor to the GRXE, based on the best understanding
of the space density and the XLF of dwarf novae available at the time.
More importantly, they recognized that this was only plausible
when the realistic Galactic distribution of \cvs\ were considered.
This was later shown to be essential, since \cite{MRetal2006} showed
that the GRXE intensity traces the Galactic stellar mass distribution.

\cite{Eetal2001b} performed a deep \chandra\ observation of a
Galactic plane field and concluded that resolved sources with fluxes
above $3 \times 10^{-15}$ \epcs, corresponding to $4 \times 10^{31}$
\eps\ at 10 kpc, made up about 10\% of the total GRXE flux in this field.
They also suggested that \cvs\ were not major contributors to the
GRXE, but this is true only when neglecting the Galactic distribution
of \cvs. A later, deeper \chandra\ observation of a Galactic bulge field
resolved the majority (up to 88$\pm$12\% in the 6.5--7.1 keV band)
of the GRXE into point sources \citep{Retal2009}. It is now thought
likely that faint, point sources including \cvs\ contribute the
majority of apparently diffuse X-rays from the ridge, the bulge,
and the Galactic center regions (see, e.g., \citealt{Yetal2016}).

However, the specific types of \cvs\ that significantly contribute
to these emission is an open question. For example, \cite{MPMetal2003}
detected 2357 sources in the 17$' \times 17'$ field around Sgr A*, and found
more than half of the sources brighter than $4 \times 10^{-15}$
\epcs\ (2--8 keV; 3 $\sim 10^{31}$ \eps\ at 8 kpc) to be hard,
perhaps indicating them to be magnetic \cvs. \cite{Yetal2012}
studied the spectral properties of the ridge and bulge emissions
and found them to be consistent with those of IPs with an
average white dwarf mass of 0.66$^{+0.09}_{-0.07}$ M$_\odot$.
This is lower than those of IPs in the solar neighborhood
\citep{Yetal2010}. Moreover, for this population to explain the
apparently diffuse emission, the XLF of IPs must extend smoothly
down to $\sim 10^{30}$ \eps\ to explain the bulk of the observed
X-rays, in contradiction with the observed XLF \citep{PM2014}.
In contrast, \cite{XWL2016} studied the Fe K lines of \cvs\ and
the GRXE, and found the properties of the latter to be consistent
with those of dwarf novae. \cite{Hetal2016}, on the other hand,
found that the Galactic center diffuse X-rays to have spectral
characteristics similar to those of field IPs, but again required
IP XLF to extend to lower luminosities than found in the solar
neighborhood.

\section{Current Issues and Future Prospects}

In this section, I discuss major unresolved questions and
topics that are under active investigation.

\subsection{Extreme Objects}

It is perhaps unavoidable that our knowledge of the faintest \cvs\ is
incomplete. In the context of X-ray emissions, X-ray dim WZ~Sge type
dwarf novae must exist in large numbers \citep{Retal2013} and there
may well be numerous LLIPs \citep{PM2014}. The discovery that SU~Lyn
was a previously unrecognized symbiotic star prompted \cite{Metal2016}
to speculate that there probably are a large number of similar systems.

On the other end, our knowledge of most X-ray luminous \cvs\ is
also incomplete, because they are rare. The recurrent nova V2487~Oph
\citep{HS2002} is a luminous X-ray source in quiescence which is
detected also with \integral\ and with \swift\ BAT (\citealt{EJBetal2006}
and Table\,\ref{bat70tab}) with an estimated luminosity of $\sim 10^{35}$
\eps. Despite multiple X-ray observations, no sign of spin modulation
has ever been found (cf. \S 4), although it has complex absorber that
Hernanz and co-workers considered to be the signature of an IP. V2491~Cyg
is another nova with quiescent X-ray luminosity above 10$^{34}$
\eps\ \citep{ZMO2015}. In this case, an X-ray modulation on a time
scale of $\sim$38 min has been seen, but it may not be strictly periodic.
Clearly, one issue for future study is how X-ray luminous accreting white dwarfs can be,
both in magnetic and non-magnetic cases, as highest luminosity systems
should have non-negligible optical depths. Another is the evolutionary
status of these extremely luminous \cvs. There are indications that
both V2487~Oph \citep{Detal2013} and V2491~Cyg \citep{Detal2011} harbor
evolved mass donors.

The other issue is the birth rate ($BR$) of such objects, which is
of interest in addition to the space density ($SD$).  The two are
connected by the relation $SD = BR \times LT$, where $LT$ is the life time
of the binary as that subclass.  $LT$ is limited by the mass of the
donor divided by the mass transfer rate.  Thus, two subclasses of
\cvs\ with an identical $BR$ can have two different present-day $SD$
because the differences in $LT$ --- one luminous and short-lived, one
dimmer and longer lasting. Currently available data allow the possibility
that the very luminous \cvs\ have a birth rate similar to those of more
familiar cousins. This is of particular interest in the context of
single degenerate channel of the progenitors of SNIa. This leads to the
open question of whether these high luminosity white dwarf binaries harbor especially
massive white dwarfs.

In addition, there may be a population of white dwarfs accreting from
early-type stars. There have long been speculations about Be-white dwarf
binaries, including $\gamma$ Cas and $\gamma$ Cas-like objects
\citep{LdOetal2006}. There are apparent UV counterparts to \ssss\ in
M31 that may indicate shell-burning white dwarfs in orbit around early
type stars \citep{Oetal2010}. There is CI~Cam, unusual X-ray transient
whose nature has been disputed. \cite{IMU2004} proposed a nova interpretation,
which is now favored given the {\sl Gaia\/} distance of 1.4$^{+0.7}_{-0.4}$
kpc \citep{Wetal2016}.  Finally, there is MAXI~J0158$-$744, a transient
supersoft source in the SMC that has been interpreted as a Be-white dwarf
binary \citep{Lietal2012}.

\subsection{Asynchronous Polars}

Of the roughly 100 polars known to date, four are definitely asynchronous,
with small ($<$1\%) but significant differences between their spin and
orbital periods. In the case of the prototype, V1500~Cyg, it was almost
certainly the 1975 nova eruption that desynchronized it, and it was seen
to be synchronizing on a timescale of $\sim$150 yrs \citep{SS1991}.
For the other three, no evidence for recent nova eruptions have
been found so far (see, e.g., \citealt{PZ2016}), although the failure
to do so does not disprove the nova origin of asynchronism. More
troublingly, \cite{Warner2002} noted that a mean synchronization time of
$\sim$300 yrs implied a nova recurrence time of all polars needed to
be of order 5000 yrs to explain the presence of four asynchronous
systems among 68 polars known at the time. Such a short recurrence
time would be very surprising, suggesting that either asynchronous
polars are over-represented in our current sample, or some are
asynchronous for reasons unrelated to nova eruptions.

X-ray properties may provide a clue. Of the $\sim$100 known or
suspected polars, 14 have been detected in the \swift\ BAT survey
(Table\,\ref{bat70tab}). These 14 include three of the four known
asynchronous polars (BY~Cam, V1432~Aql, and CD~Ind). What physical
properties distinguish the BAT-detected polars from the rest?
Does asynchronism make polars X-ray bright, or do the underlying
properties that make them X-ray bright also make them more prone
to asynchronism? Are all asynchronous polars the results of recent
nova eruptions, or is there another path to asynchronism? These are
all interesting questions without no definitive answers so far.

\subsection{The Accretion Disk and X-ray Orbital Modulations}

X-ray emissions from non-magnetic \cvs\ and \syss\ arise very close
to the white dwarf, where the accretion disk ought to be axisymmetric.
It appears likely that X-ray orbital modulation is limited to high
inclination systems, via eclipses or dips. We have already discussed
X-ray eclipses, and inferences that we can derive from them.

Dips are often seen in low-mass X-ray binaries (LMXBs), and are caused by
azimuthal structures in the outer accretion disks, probably associated
with the impact region where the ballistic stream from the secondary
hits the outer disk rim (see, for example, \citealt{Petal1986}).
In the dipping LMXBs, while the average behavior is often predictable,
individual dips are not repeatable from cycle to cycle.

Among \cvs, dips have been observed in U~Gem \citep{Setal1996}
and in WZ~Sge \citep{Petal1998} in quiescence. Both are high
inclination systems in which the bright spot, but not the white
dwarf, is eclipsed by the secondary. \cite{NlD1997} studied
the UV and X-ray data on U~Gem in outburst and concluded that
the column density along the line of sight, some 25$^\circ$
above the orbital plane, varied from $< 10^{19}$ to $\sim 10^{22}$
cm$^{-2}$ in this case. Another example is IX~Vel, a nova-like
system, which appears to show X-ray and EUV orbital modulations
\citep{vTetal1995}. The inclination is now thought to be $\sim$57$^\circ$
\citep{Letal2007}, which may define the current low inclination limit for
systems in which X-ray orbital modulation is observed. The only
exception is the mysterious recurrent nova, T~Pyx, which exhibit
X-ray orbital modulation in quiescence \citep{Balman2014} even
though it is thought to be a low inclination system ($\sim 10^\circ$
according to \citealt{UKS2010}).

Many IPs also exhibit X-ray orbital modulation: \cite{PNM2005}
found statistically significant modulation in seven of sixteen
systems they studied. Because the depth of orbital modulation is
deeper at lower energies, the likely origin is absorption by azimuthal
structures in the outer accretion disk. From the fraction of systems
exhibiting modulations, they estimated that such structures existed
up to $\sim 30^\circ$ above the orbital plane. Moreover, they found
such structures come and go from epoch to epoch for individual systems.

It might seem surprising that the outer disk has sufficient matter
30$^\circ$ above the orbital plane to be able to cause X-ray orbital
modulation, compared to the theoretical models of accretion disk:
after all, these are supposed to be optically thick, geometrically
thin disks. However, note that the theory uses exponential scale heights,
and we do expect the presence of lower density matter several scale heights
above the mid-plane. Moreover, soft X-ray dips require modest (up to
$\sim 10^{22}$ cm$^{-2}$ for U~Gem in outburst) column densities
along lines of sight that are almost parallel to the disk surface,
much less than that required to reach electron scattering optical depth of
$\sim 1$ ($\sim 10^{24}$ cm$^{-2}$) which presumably defines the disk
surface. Therefore, it is not immediately obvious if there is a
conflict between theory and observation, except in the case of T~Pyx.
On the other hand, it is possible that the disks are geometrically
thicker than commonly thought. While the luminosity of CV disks are
well below the Eddington limit, they are never fully ionized, so the potential
effect of radiation pressure on the disk structure should be investigated.
After all, radiation-driven accretion disk wind is a common phenomenon among
dwarf novae in outburst and nova-like \cvs\ \citep{DP2000}.

The nature of the intrinsic absorber in \syss\ \citep{Ketal2009}
remains an unsettled issue. If systems like V426~Oph \citep{Retal2008}
and V2487~Oph \citep{HS2002} are indeed non-magnetic, their absorbers
require an explanation, too. \cite{MZS2009} argued that the inner
accretion disk, just outside the boundary layer, can provide partial
covering absorbers if the system is highly inclined. Another attractive
possibility is the accretion disk wind \citep{DP2000}. Quantitative
studies are needed to evaluate if the observed intrinsic absorbers
can be explained as due to accretion disk wind.

\subsection{The Constant Temperature Problem}

The soft, blackbody-like component from the optically thick
boundary layer is rarely observed, which was considered to be
the `missing boundary layer' problem \citep{Fetal1982}. This
appears to be largely due to theories overestimating the
boundary layer temperature. A different problem that emerged
is the apparent constancy in the boundary layer temperature
\citep{Metal1995}.

An analogous problem has been found in a completely different context.
During the supersoft phase of the nova V2491~Cyg, \cite{Petal2010}
found the supersoft component to stabilize at about $\sim$70 eV
after an initial rise. Around day 160, the inferred luminosity of the
supersoft component drooped by over an order of magnitude, but the
temperature remained the same (see their Figure 6). Moreover,
\cite{ZMO2015} found a blackbody-like component with a similar
temperature but at a luminosity that was several orders of magnitude 
lower still in their quiescent \suzaku\ observation. Perhaps the
temperature was pegged by either the Eddington or the atmosphere
limit \citep{WKB1987}.

\subsection{Advanced X-ray Diagnostics}

As mentioned in \S 6.6, X-ray spectra allow us to measure relative
elemental abundances, most commonly that of Fe.  It is often found
to be sub-solar (see, e.g., \citealt{Betal2010}). In contrast,
\cite{MO2005} measured an apparent overabundance of Ne in the old nova,
V603~Aql. It is not yet clear what the true astrophysical implications
are of these findings, but it is essential to keep this possibility
in mind when fitting high quality data that are potentially sensitive
to elemental abundances.

The ratios of forbidden to intercombination lines of He-like triplets
can be used to constrain the density of emitting plasma in principle.
However, as is well known in the case of X-rays from early type stars
\citep{Letal2006}, photoexcitation by UV photons can also change these
ratios. Ratios of certain pairs of Fe L lines can also be to constrain
density \citep{MLF2001,MLF2003}, and the latter (ratio of intensities
of Fe XXII 11.92\AA\ and 11.77\AA\ lines) in particular are less
sensitive to photoexcitation. Unfortunately, suitable data with high
spectral resolution and high signal-to-noise currently exist only for
EX~Hya. The story is similar for the He-like triplets: most existing
grating data are not of sufficient quality to be useful in this regard
\citep{Setal2014}. In addition, resonant scattering may play a
significant role in shaping the line spectra of some magnetic
\cvs\ \citep{Tetal2001,TIM2004}.

Dynamical studies of X-ray emitting plasma have the potential to
advance our knowledge of \cvs\ and \syss\ significantly. X-ray radial velocity study
has so far successfully been performed only for EX~Hya \citep{HBM2004}
and perhaps marginally for AM~Her \citep{GRS2007}. The 500 ks HETG
observation of EX~Hya allowed \cite{Letal2010} to identify a narrow
and a broad component in some emission lines. They interpreted
the broad component as due to photoionization of the pre-shock flow.
The emission lines in SS~Cyg and U~Gem are found to become broader
during outburst \citep{Metal2005}. In contrast, the lines are narrow
in the old nova V603~Aql \citep{MO2005} and the nova-like system TT~Ari
\citep{Zetal2014}. Some of this may be due to inclination
effect, since V603~Aql is supposed to be a low inclination angle
system. However, those of TT~Ari and SS~Cyg are not very different.
This implies that, perhaps, there is a difference between nova like
systems and dwarf novae in outbursts, despite the common perception
that they are both non-magnetic \cvs\ with high accretion rates.

With the advent of \nustar, reflection has now securely been
detected in magnetic \cvs\ \citep{Metal2015}. There are efforts
under way to study reflection in non-magnetic \cvs\ and in
\syss. These observations should securely measure the reflection
amplitudes in the target objects, and allow us to infer if an
optically thick disk extends down to the white dwarf surface
and/or where the X-ray emission region is located. While such
a study was pioneered for SS~Cyg by \cite{Ietal2009} using
\suzaku\ HXD data, it is worth checking their conclusions with
the imaging \nustar\ data that are not subject to the systematic
uncertainty in the background level of the non-imaging HXD data.
Moreover, \edit2{future} observations of magnetic \cvs\ may allow us
to study the inclination angle dependence of reflection off the
white dwarf surface.

\cite{Letal2015} performed a comprehensive spectral analysis of
the long \chandra\ HETG observation of EX~Hya. Seen in detail,
they found the cooling flow model to be lacking in several respects.
In particular, while the model can reproduce the gross features of
the spectrum, H-like to He-like line ratios in particular deviated
from model predictions. It is clear that we do not yet have a full
understanding of the post-shock plasma in magnetic \cvs.

Finally, all X-ray emissions convincingly detected from accreting white dwarf binaries so
far are thermal emissions, whether they are optically thick or thin.
However, non-thermal processes do operate in \cvs\ and \syss. Radio emissions
have been detected from magnetic \citep{MG2007} and non-magnetic
\citep{Cetal2016} \cvs. $\gamma$-rays from novae during eruption
with {\sl Fermi\/} \citep{FermiNovaI,FermiNovaII} are strong evidence
of particle acceleration in shocks in novae. However, \cite{Ketal2014} did
not confirm the earlier report of non-thermal X-rays in AE~Aqr
made using non-imaging detectors. One nova, V2491~Cyg that erupted
before the launch of {\sl Fermi\/}, was probably detected above 15 keV
using \suzaku\ HXD, and, if confirmed, the origin is likely to be non-thermal
\citep{Tetal2009}. There was also a transient source in the vicinity of
V382~Vel (Nova Velorum 1999) in the {\sl Compton\/} Gamma Ray Observatory
data in the 50--300 keV range \citep{Teddyetal2016} that is now suggested
to have been the non-thermal emission from this nova. A secure detection
of non-thermal X-ray emission would likely allow us to constrain the
particle acceleration processes in novae.

\section{Summary and Future Prospects}

In this review, I have presented the large body of X-ray observations
of accreting white dwarfs, along with introductory materials. The number of objects
and subtypes is large to the point of being bewildering to newcomers
to the field. The body literature is large, and it is easy for newcomers
to miss important papers of the past. I hope this review serves as
a useful guide.

I hope to have convinced the readers that the X-ray studies of \cvs\ and \syss\ have
a lot to teach us about the physics of accretion, including that of
accretion disks and the interaction between accretion flow and magnetic field;
that they are an important part of the study of the Galaxy we live in,
including its collective X-ray emission; that they allow us valuable
insight into the evolution of close binaries. Although existing observations
have already taught us a lot, there are obviously a lot more to learn.
Upcoming releases of \gaia\ distances may also force us to
revise our old conclusions.

Existing X-ray missions are still extremely useful, including
the flagship missions \chandra\ and \xmm; they allow us to reach
low flux levels on the one hand and high resolution spectroscopy of the
bright sources on the other. \swift\ continues to provide an essential
service using its flexible observing capability for \cvs\ and \syss, particularly novae.
We have only scratched the surface of what \nustar\ can do, and have
not heard any results on accreting white dwarfs using {\sl Astrosat} yet.

The number of known accreting white dwarf binaries will continue to increase. Some will come
from X-ray surveys, including the continued use of \integral\ and \swift\ BAT
hard X-ray surveys. Since some of them are highly variable, time-resolved
hard X-ray surveys need to be utilized (see, e.g., the case of SU~Lyn:
\citealt{Metal2016}. The forthcoming \erosita\ mission should discover
many new \cvs\ in the soft and medium energy X-rays.
The serendipitous source catalogs from \xmm\ \citep{XMMcat} and
\swift\ \citep{SXPS} contain a large number of sources, which have
not been fully utilized by the community yet. It might require new
approaches such as machine learning focused on variable X-ray sources
to select \cvs\ and \syss\ for more in-depth studies (see, e.g., \citealt{Loetal2014}).
One could perform short, pointed X-ray observations of objects that
may turn out to be accreting white dwarf binaries. For example, \cite{Setal2015} observed
asymptotic giant branch stars with far UV excess and detected six,
which implies they are previously unrecognized \syss.
Any projects in the area of time-domain astronomy are guaranteed
to discover \cvs\ and \syss; here, again, the problem may well be how to
manage and exploit the huge volume of data.

Finally, there will be future X-ray missions with unprecedented
capabilities. It is highly likely that there will be an X-ray
polarimetry mission in the near future: The {\sl IXPE\/} mission
has been approved in the US, and the {\sl XIPE\/} mission is
being considered in Europe. A sensitive X-ray polarimetry mission
has the potential to provide unique insights into the accretion
columns of magnetic \cvs\ \citep{Matt2004}. In terms of X-ray spectroscopy,
the Japanese and US groups hope to re-fly the ill-fated \hitomi\ mission
with an X-ray microcalorimeter, an instrument that provides $\sim$5 eV
spectral resolution in the Fe K band. This mission will enable dynamical
studies and density diagnostics using atomic lines above 2 keV for
accreting white dwarfs. Further in the future, \athena\ will
bring the era of even higher resolution X-ray spectroscopy as a routine
tool for the study of \cvs\ and \syss.

\section*{Acknowledgments}

I thank my frequent collaborators Drs. Jennifer Sokoloski, Thomas Nelson,
and Juan Luna,  with whom I have had many fruitful discussions. Many others
researchers helped me over the years through their own research papers and
through more informal discussion. Such colleagues are too numerous to list,
but I would like to single out Dr. Christopher Mauche, whose thoughtful
papers still have a lot to teach. Specific comments on the earlier
versions of this manuscript by Drs. Sokoloski and Luna, as well as by
Dr. John Cannizzo on matters related to disk instability have hopefully
improved it, although any remaining inaccuracy are my responsibility.
In addition to these professional astronomers, I thank the countless
volunteer astronomers, many of whom are anything but amateurish, who
have contributed to the studies summarized here through the data and
the alerts provided through American Association of Variable Star
Observers (AAVSO), Center for Backyard Astrophysics (CBA) and other
worthy organizations. I acknowledge the financial support NASA under
Astrophysics Data Analysis Program grant NNX13AJ13G during the writing
of this review.


\begin{thebibliography}{}

\bibitem[Abdo et al. (2010)]{FermiNovaI}
         Abdo, A.A., Ackermann, M., Ajello, M., et al. 2010, Science,
	 329, 817

\bibitem[Ackermann et al. (2011)]{FermiNovaII}
         Ackermann, M., Ajello, M., Albert, A., et al. 2014, Science,
	 345, 554

\bibitem[Aizu (1973)]{Aizu1973}
         Aizu, K. 1973, Prog. Theoret. Phys., 49, 1184

\bibitem[Allan et al. (1998)]{Aetal1998}
         Allan, A., Hellier, C. \& Beardmore, A. 1998, MNRAS, 295, 167

\bibitem[Angelini \& Verbunt (1989)]{AV1989}
         Angelini, L. \& Verbunt, F. 1989, MNRAS, 238, 697

\bibitem[Anzolin et al. (2008)]{Aetal2008}
         Anzolin, G., de Martino, D., Bonnet-Bidaud, J.-M., Mouchet, M.,
	 G\"ansicke, B.T., Matt, G. \& Mukai, K. 2008, A\&A, 489, 1243

\bibitem[Balman \& \"Ogelman (1999)]{BO1999}
         Balman, S. \& \"Ogelman, H.B. 1999, \apj, 518, L111

\bibitem[Balman \& Revnivtsev (2012)]{BR2012}
         Balman, S. \& Revnivtsev, M. 2014, A\&A, 546, A112

\bibitem[Balman (2014)]{Balman2014}
         Balman, S. 2014, A\&A, 572, A114

\bibitem[Baptista, Borges \& Oliveira (2016)]{BBO2016}
         Baptista, R., Borges, B.W. \& Oliveira, A.S. 2016, MNRAS, 463, 3799

\bibitem[Barlow et al. (2006)]{EJBetal2006}
         Barlow, E.J., Knigge, C., Bird, A.J., Dean, A.J., Clark, D.J.,
	 Hill, A.B., Molina, M. \& Sguera, V. 2006, \mnras, 372, 224

\bibitem[Baskill, Wheatley \& Osborne (2005)]{BWO2005}
         Baskill, D.S., Wheatley, P.J. \& Osborne, J.P. 2005, MNRAS, 357, 626

\bibitem[Baumgartner et al. (2013)]{BAT70}
         Baumgartner, W.H., Tueller, J., Markwardt, C.B., Skinner, G.K.,
         Barthelmy, S., Mushotzky, R.F., Evans, P.A. \& Gehrels, N. 2013,
	 \apjs, 207, A19

\bibitem[Beardmore et al. (1995)]{Betal1995}
         Beardmore, A.P., Done, C., Osborne, J.P. \& Ishida, M. 1995,
	 MNRAS, 272, 749

\bibitem[Beardmore et al. (1998)]{Betal1998}
	 Beardmore, A.P., Mukai, K., Norton, A.J., Osborne, J.P. \&
	 Hellier, C. 1998, MNRAS, 297, 337

\bibitem[Beuermann (1999)]{B1999}
         Beuermann, K. 1999, Proc. "Heighlights in X-ray Astronomy,"
         eds. B. Aschenbach \& M.J. Freyberg, 1999, MPE Report 272, 410

\bibitem[Beuermann, Burwitz \& Reinsch (2012)]{BBR2012}
         Beuermann, K., Burwitz, V. \& Reinsch, K. 2012, A\&A, 543, A41

\bibitem[Bitner, Robinson \& Behr (2007)]{BRB2007}
         Bitner, M.A., Robinson, E.L. \& Behr, B.B. 2007, ApJ, 662, 564

\bibitem[Bode et al. (1987)]{Betal1987}
         Bode, M.F., Seaquist, E.R., Frail, D.A., Roberts, J.A.,
	 Whittet, D.C.B., Evans, A. \& Albinson, J.S. 1987, Nature, 329, 519

\bibitem[Bode et al. (2006)]{Betal2006}
         Bode, M.F., O'Brien, T.J., Osborne, J.P., Page, K.L., Senziani, F.,
	 Skinner, G.K., Starrfield, S., Ness, J.-U., Drake, J.J., Schwarz, G.,
	 Beardmore, A.P., Darnley, M.J., Eyres, S.P.S., Evans, A., Gehrels, N.,
	 Goad, M.R., Jean, P., Krautter, J. \& Novara, G. 2006, ApJ, 652, 629

\bibitem[Bowyer \& Malina (1991)]{BM1991}
         Bowyer, S. \& Malina, R.F. 1991, Extreme Ultraviolet Astronomy,
         ed. R.F. Malina \& S. Bowyer (New York: Pergamon), 391

\bibitem[Britt et al. (2015)]{Betal2015}
         Britt, C.T., Maccarone, T., Pretorius, M.L., Hynes, R.I.,
	 Jonker, P.G., Torres, M.A.P., Knigge, C., Johnson, C.O.,
	 Heinke, C.B., Steeghs, D., Greiss, S. \& Nelemans, G., 2015,
	 MNRAS, 448, 3455

\bibitem[Buckley et al. (1997)]{Betal1997}
         Buckley, D.A.H., Haberl, F., Motch, C., Pollard, K.,
	 Schwarzenberg-Czerny, A. \& Sekiguchi, K. 1997, MNRAS, 287, 117

\bibitem[Byckling et al. (2009)]{Betal2009}
         Byckling, K., Osborne, J.P., Wheatley, P.J., Wynn, G.A.,
	 Beardmore, A., Braito, V., Mukai, K. \& West, R., 2009,
	 MNRAS, 399, 1576

\bibitem[Byckling et al. (2010)]{Betal2010}
         Byckling, K., Mukai, K., Thorstensen, J.R. \& Osborne, J.P. 2010,
	 MNRAS, 408, 2298

\bibitem[Cannizzo et al. (2012)]{Cetal2012}
         Cannizzo, J.K., Smale, A.P., Wood, M.A., Still, M.D. \& Howell, S.B.
	 2012, ApJ, 747, A117

\bibitem[Cash (2002)]{CashThesis}
        Cash, J.L. 2002, Ph.D Thesis, University of Wyoming

\bibitem[Cheng et al. (1997)]{Cetal1997}
         Cheng, F.H., Sion, E.M., Horne, K., Hubeny, I., Huang, M.
	 \& Vrtilek, S.D. 1997, AJ, 114, 1165

\bibitem[Cheung et al. (2016)]{Teddyetal2016}
         Cheung, C.C., Jean, P., Shore, S.N., Grove, J.E. \& Leising, M.
	 2016, The 34th International Cosmic Ray Conference, 30 July--6
	 August 2015 (The Hague, The Netherlands), PoS(ICRC2015) 880

\bibitem[Chomiuk et al. (2014)]{Cetal2014}
         Chomiuk, L., Nelson, T., Mukai, K., Sokoloski, J.L.,  Rupen, M.P.,
	 Page, K.L., Osborne, J.P., Kuulkers, E., Moduszewski, A.J., Roy, N.,
	 Weston, J. \& Krauss, M.I. 2014, ApJ, 788, A130

\bibitem[Coley, Corbet \& Krimm (2015)]{CCK2015}
         Coley, J.B., Corbet, R.H.D. \& Krimm, H.A. 2015, ApJ, 808, A140

\bibitem[Collins \& Wheatley (2010)]{CW2010}
         Collins, D.J. \& Wheatley, P.J. 2010, MNRAS, 402, 1816

\bibitem[Coppejans et al. (2016)]{Cetal2016}
         Coppejans, D.L., K\"ording, E.G., Miller-Jones, J.C.A., Rupen, M.P.,
	 Sivakoff, G.R., Knigge, C., Groot, P.J., Woudt, P.A., Waagen, E.A.
	 \& Templeton, M. 2016, MNRAS, 463, 2229

\bibitem[Corbet et al. (2008)]{Cetal2008}
         Corbet, R.H.D., Sokoloski, J.L., Mukai, K., Markwardt, C.B.
         \& Tueller, J. 2008, ApJ, 675, 1424

\bibitem[C\'ordova \& Mason (1984)]{CM1984}
         C\'ordova, F.A. \& Mason, K.O. 1984, MNRAS, 206, 879

\bibitem[Cropper (1990)]{Mark1990}
         Cropper, M. 1990, SSRv, 54, 195

\bibitem[Cropper, Ramsay \& Wu (1998)]{CRW1998}
         Cropper, M., Ramsay, G. \& Wu, K. 1998, \mnras, 293, 222

\bibitem[Crowley, Espey \& McCandliss (2008)]{CEMcC2008}
         Crowley, C., Espey, B.R. \& McCandliss, S.R. 2008, ApJ, 675, 711

\bibitem[Darnley et al. (2011)]{Detal2011}
         Darnley, M.J., Ribeiro, V.A.R.M., Bode, M.F. \& Munari, U.
	 2011, A\&A, 530, A70

\bibitem[Darnley et al. (2013)]{Detal2013}
         Darnley, M.J., Ribeiro, V.A.R.M., Bode, M.F., Hounsell, R.A.
	 \& Williams, R.P. 2013, ApJ, 746, A61

\bibitem[de Kool \& Wickramasinghe (1999)]{dKW1999}
         de Kool, M. \& Wickramasinghe, D. 1999, MNRAS, 307, 449

\bibitem[de Martino et al. (2001)]{dMetal2001}
         de Martino, D., Matt, G., Mukai, K., Belloni, T.,
         Bonnet-Bidaud, J.-M., Chiappetti, L., G\"ansicke, B.T., Haberl, F.
         \& Mouchet, M. 2001, \aap, 377, 499

\bibitem[de Martino et al. (2005)]{dMetal2005}
         de Martino, D., Matt, G., Mukai, K., Bonnet-Bidaud, J.-M.,
	 G\"ansicke, B.T., Gonzalez Perez, J.M., Haberl, F., Mouchet, M.
	 \& Solheim, J.-E. 2005, A\&A, 437, 935

\bibitem[de Martino et al. (2008)]{dMetal2008}
         de Martino, D., Matt, G., Mukai, K., Bonnet-Bidaud, J.-M.,
         Falanga, M., G\"ansicke, B.T., Haberl, F., Marsh, T.R.,
         Mouchet, M., Littlefair, S.P. \& Dhillon, V. 2008, \aap, 481, 149

\bibitem[Dobrotka, Mineshige \& Ness (2014)]{DMN2014}
         Dobrotka, A., Mineshige, S. \& Ness, J.-U. 2014, MNRAS, 438, 1714

\bibitem[Dolence, Wood \& Silver (2008)]{DWS2008}
         Dolence, J., Wood, M.A. \& Silver, I. 2008, ApJ, 683, 375

\bibitem[Done, Osborne \& Beardmore (1995)]{Detal1995}
         Done, C., Osborne, J.~P. \& Beardmore, A.~P. 1995, \mnras, 276, 483

\bibitem[Done \& Magdziarz (1998)]{DM1998}
         Done, C. \& Magdziarz, P. 1998, \mnras, 298, 737

\bibitem[Drew \& Proga (2000)]{DP2000}
         Drew, J.E. \& Proga, D. 2000, NewAR 44, 21

\bibitem[Ebisawa et al. (2001a)]{Eetal2001a}
         Ebisawa, K., Mukai, K., Kotani, T., Asai, K., Dotani, T., Nagase, F.,
	 Hartmann, H.W., Heise, J., Kahabka, P. \& van Teeseling, A. 2001a,
	 ApJ, 550, 1007

\bibitem[Ebisawa et al. (2001b)]{Eetal2001b}
         Ebisawa, K., Maeda, Y., Kaneda, H. \& Yamauchi, S. 2001b,
	 Science, 293, 1633

\bibitem[Edmonds et al. (2003a)]{Eetal2003a}
	 Edmonds, P.D., Gilliland, R.L., Heinke, C.O. \& Grindlay, J.E.
	 2003a, ApJ, 596, 1177

\bibitem[Edmonds et al. (2003b)]{Eetal2003b}
	 Edmonds, P.D., Gilliland, R.L., Heinke, C.O. \& Grindlay, J.E.
	 2003b, ApJ, 596, 1197

\bibitem[Evans et al. (2004)]{Eetal2004}
         Evans, P.A., Hellier, C., Ramsay, G. \& Cropper, M. 2004,
	 MNRAS, 349, 715

\bibitem[Evans, Hellier \& Ramsay (2006)]{EHR2006}
         Evans, P.A., Hellier, C. \& Ramsay, G. 2006, MNRAS, 369, 1229

\bibitem[Evans \& Hellier (2007)]{EH2007}
         Evans, P.A. \& Hellier, C. 2007, ApJ, 663, 1277

\bibitem[Evans et al. (2014)]{SXPS}
         Evans, P.A., Osborne, J.P., Beardmore, A.P., Page, K.L.,
	 Willingale, R., Mountford, C.J., Pagani, C., Burrows, D.N.,
	 Kennea, J.A., Perri, M., Tagliaferri, G. \& Gehrels, N. 2014,
	 ApJS, 210, A8

\bibitem[Ezuka, Ishida \& Makino (1998)]{EIM1998}
         Ezuka, H., Ishida, M. \& Makino, F. 1998, ApJ, 499, 388

\bibitem[Ezuka \& Ishida (1999)]{EI1999}
         Ezuka, H. \& Ishida, M. 1999, \apjs, 120, 277

\bibitem[Fabian et al. (2015)]{Fetal2015}
         Fabian, A.C., Lohfink, A., Kara, E., Parker, M.L., Vasudevan, R.
	 \& Reynolds, C.S. 2015, MNRAS< 451, 4375

\bibitem[Ferland et al. (1982)]{Fetal1982}
         Ferland, G.J., Langer, S.H., MacDonald, J., Pepper, G.H.,
	 Shaviv, G. \& Truran, J.W. 1982, ApJ, 262, L53

\bibitem[Fertig et al. (2011)]{Fetal2011}
         Fertig, D, Mukai, K., Nelson, T. \& Cannizzo, J.K. 2011,
	 PASP, 123, 1054

\bibitem[Fischer \& Beuermann (2001)]{FB2001}
         Fischer, A. \& Beuermann, K. 2001, A\&A, 373, 211

\bibitem[Galloway \& Sokoloski (2004)]{GS2004}
         Galloway, D.K. \& Sokoloski, J.L. 2004, ApJ, 613, L61

\bibitem[G\"ansicke et al. (1998)]{Getal1998}
         G\"ansicke, B.T., Hoard, D.W., Beuermann, K., Sion, E.M.
	 \& Szkody, P. 1998, A\&A, 338, 933

\bibitem[G\"ansicke et al. (2009)]{Getal2009}
         G\"ansicke, B.T., Dillon, M., Southworth, J., Thorstensen, J.R.,
	 Rodr\'iguez-Gil, P., Aungwerojwit, A., Marsh, T.R., Szkody, P.,
	 Barros, S.C.C., Casares, J., de Martino, D., Groot, P.J.,
	 Hakala, P., Kolb, U., Littlefair, S.P., Mart\'inez-Pais, I.G.,
	 Nelemans, G. \& Schreiber, M.R. 2009, MNRAS, 397, 2170

\bibitem[Gehrels \& Williams (1993)]{GW1993}
         Gehrels, N. \& Williams, E.D. 1993, ApJ, 418, L25

\bibitem[Gehrels et al. (2004)]{SWIFTREF}
         Gehrels, N. et al. 2004, ApJ, 611, 1005

\bibitem[George \& Fabian (1991)]{GF1991}
         George, I.M. \& Fabian, A.C. 1991, \mnras, 249, 352

\bibitem[Ghosh \& Lamb (1979)]{GL1979}
         Ghosh, P. \& Lamb, F.K. 1979, ApJ, 232, 259

\bibitem[Giacconi et al. (1979)]{Getal1979}
         Giacconi, R., Branduardi, G., Briel, U., Epstein, A., Fabricant, D.,
         Feigelson, E., Forman, W., Gorenstein, P., Grindlay, J., Gursky, H.,
         Harnden, F.R., Henry, J.P., Jones, C., Kellogg, E., Koch, D.,
         Murray, S., Schreier, E., Seward, F., Tananbaum, H., Topka, K.,
         Van Speybroeck, L., Holt, S.S., Becker, R.H., Boldt, E.A.,
         Serlemitsos, P.J., Clark, G., Canizares, C., Markert, T.,
         Novick, R., Helfand, D. \& Long, K. 1979, ApJ, 230, 540

\bibitem[Girish, Rana \& Singh (2007)]{GRS2007}
         Girish, V., Rana, V.R. \& Singh, K.P. 2007, ApJ, 658, 525

\bibitem[Gosnell et al. (2012)]{Getal2012}
         Gosnell, N.M., Pooley, D., Geller, A.M., Kalirai, J.,
	 Mathieu, R.D., Frinchaboy, P. \& Ramirez-Ruiz, E. 2012,
	 ApJ, 745, A57

\bibitem[Greiner et al. (2010)]{Getal2010}
         Greiner, J., Kruehler, T., Schady, P., Rau, A. \& Olivares, F.
	 2010, ATel, 2746

\bibitem[Grindlay et al. (2001)]{Getal2001}
	 Grindlay, J.E., Heinke, C., Edmonds, P.D. \& Murray, S.S. 2001,
	 Sci., 292, 2290

\bibitem[Gursky et al. (1978)]{Getal1978}
         Gursky, H., Bradt, H., Doxsey, R., Schwartz, D., Schwarz, J.,
         Dowwer, R., Fabbiano, G., Griffiths, R.E., Johnston, M., Leach, R.,
         Ramsey, A. \& Spada, G. 1978, \apj, 223, 973

\bibitem[G\"uver et al. (2006)]{Getal2006}
         G\"uver, T., Uluyazi, C., \"Ozkan, M.T. \& G\"og\"us, E.
	 2006, MNRAS, 372, 450

\bibitem[Haberl \& Motch (1995)]{HM1995}
         Haberl, F. \& Motch, C. 1995, A\&A, 297, L37

\bibitem[Hailey et al. (2016)]{Hetal2016}
         Hailey, C.J., Mori, K., Perez, K., Canipe, A.M., Hong, J.,
	 Tomsick, J.A., Boggs, S.E., Christensen, F.E., Craig, W.W.,
	 Fornasini, F., Grindlay, J.E., Harrison, F.A., Nynka, M.,
	 Rahoui, F., Stern, D., Zhang, S. \& Zhang, W.W. 2016, ApJ,
	 826, A160

\bibitem[Hakala et al. (2004)]{Hetal2004}
         Hakala, P., Ramsay, G, Wheatley, P., Harlaftis, E.T. \&
	 Papadimitriou, C. 2004, A\&A, 420, 273

\bibitem[Hameury et al. (1998)]{Hetal1998}
         Hameury, J.-M., Menou, K., Dubus, G., Lasota, J.-P. \& Hur\'e, J.-M.
	 1998, MNRAS, 298, 1048

\bibitem[Harrison et al. (2013)]{NUSTARREF}
         Harrison, F.A., Craig, W.W., Christensen, F.E., et al. 2013,
	 ApJ, A103

\bibitem[Heise et al. (1985)]{Hetal1985}
         Heise, J., Brinkman, A.C., Gronenschild, E., Watson, M.,
	 King, A.R., Stella, L. \& Kieboom, K. 1987, A\&A 148, L14

\bibitem[Hellier (1993)]{Hellier1993}
         Hellier, C. 1993, MNRAS, 265 L35

\bibitem[Hellier (1997)]{Hellier1997}
         Hellier, C. 1997, MNRAS, 291, 71

\bibitem[Hellier, Mukai \& Beardmore (1997)]{Hetal1997}
         Hellier, C., Mukai, K. \& Beardmore, A.P. 1997, MNRAS, 292, 397

\bibitem[Hellier et al. (2000)]{Hetal2000}
         Hellier, C., Kemp, J., Naylor, T. Bateson, F.M., Jones, A.,
	 Overbeek, D., Stubbings, R. \& Mukai, K. 2000, MNRAS, 313, 703

\bibitem[Hellier (2014)]{Hellier2014}
         Hellier, C. 2014, EJP Web of Conferences, 64, id.07001

\bibitem[Henze et al. (2014)]{Hetal2014}
         Henze, M., Pietsch, W., Haberl, F., Della Valle, M., Sala, G.,
	 Hatzidimitriou, D., Hofmann, F., Hernanz, M., Hartmann, D.H.
	 \& Greiner, J. 2014, A\&A, 563, A2

\bibitem[Hernanz \& Sala (2002)]{HS2002}
         Hernanz, M. \& Sala, G. 2002, Science, 298, 393

\bibitem[Hertz et al. (1990)]{Hetal1990}
         Hertz, P., Bailyn, C.D., Grindlay, J.E., Garcia, M.R., Cohn, H.
         \& Lugger, P.M. 1990, ApJ, 364, 251

\bibitem[Hickman et al. (2009)]{Hetal2009}
         Hickman, R., Wheatley, P., Marsh, T.R., Dhillon, V.S.,
	 Littlefair, S., G\"ansicke, B. 2009, Poster paper presented
	 at the European Week of Astronomy and Space Sciences,
	 20--23 April 2009, University of Hertfordshire, UK, F-P07
	 (http://star.herts.ac.uk/ewass/)

\bibitem[Hoogerwerf, Brickhouse \& Mauche (2004)]{HBM2004}
         Hoogerwerf, R., Brickhouse, N.S. \& Mauche, C.W. 2004, ApJ, 610, 411

\bibitem[Hubeny (1990)]{H1990}
         Hubeny, I. 1990, ApJ, 351, 632

\bibitem[Hubeny (1994)]{H1994}
         Hubeny, I. 1994, in ``Interacting binary stars: a symposium
	 held in conjunction with the 105th Meeting of the Astronomical
	 Society of the Pacific, San Diego State University, 13--15 July 1993,''
	 ASP Conf. Ser 56, ed. A.W. Shafter, 3

\bibitem[Ibarra et al. (2009)]{v2491pre}
         Ibarra, A., Kuulkers, E., Osborne, J.P., Page, K., Ness, J.U.,
	 Saxton, R.D., Baumgartner, W., Beckmann, V., Bode, M.F.,
	 Hernanz, M., Mukai, K., Orio, M., Sala, G., Starrfield, S. \& Wynn,
	 G.A. 2009, A\&A, 497, L5

\bibitem[Ishida (1991)]{I1991}
         Ishida, M. 1991, Ph.D. Thesis, University of Tokyo

\bibitem[Ishida, Morio \& Ueda (2004)]{IMU2004}
	 Ishida, M., Morio, K. \& Ueda, Y. 2004, ApJ, 601, 1088

\bibitem[Ishida et al. (2009)]{Ietal2009}
         Ishida, M., Okada, S., Hayashi, T., Nakamura, R., Terada, Y.
	 Mukai, K. \& Hamaguchi, K. 2009, PASJ, 61, S77

\bibitem[Itoh et al. (2006)]{Ietal2006}
         Itoh, K., Okada, S., Ishida, M. \& Kunieda, H. 2006, ApJ, 639, 397

\bibitem[Jahoda et al. (1996)]{Jetal1996}
         Jahoda, K., Swank, J.H., Giles, A.B., Stark, M.J., Strohmayer, T.,
         Zhang, W. \& Morgan, E.H. 1996, Proc SPIE, 2808, 59

\bibitem[Jansen et al. (2001)]{XMMREF}
         Jansen, F., Lumb, D., Altieri, B., Clavel, J., Ehle, M., Erd, M.,
         Gabriel, C., Guainazzi, M., Gondoin, P., Much, R., Munoz, R.,
	 Santos, M., Schartel, N., Texier, D. \& Vacanti, G. 2001,
	 A\&A, 365, L1.

\bibitem[Joshi et al. (2016)]{Jetal2016}
         Joshi, A., Pandey, J.C., Singh, K.P. \& Agrawal, P.C. 2016,
	 ApJ, 830, A56

\bibitem[Kahabka \& van den Heuvel (1997)]{KvdH1997}
         Kahabka, P. \& van den Heuvel, E.P.J. 1997, ARA\&A, 35, 69

\bibitem[Karovska et al. (2005)]{Ketal2005}
         Karovska, M., Schlegel, E., Hack, W., Raymond, J.C. \&
	 Wood, B.E. 2005, ApJ, 623, L137

\bibitem[Karovska et al. (2010)]{Ketal2010}
         Karovska, M., Gaetz, T.J., Carillli, C.L., Hack, W.,
	 Raymond, J.C. \& Lee, N.P. 2010, ApJ, 710, L132

\bibitem[Kennea et al. (2009)]{Ketal2009}
         Kennea, J.A., Mukai, K., Sokoloski, J.L., Luna, G.J.M.,
	 Tueller, J., Markwardt, C.B. \& Burrows, D.N. 2009, ApJ, 701, 1992

\bibitem[King \& Williams (1985)]{KW1985}
	 King, A.R. \& Williams, G.A. 1985, MNRAS, 215, 1p

\bibitem[King \& Watson (1987)]{KW1987}
         King, A.R. \& Watson, M.G. 1987, MNRAS, 227, 205

\bibitem[King \& Lasota (1991)]{KL1991}
         King, A.R. \& Lasota, J.-P. 1991, ApJ, 378, 674

\bibitem[King (2000)]{King2000}
         King, A.R. 2000, ApJ, 541, 306

\bibitem[Kitaguchi et al. (2014)]{Ketal2014}
         Kitaguchi, T., An, H., Beloborodov, A.M., Gotthelf, E.V.,
	 Hayashi, T., Kaspi, V.M., Rana, V.R., Boggs, S.E., Christensen, F.E.,
	 Craig, W.W., Hailey, C.J., Harrison, F.A., Stern, D. \& Zhang, W.W.
	 2014, ApJ, 782, A3

\bibitem[Knigge, Baraffe \& Patterson (2011)]{KBP2011}
         Knigge, C., Baraffe, I. \& Patterson, J. 2011, ApJS, 194, A28

\bibitem[Kraft, Mathews \& Greenstein (1962)]{KMG1962}
         Kraft, R.P., Mathews, J. \& Greenstein, J.L. 1962, ApJ, 136, 312

\bibitem[Krauss et al. (2011)]{Ketal2011}
         Krauss, m.I., Chomiuk, L., Rupen, M., Roy, N., Mioduszewski, A.J.,
	 Sokoloski, J.L., Nelson, T., Mukai, K., Bode, M.F., Eyres, S.P.S.
	 \& O'Brien, T.J. 2011, ApJ, 739, L6

\bibitem[Kuijpers \& Pringle (1982)]{KP1982}
         Kuijpers, J. \& Pringle, J.K. 1982, A\&A, 114, L4

\bibitem[Lamb \& Masers (1979)]{LM1979}
         Lamb, D.Q. \& Masters, A.R. 1979, ApJ, 234, L117

\bibitem[Lasota, Kuulkers \& Charles (1999)]{LKC1999}
         Lasota, J.-P., Kuulkers, E. \& Charles, P. 1999, MNRAS, 305, 473

\bibitem[Lasota (2001)]{DIM}
         Lasota, J.-P. 2001, NewAR, 45, 449

\bibitem[Leutenegger et al. (2006)]{Letal2006}
         Leutenegger, M.A., Paerels, F.B.S., Kahn, S.M. \& Cohen, D.H. 2006,
	 ApJ, 650, 1096

\bibitem[Li et al. (2012)]{Lietal2012}
         Li, K.L., Kong, A.K.H., Charles, P.A., Lu, T.-N., Bartlett, E.S.,
	 Coe, M.J., McBride, V., Rajoelimanana, A., Udalski, A., Masetti, N.
	 \& Franzen, T. 2012, ApJ, 761, A99

\bibitem[Liebert, Bergeron \& Holberg (2003)]{LBH2003}
         Liebert, J., Bergeron, P. \& Holberg, J.B. 2003, AJ, 125, 348

\bibitem[Linnell et al. (2007)]{Letal2007}
         Linnell, A.P., Godon, P., Hubeny, I. Sion, E.M. \& Szkody, P.
	 2007, ApJ, 662, 1204

\bibitem[Littlefield et al. (2015)]{CLetal2015}
         \edit1{Littlefield, C., Mukai, K., Mumme, R., Cain, R., Magno, K.C.,
	 Corpuz, T., Sandefur, D., Boyd, D., Cook, M., Ulowetz, J. \&
	 Martinez, L. 2015, MNRAS, 449, 3107}

\bibitem[Liu et al. (2008)]{Letal2008}
         Liu, F.K., Meyer, F., Meyer-Hofmeister, E. \& Burwitz, V. 2008,
	 A\&A, 483, 231

\bibitem[Lo et al. (2014)]{Loetal2014}
         Lo, K.K., Farrell, S., Murphy, T. \& Gaensler, B.M. 2014, ApJ, 786, 20

\bibitem[Lopes de Oliveira et al. (2006)]{LdOetal2006}
         Lopes de Oliveira, R., Motch, C., Haberl, F., Negueruela, I.
	 \& Janot-Pacheco, E. 2006, A\&A, 454, 265

\bibitem[Lubow \& Shu (1975)]{LS1975}
         Lubow, S.H. \& Shu, F.H. 1975, ApJ, 198, 383

\bibitem[Luna \& Costa (2005)]{LC2005}
        Luna, G.J.M. \& Costa, R.D.D. 2005, A\&A, 435, 1087

\bibitem[Luna \& Sokoloski (2007)]{LS2007}
         Luna, G.J.M. \& Sokoloski, J.L. 2007, ApJ, 671, 741

\bibitem[Luna et al. (2009)]{Letal2009}
         Luna, G.J.M., Montez, R., Sokoloski, J.L., Mukai, K. \& Kastner, J.H.
	 2009, ApJ, 707, 1168

\bibitem[Luna et al. (2010)]{Letal2010}
         Luna, G.J.M., Raymond, J.C., Brickhouse, N.S., Mauche, C.W.,
	 Proga, D., Steeghs, D. \& Hoogerwerf, R. 2010, ApJ, 711, 1333

\bibitem[Luna et al. (2013)]{Letal2013}
         Luna, G.J.M., Sokoloski, J.L., Mukai, K. \& Nelson, T. 2013,
	 A\&A, 559, A6

\bibitem[Luna et al. (2015)]{Letal2015}
         Luna, G.J.M., Raymond, J.C., Brickhouse, N.S., Mauche, C.W.
	 \& Suleimanov, V. 2015, A\&A, 578, A15

\bibitem[Marsh \& Steeghs (2002)]{MS2002}
         Marsh, T.R. \& Steeghs, D. 2002, MNRAS, 331, L7

\bibitem[Marsh et al. (2016)]{Marshetal2016}
	 Marsh, T.R., G\"ansicke, B.T., H\"ummerich, S., Hambsch, F.-J.,
	 Bernhard, K., Lloyd, C., Breedt, E., Stanway, E.R., Steeghs, D.T.,
	 Parsons, S.G., Toloza, O., Schreiber, M.R., Jonker, P.G.,
	 van Roestel, J., Kupfer, T., Pala, A.F., Dhillon, V.S., Hardy, L.K.,
	 Littlefair, S.P., Aungwerojwit, A., Arjyotha, S., Koester, D.,
	 Bochinski, J.J., Haswell, C.A., Frank, P. \& Wheatley, P.J. 2016,
	 Nature, 537, 374

\bibitem[Mason (1985)]{Mason1985}
         Mason, K.O. 1985, SpSciRev, 40, 99

\bibitem[Mason \& Gray (2007)]{MG2007}
         Mason, P.A. \& Gray, C.L. 2007, ApJ, 660, 662

\bibitem[Matt et al. (2000)]{Metal2000}
         Matt, G., de Martino, D., G\"ansicke, B.T., Negueruela, I.,
         Silvotti, R., Bonnet-Bidaud, J.-M., Mouchet, M. \& Mukai, K. 2000,
	 \aap, 358, 177

\bibitem[Matt (2004)]{Matt2004}
         Matt, G. 2004, A\&A, 423, 495

\bibitem[Mauche et al. (1995)]{Metal1995}
         Mauche, C.W., Raymond, J.C. \& Mattei, J.A. 1995, \apj, 446, 842

\bibitem[Mauche (1999)]{Mauche1999}
         Mauche, C.W. 1999, in ``Annapolis Workshop on Magnetic
	 Cataclysmic Variables,'' ASP Conf Ser 157, eds. C. Hellier \&
	 K. Mukai, 157

\bibitem[Mauche \& Raymond (2000)]{MR2000}
         Mauche, C.W. \& Raymond, J.C. 2000, ApJ, 541, 924

\bibitem[Mauche, Liedahl \& Fournier (2001)]{MLF2001}
         Mauche, C.W., Liedahl, D.A. \& Fournier, K.B. 2001, ApJ, 560, 992

\bibitem[Mauche \& Robinson (2001)]{MR2001}
         Mauche, C.W. \& Robinson, E.L. 2001, ApJ, 562, 508

\bibitem[Mauche (2002)]{M2002}
         Mauche, C.W. 2002, in ``Continuing the Challenge of EUV Astronomy:
	 Current Analysis and Prospects for the Future,'' ASP Conf Ser. 264,
	 eds. S. Howell, J. Dupuis, D. Golombek, F. Walter \& J. Cullison, 75

\bibitem[Mauche, Liedahl \& Fournier (2003)]{MLF2003}
         Mauche, C.W., Liedahl, D.A. \& Fournier, K.B. 2003, ApJ, 588, L101

\bibitem[Mauche (2004)]{M2004}
         Mauche, C.W. 2004, ApJ, 610, 422

\bibitem[Mauche et al. (2005)]{Metal2005}
         Mauche, C.W., Wheatley, P.J., Long, K.S., Raymond, J.C. \& Szkody, P.
	 2005, in ``The Astrophysics of Cataclysmic Variables and Related
	 Objects,'' ASP Conf Ser. 330, eds. J.-M. Hamury \& J.-P. Lasota, 355

\bibitem[Mauche (2006)]{Mauche2006}
         Mauche, C.W. 2006, MNRAS, 369, 1983

\bibitem[Mauche (2009)]{Mauche2009}
         Mauche, C.W. 2009, ApJ, 706, 130

\bibitem[McGowan, Priedhorsky \& Trudolyubov (2004)]{MPT2004}
         McGowan, K.E., Priedhorsky, W.C. \& Trudolyubov, S.P. 2004, ApJ,
	 601, 1100

\bibitem[Meyer \& Meyer-Hofmeister (1994)]{MMH1994}
         Meyer, F. \& Meyer-Hofmeister, E. 1994, A\&A, 288, 175

\bibitem[Miller-Jones et al. (2013)]{MJea2013}
         Miller-Jones, J.C.A., Sivakoff, G.R., Knigge, C., C\"ording, E.G.,
         Templton, M. \& Waagen, E.O. 2013, Science, 340, 950

\bibitem[Mitsuda et al. (2007)]{SUZAKUREF}
         Mitsuda, K. et al. 2007, PASJ, 59, S1

\bibitem[Mohamed \& Podsiadlowski (2007)]{MP2007}
         Mohamed, S. \& Podsiadlowski, Ph. 2007, ASPC, 372, 397.

\bibitem[Mukai (1988)]{M1988}
         Mukai, K. 1988, MNRAS, 232, 175

\bibitem[Mukai \& Shiokawa (1993)]{MS1993}
         Mukai, K. \& Shiokawa, K. 1993, ApJ, 418, 863

\bibitem[Mukai et al. (1997)]{Metal1997}
         Mukai, K., Wood, J.H., Naylor, T., Schlegel, E.M. \& Swank, J.H.
	 1997, ApJ, 475, 812

\bibitem[Mukai (1999)]{Mukai1999}
         Mukai, K. 1999, ASP Conf. Ser. 157, 33

\bibitem[Mukai \& Ishida (2001)]{MI2001}
         Mukai, K. \& Ishida, M. 2001, ApJ, 551, 1024

\bibitem[Mukai et al. (2001)]{Metal2001}
         Mukai, K., Kallman, T., Schlegel, E., Bruch, A., Handler, G. \&
	 Kemp, J. 2001, ASP Conf. Ser. 251, 90

\bibitem[Mukai et al. (2003)]{Metal2003}
         Mukai, K., Kinkhabwala, A., Peterson, J.~R., Kahn, S.~M. \&
         Paerels, F. 2003, \apj, 586, L77

\bibitem[Mukai \& Orio (2005)]{MO2005}
         Mukai, K. \& Orio, M. 2005, ApJ, 622, 602

\bibitem[Mukai et al. (2007)]{Metal2007}
         Mukai, K., Ishida, M., Kilbourne, C. et al. 2007, PASJ, 59, 177

\bibitem[Mukai, Orio \& Della Valle (2008)]{MOD2008}
	 Mukai, K., Orio, M. \& Della Valle, M. 2008, ApJ, 667, 1248

\bibitem[Mukai, Zietsman \& Still (2009)]{MZS2009}
	 Mukai, K., Zietsman, E. \& Still, M. 2009, ApJ, 707, 652

\bibitem[Mukai et al. (2015)]{Metal2015}
         Mukai, K., Rana, V., Bernardini, F. \& de Martino, D. 2015, ApJLett,
	 807, L30

\bibitem[Mukai et al. (2016)]{Metal2016}
         Mukai, K., Luna, G.J.M., Cusumano, G., Segreto, A., Munari, U.,
	 Sokoloski, J.L., Lucy, A.B., Nelson, T. \& Nu\~nez, N.E. 2016,
	 MNRASL, 461, L1

\bibitem[Munari, Dallaporta \& Cherini (2016)]{MDC2016}
         Munari, U., Dallaporta, S. \& Cherini, G. 2016, NewA, 47, 7

\bibitem[Muno et al. (2003)]{MPMetal2003}
         Muno, M.P.,  Baganoff, F.K., Bautz, M.W., Brandt, W.N., Broos, P.S.,
	 Feigelson, E.D., Garmire, G.P., Morris, M.R., Ricker, G.R.
	 \& Townsley, L.K. 2003, ApJ, 589, 225

\bibitem[Muno et al. (2009)]{Metal2009}
         Muno, M.P. et al. 2009, ApJS, 181, 110

\bibitem[M\"urset, Wolff \& Jordan (1997)]{MWJ1997}
         M\"urset, U., Wolff, B. \& Jordan, S. 1997, A\&A, 319, 201

\bibitem[Mushotzky \& Szymkowiak (1988)]{MS1988}
         Mushotzky, R.F. \& Szymkowiak, A.E. 1988, in ``Cooling flows in
	 clusters and galaxies; Proceedings of the NATO Advanced Research
	 Workshop, Cambridge, England, June 22-26, 1987,'' Kluwer Academic
	 Publishers, Dordrecht, Netherlands, p53

\bibitem[Narayan \& Popham (1993)]{NP1993}
         Narayan, R. \& Popham, R. 1993, Nature, 362, 820

\bibitem[Naylor et al. (1988)]{Netal1988}
         Naylor, T., Bath, G.T., Charles, P.A., Hassall, B.J.M.,
	 Sonneborn, G., van der Woerd, H. \& van Paradijs, J. 1988,
	 MNRAS, 231, 237

\bibitem[Naylor \& la Dous (1997)]{NlD1997}
         Naylor, T. \& la Dous, C. 1997, MNRAS, 290, 160

\bibitem[Nelson et al. (2008)]{TNetal2008}
         Nelson, T., Orio, M., Cassinelli, J.P., Still, M., Leibowitz, E.
	 \& Mucciarelli, P. 2008, ApJ, 673, 1067

\bibitem[Nelson et al. (2011)]{TNetal2011}
         Nelson, T., Mukai, K., Orio, M., Luna, G.J.M. \& Sokoloski, J.L.
	 2011, ApJ, 737, A7

\bibitem[Nelson et al. (2012)]{Netal2012}
         Nelson, T., Donato, D., Mukai, K., Sokoloski, J. \& Chomiuk, L.
	 2012, ApJ, 748, A43

\bibitem[Nelson et al. (2014)]{Netal2014}
         Nelson, T., Chomiuk, L., Roy, N., Sokoloski, J.L., Mukai, K.,
	 Krauss, M.I., Moduszewski, A.J., Rupen, M.P. \& Weston, J. 2014,
	 ApJ, 785, A78

\bibitem[Ness et al. (2003)]{JUNetal2003}
         Ness, J.-U., Starrfield, S., Burwitz, V., Wichmann, R.,
	 Hauschildt, P., Drake, J.J., Wagner, R.M., Bond, H.E., Krautter, J.,
	 Orio, M., Hernanz, M., Gehrz, R.D., Woodward, C.E., Butt, Y.,
	 Mukai, K., Balman, S. \& Truran, J.W. 2003, ApJ, 594, L127

\bibitem[Ness et al. (2011)]{JUNetal2011}
         Ness, J.-U., Osborne, J.P., Dobrotka, A., Page, K.L., Drake, J.J.,
	 Pinto, C., Detmers, R.G., Schwarz, G., Bode, M.F., Beardmore, A.P.,
	 Starrfield, S., Hernanz, M., Sala, G., Krautter, J. \& Woodward, C.E.
	 2011, ApJ, 733, A70

\bibitem[Ness et al. (2013)]{JUNetal2013}
         Ness, J.-U., Osborne, J.P., Henze, M., Dobrotka, A., Drake, J.J.,
	 Ribeiro, V.A.R.M., Starrfield, S., Kuulkers, E., Behar, E.,
	 Hernanz, M., Schwarz, G., Page, K.L. Beardmore, A.P. \& Bode, M.F.
	 2013, A\&A, 559, A50

\bibitem[Ness et al. (2015)]{JUNetal2015}
         Ness, J.-U., Beardmore, A.P., Osborne, J.P., Kuulkers, E., Henze, M.,
	 Piero, A.L., Drake, J.J., Dobrotka, A., Schwarz, G., Starrfield, S.,
	 Kretschmar, P., Hirsch, M. \& Wilms, J. 2015, A\&A, 578, A39

\bibitem[Nichols et al. (2007)]{Netal2007}
         Nichols, J.S., DePasquale, J., Kellogg, E., Anderson, C.S.,
	 Sokoloski, J. \& Pedelty, J. 2007, ApJ, 660, 651

\bibitem[Norton \& Watson (1989)]{NW1989}
         Norton, A.J. \& Watson, M.G. 1989, \mnras, 237, 853

\bibitem[Norton, Beardmore \& Taylor (1996)]{NBT1996}
         \edit2{Norton, A.J., Beardmore, A.P., \& Taylor, P. 1996, MNRAS, 280, 937}

\bibitem[Norton et al. (1999)]{Netal1997}
         Norton, A.J., Hellier, C., Beardmore, A.P., Wheatley, P.J.,
	 Osborn, J.P. \& Taylor, P. 1997, MNRAS, 289, 362

\bibitem[Norton et al. (2008)]{Netal2008}
         Norton, A.J., Butters, O.W., Parker, T.L.  \& Wynn, G.A. 2008,
         ApJ, 672, 524

\bibitem[Nucita et al. (2009)]{Netal2009}
	 Nucita, A.A., Maiolo, B.M.T., Carpano, S., Belanger, G.,
	 Coia, D., Guainazzi, M., de Paolis, F. \& Ingrosso, G.,
	 2009, A\&A, 504, 973

\bibitem[Nucita et al. (2011)]{Netal2011}
	 Nucita, A.A., Kuulkers, E., Maiolo, B.M.T., de Paolis, F.,
	 Ingrosso, G., Vetrugno, D 2011, A\&A, 536, A75

\bibitem[Nu\~nez et al. (2016)]{Netal2016}
         Nu\~nez, N.E., Nelson, T., Mukai, K., Sokoloski, J.L.
	 \& Luna, G.J.M. 2016, ApJ, 824, A23

\bibitem[Orio et al. (2010)]{Oetal2010}
         Orio, M., Nelson, T., Bianhini, A., Di Mille, F. \& Harbeck, D.
	 2010, ApJ, 717, 739

\bibitem[Orio et al. (2013)]{Oetal2013}
         Orio, M., Behar, E., Gallagher, J., Bianchini, A., Chiosi, E.,
	 Luna, G.J.M., Nelson, T., Rauch, T. Schaefer, B.E. \&
	 Tofflemire, B. 2013, MNRAS, 429, 1342

\bibitem[Orio, Mukai \& Della Valle (2016)]{OMD2016}
         Orio, M., Mukai, K. \& Della Valle, M. 2016, ATel, 8625


\bibitem[Osborne et al. (1986)]{Oetal1986}
         Osborne, J.P., Bonnet-Bidaud, J.-M., Bowyer, S., Charles, P.A.,
	 Chiappetti, L., Clarke, J.T., Henry, J.P., Hill, G.J., Kahn, S.,
	 Maraschi, L., Mukai, K., Treves, A. \& Vrtilek, S. 1986,
	 MNRAS, 221, 823

\bibitem[Osborne et al. (1987)]{Oetal1987}
         Osborne, J.P., Beuermann, K., Charles, P., Maraschi, L., Mukai, K.
         \& Treves, A. 1987, \apj, 315, L123

\bibitem[Osborne et al. (2011)]{Oetal2011}
         Osborne, J.P., Page, K.L., Beardmore, A.P., Bode, M.F., Goad, M.R.,
	 O'Brien, T.J., Starrfield, S., Rauch, T., Ness, J.-U., Krautter, J.,
	 Schwarz, G., Burrows, D.N., Gehrels, N., Drake, J.J., Evans, A. \&
	 Eyres, S.P.S. 2011, ApJ, 727, A124

\bibitem[Page et al. (2010)]{Petal2010}
         Page, K.L., Osborne, J.P., Evans, P.A., Wynn, G.A., Beardmore, A.P.,
	 Starling, R.L.C., Bode, M.F., Ibarra, A., Kuulkers, E., Ness, J.-U.
	 \& Schwarz, G.J. 2010, MNRAS, 401, 121

\bibitem[Page et al. (2013)]{Petal2013}
         Page, K.L., Osborne, J.P., Wagner, R.M., Beardmore, A.P., Shore, S.N.,
	 Starrfield, S. \& Woodward, C.E. 2013, ApJ, 768, L26

\bibitem[Page et al. (2015)]{Petal2015}
         Page, K.L., Osborne, J.P., Kuin, N.P.M., Henze, M., Walter, F.M.,
	 Beardmore, A.P., Bode, M.F., Darnley, M.J., Delgado, L., Drake, J.J.,
	 Hernanz, M., Mukai, K., Nelson, T., Ness, J.-U., Schwarz, G.J.,
	 Shore, S.N., Starrfield, S. \& Woodward, C.E. 2015, MNRAS, 454, 3108

\bibitem[Pagnotta \& Zurek (2016)]{PZ2016}
         Pagnotta, A. \& Zurek, D. 2016, MNRAS, 458, 1833

\bibitem[Pandel et al. (2002)]{Petal2002}
         Pandel, D., C\'ordova, F.A., Shirey, R.E., Ramsay, G., Cropper, M.,
	 Mason, K.O., Much, R. \& Kilkenney, D. 2002, MNRAS, 332, 116

\bibitem[Pandel et al. (2005)]{DPetal2005}
         Pandel, D., C\'ordova, F.A., Mason, K.O. \& Priedhorsky, W.C.
	 2005, ApJ, 629, 396

\bibitem[Parker, Norton \& Mukai (2005)]{PNM2005}
         Parker, T.L., Norton, A.J. \& Mukai, K. 2005, A\&A, 439, 213

\bibitem[Parmar et al. (1986)]{Petal1986}
	 Parmar, A.N., White, N.E., Giommi, P. \& Gottwald, M. 1986,
	 ApJ, 308, 199

\bibitem[Patterson (1981a)]{P1981a}
         Patterson, J. 1981a, Nature, 292, 810

\bibitem[Patterson (1981b)]{P1981b}
         Patterson, J. 1981b, ApJS, 45, 517

\bibitem[Patterson et al. (1984)]{Petal1984}
         Patterson, J., Beuermann, K., Lamb, D.Q., Fabbiano, G., Raymond, J.C.,
	 Swank, J. \& White, N.E. 1984, ApJ, 279, 785

\bibitem[Patterson \& Raymond (1985a)]{PR1985a}
         Patterson, J. \& Raymond, J.C. 1985a, ApJ, 292, 535

\bibitem[Patterson \& Raymond (1985b)]{PR1985b}
         Patterson, J. \& Raymond, J.C. 1985b, ApJ, 292, 550

\bibitem[Patterson (1994)]{Joe1994}
         Patterson, J 1994, PASP, 106, 209

\bibitem[Patterson et al. (1998)]{Petal1998}
         Patterson, J., Richman, H., Kemp, J. \& Mukai, K. 1998,
	 PASP, 110, 403

\bibitem[Patterson \& Knigge (2017)]{PK2017}
         Patterson, J. \& Knigge, C. 2017, PASP, in preparation

\bibitem[Pietsch et al. (2005)]{Petal2005}
         Pietsch, W., Fliri, J., Freyberg, M.J., Haberl, F.,
	 Riffeser, A. \& Sala, G. 2005, A\&A, 442, 879

\bibitem[Pinto et al. (2012)]{Petal2012}
	 Pinto, C., Ness, J.-U., Verbunt, F., Kaastra, J.S.,
	 Costantini, E. \& Detmers, R.G. 2012, A\&A, 543, A134

\bibitem[Popham \& Narayan (1995)]{PN1995}
         Popham, R. \& Narayan, R. 1995, ApJ, 442, 337

\bibitem[Pratt et al. (1999)]{Petal1999}
         Pratt, G.W., Hassall, B.J.M., Naylor, T. \& Wood, J.H.
	 1999, MNRAS, 307, 413

\bibitem[Pratt et al. (2004)]{Petal2004}
         Pratt, G.W., Mukai, K., Hassall, B.J.M., Naylor, T.
	 \& Wood, J.H. 2004, MNRAS, 348, L49

\bibitem[Predehl \& Schmitt (1995)]{PS1995}
         Predehl, P. \& Schmitt, J.H.M.M. 1995, A\&A, 293, 889

\bibitem[Press \& Rybicki (1989)]{PR1989}
         Press, W.H. \& Rybicki, G.B. 1989, ApJ, 338, 277

\bibitem[Pretorius et al (2007)]{Petal2007}
         Pretorius, M.L., Knigge, C., O'Donoghue, D., Henry, J.P.,
	 Gioia, I.M. \& Mullis, C.R. 2007, MNRAS, 382, 1279

\bibitem[Pretorius \& Knigge (2012)]{PK2012}
         Pretorius, M.L. \& Knigge, C. 2012, MNRAS, 419, 1442

\bibitem[Pretorius, Knigge \& Schwope (2013)]{PKS2013}
         Pretorius, M.L., Knigge, C. \& Schwope, A.D. 2013, MNRAS, 432, 570

\bibitem[Pretorius \& Mukai (2014)]{PM2014}
         Pretorius, M.L. \& Mukai, K. 2014, MNRAS, 442, 2580

\bibitem[Priedhorsky, Marshall \& Hearn (1987)]{PMH1987}
         Priedhorsky, W., Marshall, F.J. \& Hearn, D.R. 1987, A\&A, 173, 95

\bibitem[Pringle \& Webbink (1975)]{PW1975}
         Pringle, J.E. \& Webbink, R.F. 1975, MNRAS, 172, 493

\bibitem[Pringle (1977)]{P1977}
         Pringle, J.E. 1977, MNRAS. 178, 195

\bibitem[Ramsay et al. (1994)]{Retal1994}
         Ramsay, G., Mason, K.O., Cropper, M., Watson, M.G. \&
	 Clayton, K.L. 1994, MNRAS, 270, 692

\bibitem[Ramsay et al. (1998)]{Retal1998}
         Ramsay, G., Cropper, M., Hellier, C. \& Wu, K. 1998, MNRAS, 297, 1269

\bibitem[Ramsay et al. (2001)]{Retal2001}
         Ramsay, G., Poole, T., Mason, K., C\'ordova, F., Priedhorsky, W.,
	 Breeveld, A., Much, R., Osborne, J. Pandel, D., Potter, S.,
	 West, J. \& Wheatley, P. 2001, A\&A, 365, L288

\bibitem[Ramsay \& Cropper (2004)]{RC2004}
         Ramsay, G. \& Cropper, M. 2004, MNRAS, 347, 497

\bibitem[Ramsay et al. (2004)]{Retal2004}
         Ramsay, G., Cropper, M., Wu, K., Mason, K.O., C\'ordova, F.A., \&
	 Priedhorsky, W. 2004, MNRAS, 350, 1373

\bibitem[Ramsay et al. (2005)]{Retal2005}
         Ramsay, G., Hakala, P., Wu, K., Cropper, M., Mason, K.O.,
	 C\'ordova, F.A. \& Priedhorsky, W. 2005, MNRAS, 357, 49

\bibitem[Ramsay et al. (2006)]{Retal2006}
         Ramsay, G., Groot, P.J., Marsh, T., Nelemans, G., Steeghs, D.
	 \& Hakala, P. 2006, A\&A, 457, 623

\bibitem[Ramsay \& Cropper (2007)]{RC2007}
         Ramsay, G. \& Cropper, M. 2007, MNRAS, 379, 1209

\bibitem[Ramsay (2008)]{R2008}
         Ramsay, G., 2008, MNRAS, 384, 687

\bibitem[Ramsay et al. (2008)]{Retal2008}
         Ramsay, G., Wheatley, P.J., Norton, A.J., Hakala, P.
	 \& Baskill, D. 2008, MNRAS, 387, 1157

\bibitem[Ramsay et al. (2016)]{Retal2016}
         Ramsay, G., Sokoloski, J.L., Luna, G.J.M. \& Nu\~nez, N.E.,
	 2016, MNRAS, 461, 3599

\bibitem[Reimer et al. (2008)]{TWRetal2008}
         Reimer, T.W., Welsh, W.F., Mukai, K. \& Ringwald, F.A. 2008,
	 ApJ, 678, 276

\bibitem[Reis et al. (2013)]{Retal2013}
         Reis, R.C., Wheatley, P.J., G\"ansicke, B.T. \& Osborne, J.P.
	 2013, MNRAS, 430, 1994

\bibitem[Revnivtsev et al. (2006)]{MRetal2006}
         Revnivtsev, M., Sazonov, S., Gilfanov, M., Churazov, E.,
	 \& Sunyaev, R. 2006, A\&A 452, 169

\bibitem[Revnivtsev et al. (2009)]{Retal2009}
         Revnivtsev, M., Sazonov, S., Churazov, E., Forman, W.,
	 Vikhlinin, A. \& Sunyaev, R. 2009, Nature, 458, 1142

\bibitem[Rosen et al. (1991)]{Retal1991}
	 Rosen, S.R., Mason, K.O., Mukai, K. \& Williams, O.R. 1991, MNRAS,
	 249, 417

\bibitem[Rosen (1992)]{Rosen1992}
         Rosen, S.R. 1992, MNRAS, 254, 493

\bibitem[Rosen et al. (2016)]{XMMcat}
         Rosen, S.R., Webb, N.A., Watson, M.G. et al. 2016, A\&A, 590, A1

\bibitem[Rothschild et al. (1981)]{Retal1981}
         Rothschild, R.~E., Gruber, D.~E., Knight, F.~K., Matteson, J.~L.,
         Nolan, P.~L., Swank, J.~H., Holt, S.~S., Serlemitsos, P.~J.,
	 Mason, K.~O. \& Tuohy, I.~R. 1981, \apj, 250, 723

\bibitem[Sahai et al. (2015)]{Setal2015}
         Sahai, R., Sanz-Forcada, J., S\'anchez Contreras, C. \&
	 Stute, M. 2015, ApJ, 810, A77

\bibitem[Sambruna et al. (1994)]{Setal1994}
         Sambruna, R.M., Parmar, A.N., Chiappetti, L., Maraschi, L.
	 \& Treves, A. 1994, ApJ, 424, 947

\bibitem[Scargle (1982)]{Scargle1982}
         Scargle, J.D. 1982, ApJ, 263, 835

\bibitem[Schlegel et al. (2014)]{Setal2014}
         Schlegel, E.M., Shipley, H.V., Rana, V.R., Barrett, P.E.
	 \& Singh, K.P. 2014, ApJ, 797, A38

\bibitem[Schmidt \& Stockman (1991)]{SS1991}
         Schmidt, G.D. \& Stockman, H.S. 1991, ApJ, 371, 749

\bibitem[Schreiber, Hameury \& Lasota (2003)]{SHL2003}
         Schreiber, M.R., Hameury, J.-M. \& Lasota, J.-P. 2003,
	 A\&A, 410, 239

\bibitem[Schwarz et al. (2007)]{Setal2007}
         Schwarz, R., Schwope, A.D., Staude, A., Rau, A., Hasinger, G.,
	 Urrutia, T. \& Motch, C. 2007, A\&A, 473, 511

\bibitem[Schwarz et al. (2011)]{Setal2011}
         Schwarz, G.J., Ness, J.-U., Osborne, J.P., Page, K.L., Evans, P.A.,
	 Beardmore, A.P., Walter, F.M., Hilton, L.A., Woodward, C.E., Bode, M.,
	 Starrfield, S. \& Drake, J.J. 2011, ApJS, 197, A31

\bibitem[Schwope et al. (2001)]{Setal2001}
         Schwope, A., Schwarz, R., Sirk, M. \& Howell, S.B. 2001, A\&A, 375, 419

\bibitem[Shaviv \& Wehrse (1986)]{SW1986}
         Shaviv, G. \& Wehrse, R. 1986, A\&A, 159, L5

\bibitem[Silber (1992)]{S1992}
         Silber, A.D. 1992, Ph.D. Thesis, Massachusetts Institute of Technology

\bibitem[Sion et al. (1996)]{Sionetal1996}
         Sion, E.M., Cheng, F.-H., Huang, M., Hubeny, I. \& Szkody, P.
	 1996, ApJ, 471, L41

\bibitem[Smith \& Hughes (2010)]{SH2010}
         Smith, R.K. \& Hughes, J.P. 2010, ApJ, 718, 583

\bibitem[Sokoloski \& Bildsten (1999)]{SB1999}
	 Sokoloski, J.L. \& Bildsten, L. 1999, ApJ, 517, 919

\bibitem[Sokoloski et al. (2006a)]{Setal2006}
         Sokoloski, J.L., Kenyon, S.J., Epsey, B.R., Keyes, C.D.,
	 McCandliss, S.R., Kong, A.K.H., Aufdenberg, J.P., Filippenko, A.V.,
	 Li, W., Brocksopp, C., Kaiser, C.R., Charles, P.A., Rupen, M.P.
	 \& Stone, R.P.S. 2006, ApJ, 636, 1002

\bibitem[Sokoloski et al. (2006b)]{SLMK2006}
         Sokoloski, J.L., Luna, G.J.M., Mukai, K. \& Kenyon, S.J. 2006,
	 Nature, 442, 276

\bibitem[Sokoloski \& Bildsten (2010)]{SB2010}
         Sokoloski, J.L. \& Bildsten, L. 2010, ApJ, 723, 1188

\bibitem[Sokoloski et al. (2017)]{Setal2017}
         \edit1{Sokoloski, J.L., Lawrence, S., Crotts, A.P.S. \& Mukai, K. 2017,
	 arXiv:1072.05898}

\bibitem[Solheim (2010)]{AMCVN}
         Solheim, J.-E. 2010, PASP, 122, 1133

\bibitem[Strohmayer (2004)]{S2004}
         Strohmayer, T.E. 2004, ApJ, 608, L53

\bibitem[Strohmayer (2008)]{S2008}
         Strohmayer, T.E. 2008, ApJ, 679, L109

\bibitem[Suleimanov, Revnivtsev \& Ritter (2005)]{SRR2005}
         Suleimanov, V., Revnivtsev, M. \& Ritter, H. 2005, A\&A, 435, 191

\bibitem[Szkody et al. (1996)]{Setal1996}
         Szkody, P., Long, K.S., Sion, E.M. \& Raymond, J.C. 1996,
	 ApJ, 469, 834

\bibitem[Takei et al. (2009)]{Tetal2009}
         Takei, D., Tsujimoto, M., Kitamoto, S., Ness, J.-U., Drake, J.J.,
	 Takahashi, H. \& Mukai, K. 2009, ApJ, 697, L54

\bibitem[Takei et al. (2015)]{Tetal2015}
         Takei, D., Drake, J.J., Yamaguchi, H., Slane, P., Uchiyama, Y.
	 \& Katsuda, S. 2015, ApJ, 801, A92

\bibitem[Tanaka, Inoue \& Holt (1994)]{TIH1994}
         Tanaka, Y, Inoue, H. \& Holt, S.S. 1994, PASJ, 46, L37

\bibitem[Taylor et al. (1981)]{EXOSATref}
         Taylor, B.G., Andresen, R.D., Peacock, A. \& Zobl, R. 1981,
	 SSRv, 30, 479

\bibitem[Terada et al. (2001)]{Tetal2001}
         Terada, Y., Ishida, M., Makishima, K., Imanari, T., Fujimoto, R.
	 Matsuzaki, K. \& Kaneda, H. 2001, MNRAS, 328, 112

\bibitem[Terada et al. (2004)]{TIM2004}
         Terada, Y., Ishida, M. \& Makishima, K. 2004, PASJ, 56, 523

\bibitem[Thorstensen, L\'epine \& Shara (2008)]{TLS2008}
         Thorstensen, J.R., L\'epine, S. \& Shara, M. 2008, AJ, 136, 2107

\bibitem[Tr\"umper (1983)]{T1983}
         Tr\"umper, J. 1983, AdSpR, 2, 241.

\bibitem[Tueller et al. (2010)]{BAT22}
         Tueller, J., Baumgartner, W.H., Markwardt, C.B., et al. 2010,
	 ApJS, 186, 378

\bibitem[Turner et al. (1989)]{Tetal1989}
         Turner, M.J.L., Thomas, H.D., Patchett, B.E., Reading, D.H.,
         Makishima, K., Ohashi, T., Dotani, T., Hayashida, K., Inoue, H.,
         Kondo, H., Koyama, K., Mitsusa, K., Ogawara, Y., Takano, S.,
         Awaki, H., Tawara, Y \& Nakamura, N. 1989, PASJ, 41, 345

\bibitem[Uthas, Knigge \& Steeghs (2010)]{UKS2010}
         Uthas, H., Knigge, C. \& Steeghs, D. 2010, MNRAS, 409, 237

\bibitem[Valyavin et al. (2014)]{Vetal2014}
         Valyavin, G., Shulyak, D., Wade, G.A., Antonyuk, K., Zharikov, S.V.,
	 Galazutdinov, G.A., Plachinda, S., Bagnulo, S., Fox Machado, L.,
	 Alvarez, M., Clark, D.M., Lopez, L.M., Hiriat, D., Han, I.,
	 Jeon, Y.-B., Zurita, C., Mujica, R., Burlakova, T., Szeifert, T. \&
	 Burenkov, A. 2014, Nature, 515, 88

\bibitem[van den Berg et al. (2006)]{vdBetal2006}
         van den Berg, M., Grindlay, J., Laycock, S., Hong, J., Zhao, P.,
	 Koenig, X., Schlegel, E.M., Cohn, H., Lugger, P., Rich, R.M.,
	 Dupree, A.K., Smith, G.H. \& Strader, J. 2004, ApJ, 647, L135

\bibitem[van Teeseling et al. (1995)]{vTetal1995}
	 van Teeseling, A., Drake, J.J., Drew, J.E., Hoare, M.G.
	 \& Verbunt, F. 1995, A\&A, 300, 808

\bibitem[van Teeseling, Beuermann \& Verbunt (1996)]{vTBV1996}
	 van Teeseling, A., Beuermann, K. \& Verbunt, F. 1996,
	 A\&A, 315, 467

\bibitem[Vaytet, O'Brien \& Bode (2007)]{VOBB2007}
         Vaytet, N.M.H., O'Brien, T.J. \& Bode, M.F. 2007, ApJ 665, 654

\bibitem[Wada et al. (2017)]{Wetal2017}
         Wada, Q., Tsujimoto, M., Ebisawa, K. \& Hayashi, T. 2017,
	 PASJ, in press

\bibitem[Warner (1989)]{Warning}
         Warner, B. 1989, IBVS, 3383

\bibitem[Warner (1995)]{tome}
         Warner, B. 1995, Cataclysmic Variable Stars (Cambridge:
         Cambridge University Press

\bibitem[Warner (2002)]{Warner2002}
         Warner, B. 2002, in ``Classical Nova Explosions,'' AIP Conf. Proc.
	 637, eds. M. Hernanz \& J. Jos\'e, 3

\bibitem[Warwick et al. (1985)]{Wetal1985}
         Warwick, R.S., Turner, M.J.L., Watson, M.G. \& Willingale, R.
	 1985, Nature, 317, 218

\bibitem[Weisskopf, O'Dell \& van Speybroeck (1996)]{CHANREF}
         Weisskopf, M.C., O'Dell, S.L. \& van Speybroeck, L.P 1996,
         Proc. SPIE 2085, 2

\bibitem[Welsh, Horne \& Gomer (1998)]{WHG1998}
         Welsh, W.F., Horne, K. \& Gomer, R. 1998, MNRAS, 298, 285

\bibitem[Weston et al. (2016a)]{JWetal2016a}
         Weston, J.H.S., Sokoloski, J.L., Metzger, B.D., Zheng, Y.,
	 Chomiuk, L., Krauss, M.I., Linford, J.D., Nelson, T.,
	 Mioduszewski, A.J., Rupen, M.P., Finzell, T. \& Mukai, K. 2016a,
	 MNRAS, 457, 887

\bibitem[Weston et al. (2016b)]{JWetal2016b}
         Weston, J.H.S., Sokoloski, J.L., Chomiuk, L., Linford, J.D.,
	 Nelson, T., Mukai, K., Finzell, T., Mioduszewski, A.J., Rupen, M.P.
	 \& Walter, F.M. 2016b, MNRAS, 460, 2687

\bibitem[Wheatley, Mauche \& Mattei (2003)]{WMM2003}
         Wheatley, P.J., Mauche, C.W. \& Mattei, J.A. 2003, MNRAS, 345, 49

\bibitem[Wheatley \& West (2003)]{WW2003}
	 Wheatley, P.J. \& West, R.G. 2003, MNRAS, 345, 1009

\bibitem[Wijngaarden et al. (2016)]{Wetal2016}
         Wijngaarden, M.J.P., Gourdji, K., Oostrum, L.C. \& Henrichs, H.F.
	 2016, Astronomers Telegram, No. 9634

\bibitem[Williams, King \& Brooker (1987)]{WKB1987}
         Williams, G.A., King, A.R. \& Brooker, J.R.E. 1987, MNRAS, 266, 725

\bibitem[Williams (2003)]{Williams2003}
         Williams, G. 2003, PASP, 115, 618

\bibitem[Wilms, Allen \& McCray (2000)]{WAC2000}
         Wilms, J., Allen, A. \& McCray, R. 2000, ApJ, 542, 914

\bibitem[Winkler et al. (2003)]{INTEGRALREF}
         Winkler, C., Courvoisier, T.J.-L., Di Cocco, G., Gehrels, N.,
         Gim\'enez, A., Grebenev, S., Hermsen, W., Mas-Hesse, J.M.,
         Lebrun, F., Lund, N., Palumbo, G.G.C., Paul, J., Roques, J.-P.,
         Schnopper, H., Sch\"onfelder, V., Sunyaev, R., Teegarden, B.,
         Ubertini, P., Vedrenne, G. \& Dean, A.J. 2003, A\&A, 411, L1

\bibitem[Wynn \& King (1992)]{WK1992}
	 Wynn, G.A. \& King, A.R. 1992, MNRAS, 255, 83

\bibitem[Xu, Wang \& Li (2016)]{XWL2016}
         Xu, X.-J., Wang, Q.D. \& Li, X.-D. 2016, ApJ, 818, A136

\bibitem[Yamauchi et al. (2016)]{Yetal2016}
         Yamauchi, S., Nobukawa, K.K., Nobukawa, M., Uchiyama, H.
	 \& Koyama, K. 2016, PASJ, 68, A59

\bibitem[Yoshida, Inoue \& Osaki (1992)]{YIO1992}
         Yoshida, K., Inoue, H. \& Osaki, Y. 1992, PASJ, 44, 537

\bibitem[Yuasa et al. (2010)]{Yetal2010}
         Yuasa, T., Nakazawa, K., Makishima, K., Saitou, K., Ishida, M.,
         Ebisawa, K., Mori, H. \& Yamada, S. 2010, \aap, 520, A25

\bibitem[Yuasa et al. (2012)]{Yetal2012}
         Yuasa, T., Makishima, K. \& Nakazawa, K. 2012, ApJ, 753, A129

\bibitem[Zemko et al. (2014)]{Zetal2014}
         Zemko, P., Orio, M., Mukai, K. \& Shugarov, S. 2014, MNRAS, 445, 869

\bibitem[Zemko, Mukai \& Orio (2015)]{ZMO2015}
         Zemko, P., Mukai, K. \& Orio, M. 2015, ApJ, 807, A61

\end{thebibliography}
\end{document}